\documentclass[twocolumn]{aastex63}
\usepackage{amsmath}
\usepackage{graphics}
\usepackage{savesym}
\savesymbol{tablenum}
\usepackage{siunitx}
\restoresymbol{SIX}{tablenum}
\usepackage{url}
\usepackage{hyperref}

\newcommand{\del}{\partial}

\newcommand{\addsf}[1]{{\color{red} #1}} % added by SF

\renewcommand{\addsf}[1]{{\color{black} #1}} % added by SF
\received{}
\revised{\today}
\accepted{}

\shorttitle{Collapsar leaving BH-disk}
\shortauthors{Fujibayashi et al.}

\begin{document}

\title{Collapse of rotating massive stars leading to black hole formation and energetic supernovae}
% \title{Viscosity-driven explosion of collapse of massive rotating star}

\correspondingauthor{Sho Fujibayashi}
\email{sho.fujibayashi@aei.mpg.de}

\author[0000-0001-6467-4969]{Sho Fujibayashi}
\affiliation{Max-Planck-Institut f\"ur Gravitationsphysik (Albert-Einstein-Institut), Am M\"uhlenberg 1, D-14476 Potsdam-Golm, Germany}

\author[0000-0002-2648-3835]{Yuichiro Sekiguchi}
\affiliation{Center for Gravitational Physics and Quantum Information, Yukawa Institute for Theoretical Physics, Kyoto University, Kyoto, 606-8502, Japan}
\affiliation{Department of Physics, Toho University, Funabashi, Chiba 274-8510, Japan}

\author[0000-0002-4979-5671]{Masaru Shibata}
\affiliation{Max-Planck-Institut f\"ur Gravitationsphysik (Albert-Einstein-Institut), Am M\"uhlenberg 1, D-14476 Potsdam-Golm, Germany}
\affiliation{Center for Gravitational Physics and Quantum Information, Yukawa Institute for Theoretical Physics, Kyoto University, Kyoto, 606-8502, Japan}

\author[0000-0002-4759-7794]{Shinya Wanajo}
\affiliation{Max-Planck-Institut f\"ur Gravitationsphysik (Albert-Einstein-Institut), Am M\"uhlenberg 1, D-14476 Potsdam-Golm, Germany}
%\affiliation{Department of Engineering and Applied Sciences, Faculty of Science and %Technology, Sophia University, 7-1 Kioicho, Chiyoda-ku, Tokyo, 102-8554, Japan}
%\affiliation{Interdisciplinary Theoretical and Mathematical Science (iTHEMS) Research Group, RIKEN, Wako, Saitama, 351-0198, Japan}

\begin{abstract}
We explore a possible explosion scenario resulting from core collapses of rotating massive stars that leave a black hole by performing radiation-viscous-hydrodynamics simulations in numerical relativity.
We take moderately and rapidly rotating compact pre-collapse stellar models with the zero-age-main-sequence mass of $9M_\odot$ and $20M_\odot$ based on stellar-evolution calculations as the initial conditions.
We find that the viscous heating in the disk formed around the central black hole powers an outflow.
The moderately rotating models predict a small ejecta mass of order $0.1M_\odot$ and explosion energy of $\lesssim 10^{51}$\,erg.
Due to the small ejecta mass, these models may predict a short-timescale transient with the rise time 3--5\,d.
It can lead to a bright ($\sim10^{44}$\,erg/s) transient like superluminous supernovae in the presence of a dense massive circum-stellar medium.
For hypothetically rapidly rotating models that have a high mass infall rate onto the disk, the explosion energy is $\gtrsim \SI{3e51}{erg}$, which is comparable to or larger than that of typical stripped-envelope supernovae, indicating that a fraction of such supernovae may be explosions powered by black-hole accretion disks.
The explosion energy is still increasing at the end of the simulations with a rate of $>10^{50}$\,erg/s, and thus, it may reach $\sim10^{52}$\,erg.
The nucleosynthesis calculation shows that the mass of $^{56}$Ni amounts to $\gtrsim 0.1M_\odot$, which, together with the high explosion energy, may satisfy the required amount for broad-lined type Ic supernovae.
Irrespective of the models, the lowest value of the electron fraction of the ejecta is $\gtrsim 0.4$; thus, the synthesis of heavy $r$-process elements is not found in our models.
% \addms{For rapidly rotating progenitor models, not only high explosion energy is found but also the Nickel production amounts to 
% $\agt 0.1 M_\odot$, satisfying the required amount for hypernovae.}
%\addms{Implication to the latest gamma-ray burst event GRB221009 is also discussed.}

\end{abstract}


\section{Introduction}
\label{sec:intro}
At the final evolution stage of massive stars, their iron cores become gravitationally unstable and collapse. After the core bounce due to the formation of a proto-neutron star, a shock wave is formed on its surface and propagates outward.
The shock wave then stalls primarily because of the photodissociation of heavy (iron-group) nuclei in the infalling matter swept by the shock.
In the standard neutrino-driven delayed-explosion scenario, the shock is revived by the heating of neutrinos emitted by the proto-neutron star~(e.g., \citealt{Janka2012b}).

If the collapsing star has a very compact core, the proto-neutron star is likely to collapse into a black hole with no successful shock revival due to the strong ram pressure by the matter  infall~(e.g., \citealt{Oconnor2011apr}; but see \citealt{Burrows2019}).
Even in this case, there is a possibility of the explosion if a massive disk is formed around the black hole due to the rotation of the progenitor star.
In the disk, the magnetorotational instability (MRI) could amplify the magnetic field, developing a turbulent state inside the disk (e.g., \citealt{Balbus1991a}).
The turbulent motion then induces an effective viscosity in the disk, which governs the evolution of the disk through the angular momentum transport and heating.
The viscous heating rate is estimated by
\begin{align}
L_\mathrm{vis}&\sim \nu \Omega^2 M_\mathrm{disk}\notag\\
&\approx \SI{7e51}{erg/s}\, \biggl(\frac{\alpha_\mathrm{vis}}{0.03}\biggr)\biggl(\frac{c_s}{10^9\,\mathrm{cm/s}}\biggr)^2\notag\\
&~~~\times\biggl(\frac{M_\mathrm{BH}}{10M_\odot}\biggr)^{1/2}\biggl(\frac{r_\mathrm{disk}}{10^7\,\mathrm{cm}}\biggr)^{-3/2}\biggl(\frac{M_\mathrm{disk}}{0.1M_\odot}\biggr),\label{eq1}
\end{align}
where $M_\mathrm{BH}$ is the mass of the black hole, $M_\mathrm{disk}$ and $r_\mathrm{disk}$ are the mass and typical radius of the disk, $c_s$ is the sound speed, and $\Omega$ is the local angular velocity. Equation~(\ref{eq1}) is a radially integrated form of viscous heating rate per unit area. See, e.g., \cite{Frank2002}. 
Here, we assumed a Keplerian rotation and the Shakura-Sunyaev-type alpha-viscosity model for the kinetic viscous coefficient $\nu=\alpha_\mathrm{vis}c_s H$ \citep{Shakura1973a} with the disk scale height $H=c_s/\Omega$.
$\alpha_\mathrm{vis}$ is the so-called alpha parameter, which is likely to be of order $10^{-2}$ in the presence of MRI turbulence (e.g.,  \citealt{Balbus:1998ja,Hawley:2013lga,Suzuki:2013rka,Shi:2015mvh,kiuchi2018a,Held:2022gds}). 
Accretion disks formed around black holes in the collapsar scenario \citep{Macfadyen1999} often become a neutrino-dominated accretion disk (e.g., \citealt{Kohri:2002kz,DiMatteo:2002iex}) in which internal energy generated by the viscous heating is released primarily by the neutrino emission.
However, the latest studies for the systems of a black hole and a compact accretion disk have shown that in the late stage of the viscous evolution, the neutrino cooling rate drops due to the viscous angular momentum transport and subsequent expansion of the disk \citep{Fernandez2013a,Just2015a,Fujibayashi2020a,Fujibayashi2020b,Just2022jan}.
In such a stage, the viscous heating can be used for launching a strong outflow from the disk in a viscous timescale estimated by
%For the advection-dominated accretion flow \addms{!!! COMMENT: This is wrong. !!!}, most of the internal energy generated by the viscous heating can be used for the expansion of the disk, which can subsequently lead to an outflow. Thus, the viscous heating rate estimated above and the viscous timescale of the disk
\begin{align}
t_\mathrm{vis} \sim \frac{r_\mathrm{disk}^2}{\nu} &\approx \SI{4}{s}\,\biggl(\frac{\alpha_\mathrm{vis}}{0.03}\biggr)^{-1}\biggl(\frac{c_s}{10^9\,\mathrm{cm/s}}\biggr)^{-2}\notag\\
&~~~\times\biggl(\frac{M_\mathrm{BH}}{10M_\odot}\biggr)^{1/2}\biggl(\frac{r_\mathrm{disk}}{10^7\,\mathrm{cm}}\biggr)^{1/2}. 
\end{align}
%indicate that a viscosity-driven explosion may be possible with 
The order of magnitude of the explosion energy generated by the viscous heating is estimated by $\sim L_\mathrm{vis}t_\mathrm{vis}$, which is comparable to or even larger than that of typical supernovae ($\sim 10^{51}$\,erg) for plausible values of $M_\mathrm{BH}$, $M_\mathrm{disk}$, and $r_\mathrm{disk}$.
This motivates us to explore a scenario of the explosion from a massive accretion disk around a spinning black hole formed during the rotating stellar core collapse.

The sub-relativistic outflow from the disk is of importance for several aspects.
First, it can be an essential energy source to power a supernova-like explosion associated with long-duration gamma-ray bursts~(e.g., \citealt{Macfadyen1999}, \citealt{Pruet2003apr}, \citealt{Nagataki2007apr}, \citealt{Surman2006jun}, and \citealt{Hayakawa2018feb}), for which the promising central engine of the gamma-ray bursts is likely to be a spinning black hole penetrated by a strong magnetic field.
In this scenario, however, we additionally need the supernova component.
\cite{Eisenberg2022nov} suggested, based on their simulations, that the observationally inferred velocity distribution of the supernova-component is not likely reproduced only by the relativistic jet.
This indicates that there has to be another energy source to drive the supernova component in addition to the relativistic jet accounting for the gamma-ray burst.
% Blandford-Znajek mechanism \citep{Blandford1977}.
\cite{Kohri2005aug} applied a disk explosion scenario to normal supernovae by analytically solving a stationary neutrino-cooled accretion-disk model and indicated that an energetic outflow could be driven from the collapse of rotating stars when the accretion flow is advection-dominated.\footnote{
Note that it is still possible to achieve successful neutrino-driven explosion in proto-neutron star phases even for collapses for compact progenitor stellar cores (see a series of work \citealt{Obergaulinger2020,Aloy2021jan,Obergaulinger2021, Obergaulinger2022may} and also \citealt{fujibayashi2021oct}).}

% A series of work by Obergaulinger and Aloy \citep{Obergaulinger2020,Aloy2021jan,Obergaulinger2021, Obergaulinger2022may} numerically studied possible fates of the collapse of compact massive stars by magnetohydrodynamics simulations with neutrino radiation transport.
% They showed that, depending on the density, angular momentum, and magnetic field profiles, some of their models exhibit explosions either by neutrino heating or by the magnetorotational effect in proto-neutron star phases.

Second, it has been speculated that the matter in the disk outflow could be neutron-rich, and thus, the outflow may be a site for $r$-process nucleosynthesis \citep{Surman2006jun,Pruet2003apr,Kohri2005aug}.
\cite{Siegel2019may} suggested, based on their magnetohydrodynamics simulations with an approximate neutrino treatment in a fixed black-hole spacetime, that neutron-rich matter may be ejected from the disk cooled by neutrinos and the heavy nuclei up to third peak of the $r$-process elements may be synthesized.
In a similar setup but with Monte-Carlo neutrino transfer, however, \cite{Miller2020oct} pointed out that the electron fraction ($Y_\mathbf{e}$) of the ejecta is higher than 0.3, and thus, nuclei only up to the second peak of $r$-process elements are synthesized.
\cite{Just2022aug} performed viscous hydrodynamics simulations in Newtonian gravity with general relativistic corrections incorporating moment-based neutrino radiation transfer for the collapse of a rotating massive star and showed that the outflow from the disk formed around the black hole has the electron fraction higher than 0.4.
Therefore, the speculation in this field has not converged yet.
Moreover, no fully general relativistic work, which self-consistently takes into account the self gravity of the collapsing star and the formed black hole in general relativistic manner, has been carried out.
Obviously more detailed studies are required.

Third, recent high-cadence transient surveys have shown that there is a variety of optical transients that are not canonical supernovae.
Those with timescales of a few days, which are much shorter than that of normal supernovae ($>10$\,d), are such examples (e.g., \citealt{Drout2014oct}, \citealt{Prentice2018sep}, and \citealt{Tampo2020may}).
Despite intensive photometric and spectral observations, the progenitors of the transients different from the canonical supernovae are still not clear.
There are several scenarios in which a collapse of a massive star leading to black-hole formation plays a central role (e.g., \citealt{Margutti2019feb}, \citealt{Perley2019mar}).
However, the previous studies are limited only to the ones based on simplified models (see \citealt{Piran2019feb} and \citealt{Gottlieb2022jul} for a recent simulation-based model).
Thus, it is important to provide predictions based on reliable numerical simulations for interpreting the observation and confirming the origins of the mysterious transients.

Motivated by these current situations, in this paper, we explore the long-term evolution of the collapse of rotating massive stars by fully general relativistic radiation-viscous-hydrodynamics simulations with an approximate neutrino transfer.
% \delsf{The numerical simulations are performed employing two types of the initial condition.
% The first one is based on rotating pre-collapse stellar models calculated by \cite{Aguilera-Dena2020oct}.
% In addition, we employ two rapidly rotating models to explore the effect of the high angular momentum of the system.
% One is based on a non-rotating pre-collapse stellar model evolved from a $20M_\odot$ helium star \citep{Takahashi2018}, for which we add a hypothetically rapid stellar rotation.
% The second one is prepared by artificially enhancing the angular velocity for one of the models of \cite{Aguilera-Dena2020oct}.
% For these initial conditions, we explore how the disk is formed and evolved around the black hole during rotating stellar collapse, leading to a supernova-like stellar explosion, and these dependence of the processes on the degree of the stellar rotation.}

This paper is organized as follows.
In \S~\ref{sec:method}, our method of the simulations is briefly described.
We also introduce the pre-collapse stellar models which we employ from the stellar evolution calculations.
Then, in \S~\ref{sec:result}, the results of our numerical-relativity simulations and the nucleosynthesis calculations are presented.  
We discuss the possible optical transients and implications to broad-lined type Ic supernovae and gamma-ray bursts based on our results in \S~\ref{sec:discussion}.
We also discuss the possible production of light $r$-process nuclei and effects on the optical transient.
\S~\ref{sec:summary} is devoted to a summary.
Throughout this paper, $G$, $c$, and $k_\mathrm{B}$ denote the gravitational constant, speed of light, and Boltzmann's constant, respectively.

\section{Method}
\label{sec:method}

\subsection{Numerical code}

Numerical-relativity simulations are performed with our latest axisymmetric neutrino-radiation viscous-hydrodynamics code.
The detail of the code is described in \cite{fujibayashi2017a,Fujibayashi2020c}.
In this code, Einstein's equation is solved in the original version of the Baumgarte-Shapiro-Shibata-Nakamura formalism \citep{shibata1995a,baumgarte1998a} with a constraint propagation prescription to make the constraint violation to propagate outward~\citep{Hilditch2013a}.
A dynamical gauge condition described in \cite{fujibayashi2017a} is employed.
To impose the axisymmetry for the geometrical variables, the so-called cartoon method~\citep{Alcubierre2001a,Shibata2000a} with the fourth-order Lagrange interpolation is implemented.

The neutrino radiation transfer equations are approximately solved using a leakage scheme together with the truncated moment formalism (\citealt{fujibayashi2017a}; see also \citealt{Sekiguchi2010a}). In this formulation, 
the neutrino field is split into two components; \textit{trapped} and \textit{free-streaming} neutrinos. The trapped neutrinos are assumed to be tightly coupled with the fluid and have the same local temperature and velocity as those of the fluid.
This component is treated as a part of the fluid and contributes to the internal energy and pressure.
It becomes the free-streaming component with the generation rate controlled by the local diffusion rate of neutrinos.
The free-streaming neutrinos are assumed to obey radiation transfer equations, which are  solved by a truncated moment formalism with the M1 closure relation \citep{Thorne1981a,shibata2011a}.
Following our previous work \citep{Fujibayashi2020a,Fujibayashi2020b,Fujibayashi2020c}, we solve the equations for the frequency-integrated energy and momentum density for the three neutrino radiation fields (electron, electron anti-, and other neutrinos). 

\begin{table*}[t]
    \centering
    \caption{Model description. The columns provide from left to right: model name, progenitor model name, mass of the progenitor star, angular velocity profile, equation of state, compactness just prior to the collapse, innermost grid spacing, the values of $\delta$ and $N$, and the location of  the outer boundaries along each axis. The last column shows a model for the black hole-disk system.
    }
    \begin{tabular}{lllccccccc}
    \hline\hline
        Model & Progenitor & Progenitor star & $\Omega$ profile & EOS &$\xi_{2.5}$ & $\Delta x_0$ (m) & $\delta$ & $N$ & $L$ (cm) \\
        \hline
        \texttt{AD09x1} & \texttt{AD09} & $M_\mathrm{ZAMS}=9M_\odot$ & Original $\times 1$ & DD2 & 0.68 & 175 & 0.01 & 975 & \SI{1.0e10}{}\\
        \texttt{AD20x1} & \texttt{AD20} & $M_\mathrm{ZAMS}=20M_\odot$ & Original $\times 1$ & DD2 & 0.66 & 175 & 0.01 & 975 & \SI{1.0e10}{} \\
        \texttt{AD20x2} & \texttt{AD20} & & Original $\times 2$ & & \\
%        \texttt{T20}    & \texttt{T20}$^\mathrm{a}$  & $M_\mathrm{He}=20M_\odot$ & Eq.~\eqref{eq:omega_ana} & SFHo & 0.74 & 150 & 0.01 & 991 & \SI{1.0e10}{}\\
        % \hline
        % \texttt{T20l}   & \texttt{T20}$^\mathrm{a}$   & lower-res. \texttt{T20}    &  & SFHo & 0.74 & 150 & 0.013 & 807 & \SI{1.0e10}{} \\
        % \texttt{AD09x1l} & \texttt{AD09}$^\mathrm{b}$ & lower-res. \texttt{AD09x1} &  & DD2 & 0.68 & 175 & 0.013 & 793 & \SI{1.0e10}{}\\
        \hline
        \texttt{BHdisk} & -- & $10M_\odot$ BH-$3M_\odot$ disk & -- & DD2 & -- & 220 & 0.01 & 801 & \SI{3.6e9}{}\\
        
        \hline
    \end{tabular}\\
    % $^\mathrm{a}$\cite{Takahashi2018},
%    $^\mathrm{b}$\cite{Aguilera-Dena2020oct}\addsw{*****this may not be needed*****}
    \label{tab:model}
\end{table*}

% viscosity switched on at:
% T20: 2.044E+06 = 3.28 s
% AD09x1: 7.096E+06 = 11.40 s
% AD20x1: 1.1046635938E+07 = 17.74 s
% AD20x2: 5.517E+06 = 8.86 s

The viscous hydrodynamics equations are solved using the formulation described in \cite{shibata2017b}, in which the energy-momentum tensor is written as
\begin{align}
T_{\mu\nu} = \rho h u_\mu u_\nu + Pg_{\mu\nu} + \nu \rho h \tau^0_{\mu\nu},
\end{align}
where $\rho:=m_\mathrm{u}n_\mathrm{b}$ is the rest-mass density, $h=c^2+\varepsilon+P/\rho$ is the specific enthalpy with the specific internal energy $\varepsilon$ and pressure $P$, $\nu$ is the kinematic viscous coefficient, and $g_{\mu\nu}$ is the spacetime metric. 
The viscous tensor $\tau^0_{\mu\nu}$ is a symmetric tensor that satisfies $\tau^0_{\mu\nu}u^\mu = 0$ \citep{Israel1979a}, and it is determined by the following equation 
\begin{align}
\mathcal{L}_u \tau^0_{\mu\nu} = -\zeta (\tau^0_{\mu\nu}-\sigma_{\mu\nu}), 
\end{align}
where $\mathcal{L}_u$ denotes the Lie derivative with respect to $u^\mu$ and $\zeta$ is a coefficient, for which we set $\zeta^{-1}=O(10\mu {\rm s})$.
Assuming the form of the shear tensor as
\begin{align}
\sigma_{\mu\nu} = \nabla_\mu u_\nu+\nabla_\nu u_\mu = \mathcal{L}_u (g_{\mu\nu}+u_\mu u_\nu),
\end{align}
where $\nabla_\mu$ is the covariant derivative with respect to $g_{\mu\nu}$, 
we obtain the evolution equation for $\tau_{\mu\nu} := \tau^0_{\mu\nu} - \zeta (g_{\mu\nu}+u_\mu u_\nu)$ as
\begin{align}
\mathcal{L}_u \tau_{\mu\nu} = -\zeta \tau^0_{\mu\nu}.
\end{align}
We only need to solve the spatial part of $\tau_{\mu\nu}$ because of the presence of the condition $\tau^0_{\mu\nu}u^\mu = 0$.
The spatial part obeys the following evolution equation (in Cartesian coordinates):
\begin{align}
&\del_t (\rho u^t \sqrt{-g} \tau_{ij}) +  \del_k (\rho u^k \sqrt{-g} \tau_{ij}) \notag\\
&+ \rho u^t \sqrt{-g}(\tau_{ik}\del_jv^k+\tau_{jk}\del_iv^k) = -\rho\sqrt{-g}\zeta \tau^0_{ij}.
\end{align}

The effective viscosity in the disk is believed to arise as a result of the turbulence induced by magnetohydrodynmical instabilities such as MRI \citep{Balbus1991a,Balbus:1998ja} and Kelvin-Helmholtz instability (e.g., \citealt{Obergaulinger2010jun}).
Following \cite{Shakura1973a}, we define the viscous coefficient by
\begin{align}
%\nu = \alpha_\mathrm{vis} c_s H_\mathrm{tur},
\nu = c_s \ell_\mathrm{tur},
\end{align}
where $\ell_\mathrm{tur}$ is the mixing length scale (or the largest eddy size) in the turbulence. 
In the alpha disk model, $\ell_\mathrm{tur}$ is written as $\alpha_\mathrm{vis}H$ \citep{Shakura1973a}. 
In this study, we assume that $\ell_\mathrm{tur}$ is proportional to the size of the black hole as
\begin{align}
\ell_\mathrm{tur} = 0.03\times \frac{2GM_\mathrm{BH}}{c^2},\label{eq:ellvis} 
\end{align}
where the black-hole mass, $M_\mathrm{BH}$, continuously increases with time due to the matter accretion.
Thus, we employ a time-varying form of $\ell_\mathrm{tur}$.
By the above definition, the mixing length scale becomes  $\ell_\mathrm{tur}\approx \SI{0.9}{km} (M_\mathrm{BH}/10M_\odot)$.
Since the disk scale height $H$ is larger than $2GM_\mathrm{BH}/c^2$ for most parts of the disk around the black hole, Equation~(\ref{eq:ellvis}) implies that we assume $\alpha_\mathrm{vis} \leq 0.03$, i.e., a conservative value of $\alpha_\mathrm{vis}$.
Even for such a conservative value of $\alpha_\mathrm{vis}$, we will find a significant effect in \S~\ref{sec:result}.
\addsf{This spatially constant mixing length scale, Eq.~\eqref{eq:ellvis}, may lead to a smaller kinetic viscous coefficient than that employed in \cite{Just2022aug} (see also \citealt{Just2022jan}).}

We note that the viscosity is incorporated just prior to the formation of the disk around the black hole.
\addsf{Specifically, we first perform simulations without viscosity until the disks are formed.
We then go back to the time slice prior to the disk formation and rerun the simulation with viscosity.
We check that there are any significant differences of the density and angular velocity profiles for infalling matter between the results of the simulations with and without viscosity.\footnote{The infalling matter at the formation of the disk is that from carbon-oxygen-neon layer of the star, which does not have significant differential rotation. In this sense, turning on the viscosity does not have a significant effect during the collapse.}}
%We note that the viscosity is incorporated just prior to the formation of the disk around the black hole. That is, for the formation and early growth stages of the black hole, we do not consider the viscous effects.

The grid structure is the same as in the 2D simulations recently performed with the same code~\citep{Fujibayashi2023jan}, in which the cylindrical coordinates $(R,z)$ are employed.
In the inner cylindrical region of $R \leq 100 \Delta x_0$ and $z \leq 100 \Delta x_0$, a uniform grid with the grid spacing of $\Delta x_0$ is prepared, while in the outer region, a non-uniform grid with an increase rate of the grid spacing of $1+\delta$ is prepared.
The value of $\delta$, grid number $N$ for each axis, and location of the outer boundaries along each axis (denoted by $L$) are listed in Table~\ref{tab:model}.
We assume the plane symmetry with respect to the $z=0$ plane (the equatorial plane).

\subsection{Models} \label{subsec:model}

\begin{figure}
\epsscale{1.17}
\includegraphics[width=0.48\textwidth]{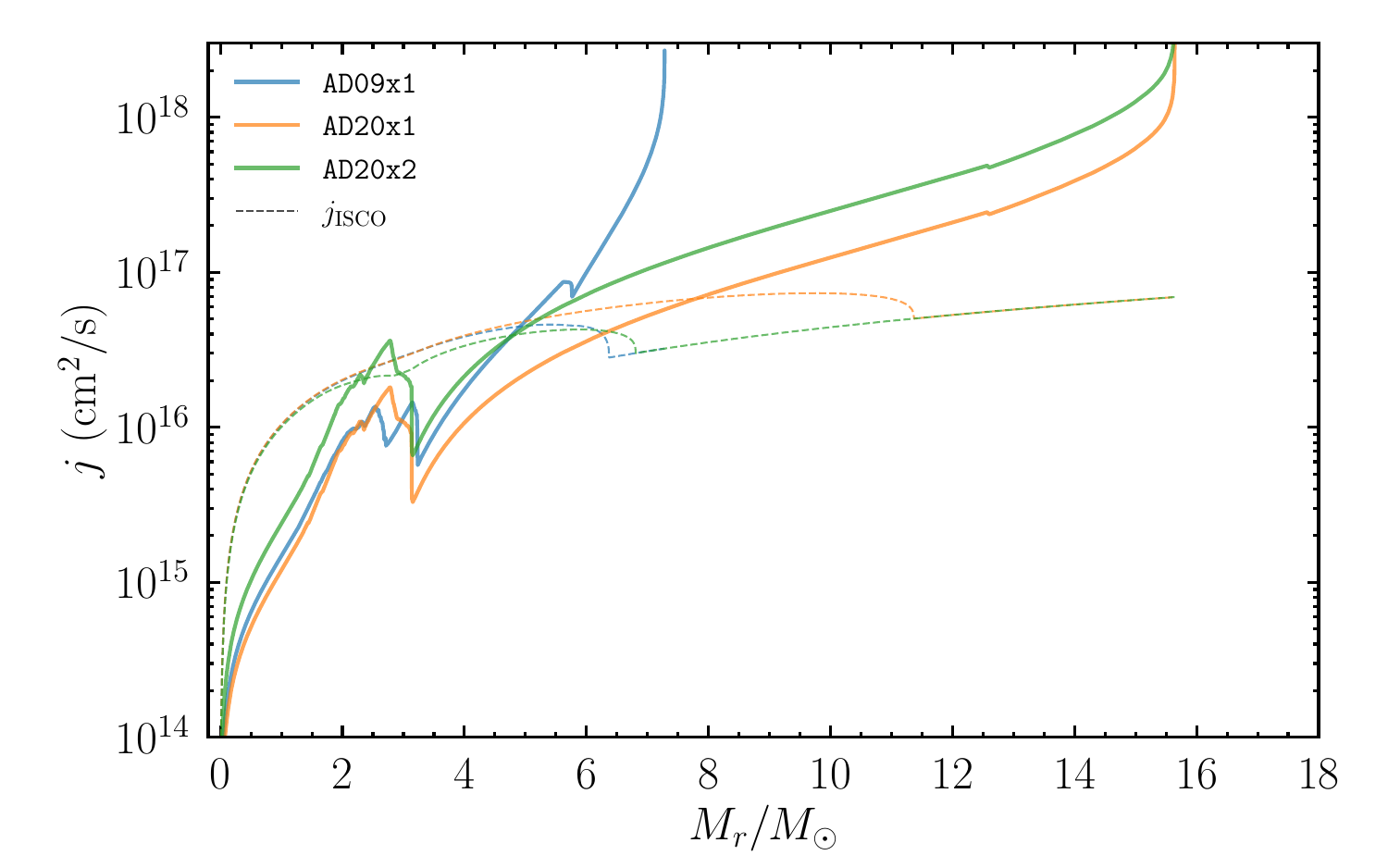}
\caption{
Specific angular momentum (solid curves) and that for innermost stable circular orbits (dashed curves) along the equatorial direction as functions of enclosed mass for each model.
}
\label{fig:ang}
\end{figure}

One of the theoretically accepted central engines of long-duration gamma-ray bursts is the system of a spinning black hole with a surrounding accretion disk~\citep[e.g.,][]{Woosley1993}.
For the formation of a massive accretion disk around the black hole, the progenitor star has to be rapidly rotating.
The progenitor models of \cite{Aguilera-Dena2020oct} may be promising for such a scenario.
% We employed in total four pre-collapse states of massive stars evolved in stellar evolution calculations as the initial conditions.
We employ, from their work, two of the rotating stars with zero-age-main-sequence (ZAMS) masses of 9 and 20$M_\odot$  (hereafter \texttt{AD09} and \texttt{AD20}, respectively).
Because of nearly chemically homogeneous evolution, the pre-collapse stars have very massive cores that are very compact.
The compactnesses at the radius with its enclosed mass $2.5M_\odot$ (referred to as $r_{2.5M_\odot})$, $\xi_{2.5}:= 2.5/(r_{2.5M_\odot}/\SI{1000}{km})$, are 0.68 and 0.66 for \texttt{AD09} and \texttt{AD20}, respectively.
This suggests that they are likely to form a black hole with no neutrino-driven explosion if it is non-rotating (see, e.g., \citealt{Oconnor2011apr} but see \citealt{Burrows2019}.)

It should be noted that the angular momentum transport via the convection, circulation, and magnetohydrodynamical interactions is taken into account only in an approximate way in current stellar evolution studies.
The original rotation profile is applied for models \texttt{AD09x1} and \texttt{AD20x1}, while for model \texttt{AD20x2}, the original angular velocity is doubled to investigate a possible case in which the rotation of the star is even faster.

Figure~\ref{fig:ang} shows the specific angular momentum along the equatorial direction as a function of enclosed mass for the models employed in this paper.
The specific angular momentum at the innermost stable circular orbit for the black hole with the enclosed mass and angular momentum, $j_\mathrm{ISCO}$, is also plotted in the dashed curves.
For each progenitor model, the enclosed mass at which the solid and dashed curves cross approximately gives the black-hole mass at which the infalling matter starts forming an accretion disk around the black hole.
This figure indicates that disks are likely to be formed when the black holes grow to 4.7, 4.9, and 7.8$M_\odot$ for models \texttt{AD20x2}, \texttt{AD09x1}, and \texttt{AD20x1}, respectively.
In \S~\ref{sec:result}, we will confirm that this prediction is approximately correct.

At the time of the disk formation, the mass infalling rates for model \texttt{AD20x2} are higher than those for \texttt{AD09x1} and \texttt{AD20x1}.
This makes a difference in the evolution process of the disk and the onset timing of the outflow from the disk as found in \S~\ref{subsec:high-mdot} and \S~\ref{subsec:low-mdot}. %\delsf{This finding motivates us to perform an additional simulation, for which the results are summarized in Appendix \ref{appA}.}

In several previous studies, the mass ejection and nucleosynthesis calculation were carried out by simply modeling collapsar remnants as a black hole-disk system~\citep{Siegel2019may,Miller2020oct}. 
The evolution process of the black hole-disk system as a result of stellar collapse may be different from that of an isolated disk around a black hole.
To clarify the possible differences, we also perform a simulation for a system of a spinning black hole and a massive disk (model \texttt{BHdisk}) as in our previous studies \citep{Fujibayashi2020a,Fujibayashi2020b}.
The initial condition is constructed using a code of \citet{Shibata2007}. 
The masses of the black hole and disk are 10 and 3$M_\odot$, respectively, and the dimensionless spin of the black hole is $\approx 0.6$. The inner and outer edges of the disk are chosen to be $4GM_\mathrm{BH}/c^2 \approx \SI{59}{km}$ and $400GM_\mathrm{BH}/c^2 \approx \SI{5900}{km}$, respectively.
The constant entropy per baryon of $s=7k_\mathrm{B}$ is simply assumed.

In this paper, we employ a tabulated equation of state (EOS) referred to as DD2 (\citealt{banik2014a}).
We extended the table down to low-density ($\rho\approx\SI{0.17}{g/cm^3}$) and low-temperature ($k_\mathrm{B}T=10^{-3}$\,MeV); see \cite{Hayashi2022} for the procedure.
% On the other hand, we employed DD2 EOS for \texttt{AD09} and \texttt{AD20} progenitors.
% Since the maximum mass of spherical neutron stars in the SFHo EOS is smaller than that in the DD2 EOS, a black hole is formed slightly earlier (i.e., the black-hole mass is smaller at its formation) in the model with the SFHo EOS.
% However, besides this difference, the dependence of the results on the EOS is weak because the maximum density of the disk is always smaller than the nuclear saturation density.

To evolve black holes in a good precision, the radius of the apparent horizon has to be resolved well (for the dependence of the black-hole mass and spin on the  grid resolution, see, e.g., \citealt{Fujibayashi2020a}).
Employing a \textit{stiff} EOS like DD2, which predicts a larg maximum mass of neutron star, is advantageous to numerically resolve black holes in a good accuracy (for a given computational resource), because the black-hole mass (i.e.,  the radius of the apparent horizon) at its formation is larger for a stiffer EOS.
Thus, for the present study, the DD2 EOS would be a better choice to save computational resources.

\begin{table*}[t]
    \centering
    % \scriptsize
    \footnotesize
    \caption{Main results. The columns provide from left to right: model name, post-bounce time of black hole formation and explosion, ejecta mass, explosion energy, mass of the matter with maximum temperature higher than \SI{5}{GK} ($\SI{1}{GK}=10^9$\,K), mass of $^{56}$Ni, estimated peak bolometric luminosity, and rise time of bolometric light curve.
    For $M_\mathrm{ej}$ and $E_\mathrm{exp}$, there are two columns with different extraction radius.
    The values in parentheses for columns of $M_\mathrm{ej}$ and $E_\mathrm{exp}$ are, respectively, the mass and binding energy of stellar matter above the extraction radius. Those for columns of $L_\mathrm{peak}$ and $t_\mathrm{rise}$ are values for the case in which a half of the mass and binding energy above the extraction radius are considered.
    }
    \begin{tabular}{lrrrrrrrrrr}
    \hline\hline
    Model & $t_\mathrm{BH}$ & $t_\mathrm{exp}$ & \multicolumn{2}{r}{$M_\mathrm{ej}$} & \multicolumn{2}{r}{$E_\mathrm{exp}$} & $M_\mathrm{>5\,GK}$ &$M_\mathrm{Ni}$ & $L_\mathrm{peak}$ & $t_\mathrm{rise}$ \\
    & (s)  & (s) & \multicolumn{2}{r}{($M_\odot$)} & \multicolumn{2}{r}{($10^{51}$\,erg)} &($M_\odot$) & ($M_\odot$) & ($10^{42}$\,erg/s) & (days) \\
    & & & $r_\mathrm{ext}=\SI{1e9}{cm}$ & $\SI{2e9}{cm}$ & $\SI{1e9}{cm}$ & $\SI{2e9}{cm}$ \\
    \hline
%    \multicolumn{4}{l}{Standard resolution models}\\
    \texttt{AD09x1} & 0.87 & 13.2 & 0.08   (1.6)  & 0.12   (1.5) & 0.57  ($-0.21$) & 0.53   ($-0.19$) & 0.04    &  0.01   & 0.50 ( 0.28) &  3.3 (13.1) \\
    \texttt{AD20x1} & 0.92 & 20.9 & 0.22   (6.0)  & 0.25   (5.6) & 1.8   ($-1.4$)  & 1.3    ($-1.4$)  & 0.14    &  0.06   & 2.6 ( 0.83) &  4.4 (27.8) \\
    \texttt{AD20x2} & 2.33 & 15.2 & 0.96   (7.9)  & 1.3    (6.6) & 3.5   ($-1.7$)  & 3.1    ($-1.6$)  & 0.63    &  0.15   & 4.1 ( 2.0) & 10.9 (26.8) \\
 %   \texttt{T20}    & 0.29 & 14.9 & $>1.4$ (7.0)  & $>1.6$ (6.1) & $>4.8$ ($-2.5$) & $>3.7$ ($-2.0$)  & $>0.71$ & $>0.56$ & 14.5 ( 7.7) & 12.0 (26.0) \\

    % \hline
    % \multicolumn{4}{l}{Lower resolution models}\\
    % \texttt{T20l}   & 0.29 & 13.0 & $>0.62$ (8.2) & $>1.1$ (7.7) & $>1.8$ ($-3.2$) & $>1.5$ ($-2.9$ ) & $>0.25$ &  --   & --  & --  \\
    % \texttt{AD09x1l}& 0.87 & 14.4 &   0.11  (1.5) &  0.21  (1.4) & 0.57  ($-0.20$) & 0.54   ($-0.18$) & 0.04    &  --   & --  & --  \\
    \hline
    \texttt{BHdisk} & 0.0  & 12.2 & \multicolumn{2}{r}{$>0.088$ (--)} & \multicolumn{2}{r}{$>0.3$ (--)} & $>0.088$ &  $>0.037$  & --  & -- \\
    \hline
    \end{tabular}
    \label{tab:results}
    
    Values with ``$>$" denote that they are still increasing at the end of the simulation.
\end{table*}

\subsection{Diagnostic} \label{subsec:diag}
We define the mass infall rate to the central region by
\begin{align}
\dot{M}_\mathrm{fall} = \int_{r=r_\mathrm{in}} \sqrt{-g}\rho u^k ds_k,
\end{align}
where $ds_i = r^2 \delta_{ir}d\cos\theta d\phi$ is the surface element of a sphere, $g=\mathrm{det}(g_{\mu\nu})$, and $r_\mathrm{in}$ is chosen in the following manner: 
In the early phase, it is the largest value among the radii of the surface of a standing accretion shock formed as a result of the core bounce; after this shock disappears due to the matter infall from the outer region which induces the collapse of a proto-neutron star to a black hole, the maximum radius of the apparent horizon $r_\mathrm{AH}$ is used for $r_\mathrm{in}$; 
after the formation of the disk, we again choose the largest value among the radii of the surface of a standing accretion shock, which is formed along the disk surface.
In the late phase, we also define the mass accretion rate to the black hole by
\begin{align}
\dot{M}_\mathrm{BH} = \int_{r=r_\mathrm{AH}} \sqrt{-g}\rho u^k ds_k.
\end{align}

% To investigate the cooling efficiency in the disk,
% the total neutrino luminosity and viscous heating rate of the disk are defined by
% \begin{align}
% L_\nu = \int_{r<r_\mathrm{in}} (q^-_\nu - q^+_\nu) \sqrt{-g}d^3x, 
% \end{align}

% The definitions of the unbound matter and explosion energy are the same as those in  \cite{fujibayashi2021oct}. 
% We first define the specific binding energy of the matter as
% \begin{align}
% e_\mathrm{bind}&:=-\frac{T_t{}^t}{\rho u^t}-(1+\varepsilon_\mathrm{min}),
% \end{align}
% where $T_t{}^t$ is the time-time component of the energy-momentum tensor and $\varepsilon_\mathrm{min}$ is the minimum specific internal energy of the employed EOS table ($\approx -0.0013c^2$ for SFHo and DD2). 
% We then define the condition for unbound matter as $e_\mathrm{bind}>0$.
% The diagnostic explosion energy is defined by integrating the positive binding energy in the computational domain as
% \begin{align}
% E_\mathrm{exp}:=\int_{e_\mathrm{bind}>0} e_\mathrm{bind}\rho u^t\sqrt{-g}d^3x.
% \end{align}

The definitions of the unbound matter and explosion energy are the same as those in  \cite{fujibayashi2021oct}. 
We first define the specific binding energy and binding energy flux density of the matter as
\begin{align}
e_\mathrm{bind}&:=-\frac{T_t{}^t}{\rho u^t}-(1+\varepsilon_\mathrm{min}), \label{eq:ebind}\\
f^i_\mathrm{bind}&:=-T_t{}^i-\rho u^i(1+\varepsilon_\mathrm{min}), \label{eq:fbind}
\end{align}
where $T_t{}^t$ and $T_t{}^i$ are the time-time and time-space components of the energy-momentum tensor and $\varepsilon_\mathrm{min}$ is the minimum specific internal energy of the employed EOS table ($\approx -0.0013c^2$ for DD2). 
We then define the condition for unbound matter as $e_\mathrm{bind}>0$.
We find however that the binding energy of the outer layer of the star in our computational domain is not accurately calculated due to a numerical error accumulated in long-term simulations.
Therefore, we decide to calculate the diagnostic explosion energy as well as the mass of the unbound matter as
\begin{align}
E_\mathrm{exp}&:=\int_{e_\mathrm{bind}>0, r<r_\mathrm{ext}} e_\mathrm{bind}\rho u^t\sqrt{-g}d^3x\notag\\
&+ \int^t \int_{e_\mathrm{bind}>0, r=r_\mathrm{ext}} f^k_\mathrm{bind}\, ds_k\, dt,\\
M_\mathrm{ej}&:=\int_{e_\mathrm{bind}>0, r<r_\mathrm{ext}} \rho u^t\sqrt{-g}d^3x\notag\\
&+ \int^t \int_{e_\mathrm{bind}>0, r=r_\mathrm{ext}} \rho u^k_\mathrm{bind}\, ds_k\, dt,
\end{align}
where $ds_k$ is the surface element of the sphere with the radius $r_\mathrm{ext}$.
These quantities are defined by the volume integral for the matter of the positive binding energy inside an extraction radius $r_\mathrm{ext}$ plus the time integral for the components of the positive binding energy flux at the radius.

A part of the stellar matter is located outside the extraction radius and even outside the computational domain.
Its mass and binding energy can contribute to the ejecta mass and explosion energy.
To estimate their contribution, we first obtain the enclosed mass at the extraction radius when the shock wave reaches the extraction radius.
We then estimate the binding energy of the matter above the radius with the same enclosed mass in the pre-collapse profile.
%, which is the estimated binding energy above the extraction radius.
We compare the explosion energies with different extraction radii in \S~\ref{subsec:exp}.

%We note that the matter in the progenitor stars is composed mainly of $^4$He, $^{12}$C, and $^{16}$O.
We note that the matter in the progenitor stars is composed mainly of $^4$He, $^{12}$C, $^{16}$O, and $^{20}$Ne.
When $^{16}$O burns into $^{56}$Ni, the rest-mass energy of \SI{0.67}{MeV} per baryon should be released into the internal energy, which can be an important energy source of the explosion of the star (if the internal energy is not carried away by the neutrino emission).
However, this effect is absent in our simulation because of the assumption of nuclear statistical equilibrium (NSE) in constructing the EOSs, for which the low-temperature matter in the computational domain is assumed to be composed mostly of $^{56}$Ni.
In addition, $^{56}$Ni is photo-dessociated into lighter nuclei if the temperature of the matter increases to $\gtrsim \SI{7}{GK}$, consuming more internal energy than in the dissociation of $^{12}$C or $^{16}$O.
\addsf{These temperature conditions are found when the shock formed near the disk around the black hole has propagated sufficiently far out within the star and, hence, an excessively large energy of 8.6\,MeV/nucleon, instead of $\approx8.0$\,MeV/nucleon, is consumed.}
%This photo-dissociation spuriously takes place when a shock is formed near the disk around a black hole and propagates in the outer region of the star.
Thus, the explosion energy in our present simulations is underestimated. This point will be discussed in \S\,\ref{subsec:Ni}. 

\addsf{We also note that the inclusion of $\varepsilon_\mathrm{min}$ in Eqs.~\eqref{eq:ebind} and \eqref{eq:fbind} is important to estimate the ejecta mass and explosion energy in a physically correct way.
As mentioned above, the low-temperature matter in the outer region of the star is assumed to be composed of iron group nuclei, which have smaller rest masses per baryon than the atomic mass unit ($m_u$; the mass of $^{12}$C divided by 12).
Due to the low-temperature, such matter has a low specific internal energy, which leads to $\varepsilon<0$ for the matter.
If we do not include $\varepsilon_\mathrm{min}$ in Eqs.~\eqref{eq:ebind} and \eqref{eq:fbind}, such matter with $\varepsilon<0$ is not recognized as ejecta, even though it gains sufficient energy to be unbound after being swept by the shock wave.
}

% MEMO: 
%  Mass excess (value from AME 2012):
%  n    8.07131714 MeV
%  p    7.28897059 MeV - 0.511 MeV
% 4He   2.42491561 MeV,  0.606 MeV/b - 0.5*0.511
% 16O  -4.73700137 MeV, -0.296 MeV/b - 0.5*0.511
% 20Ne -7.04193055 MeV, -0.352 MeV/b - 0.5*0.511
% 56Ni -53.906909  MeV, -0.963 MeV/b - 0.5*0.511

\begin{figure}
\epsscale{1.17}
\includegraphics[width=0.48\textwidth]{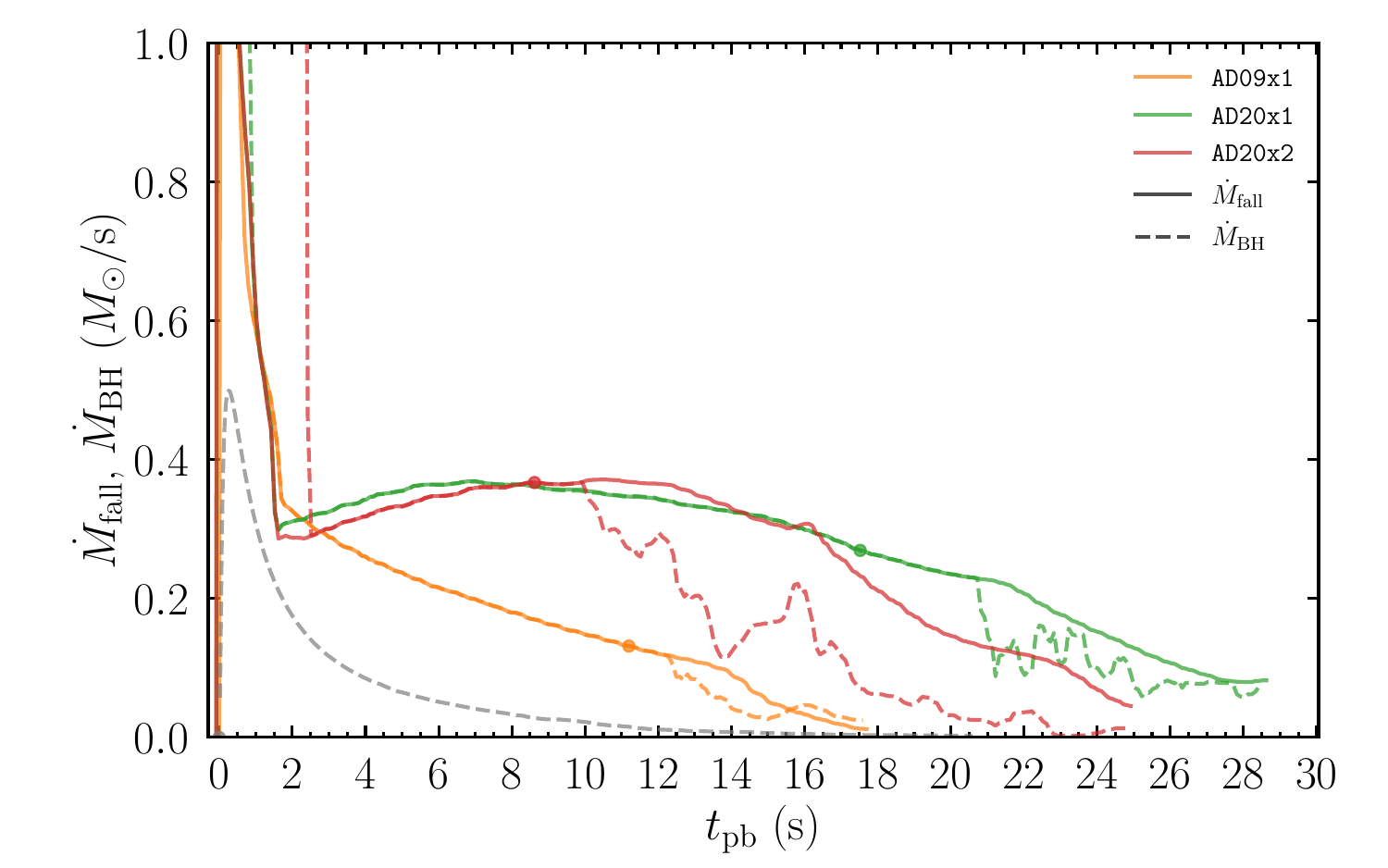}
\includegraphics[width=0.48\textwidth]{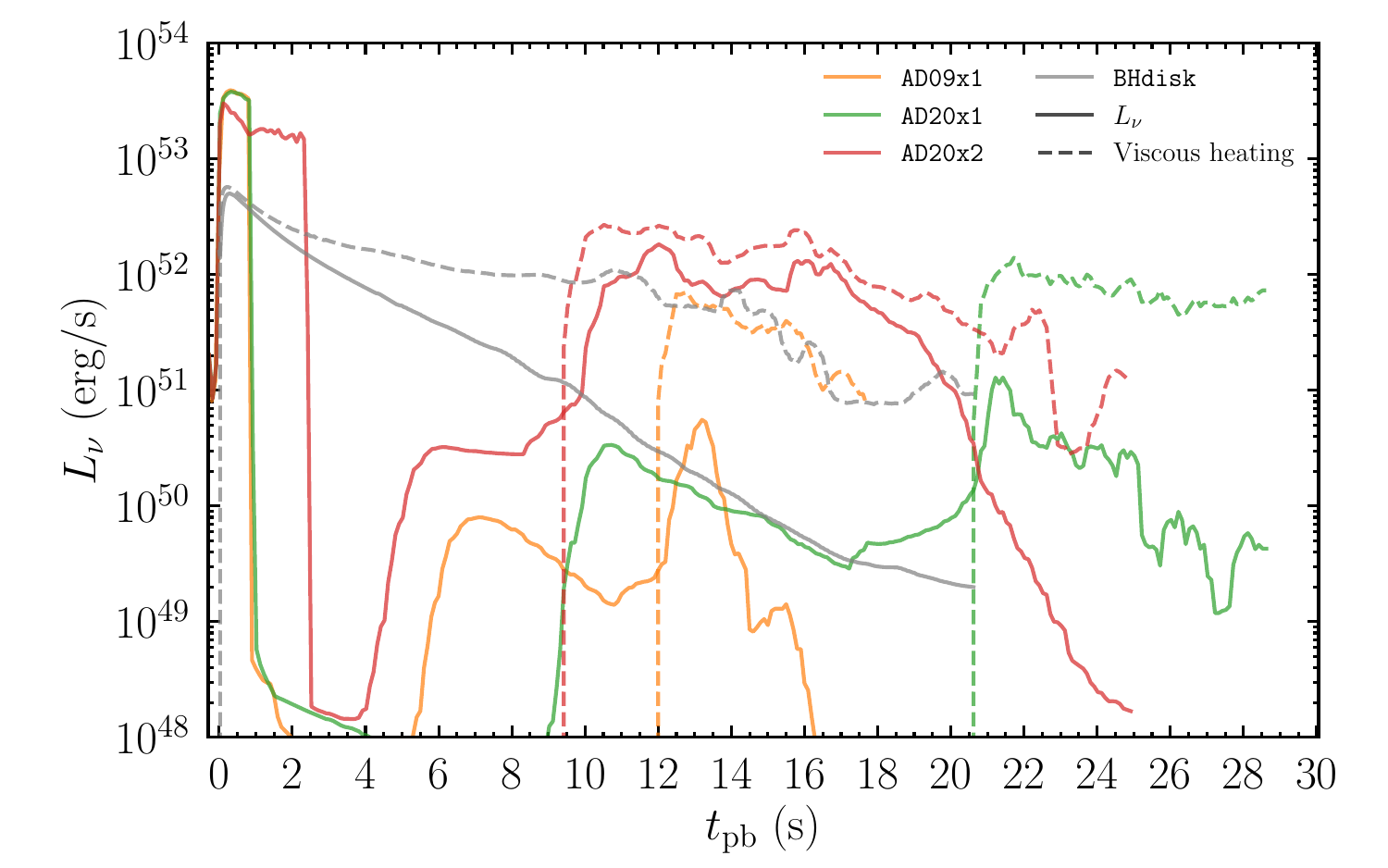}
\includegraphics[width=0.48\textwidth]{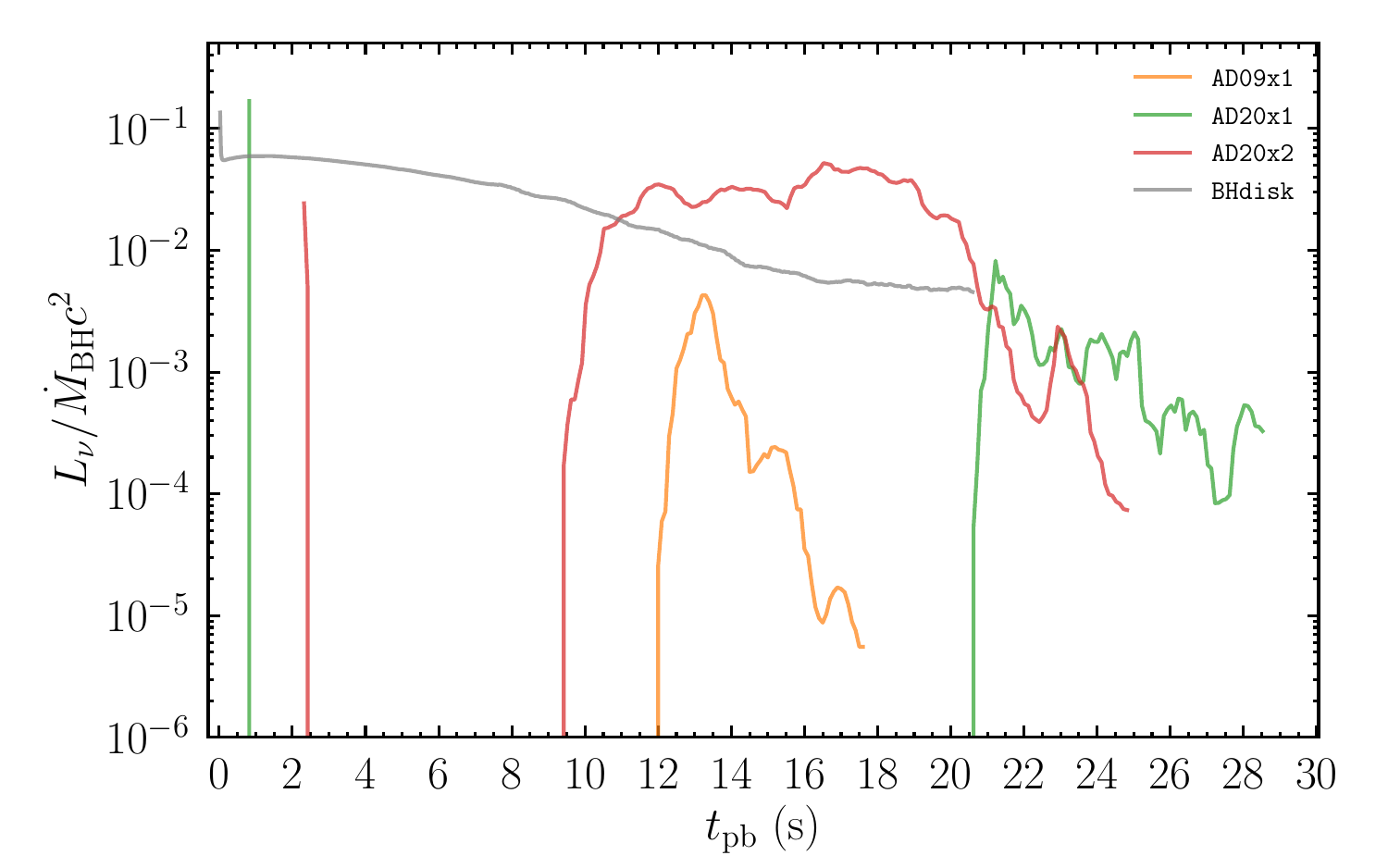}
\caption{
Top: Mass infall rate across the surface of $r=r_\mathrm{in}$ (solid curves) and mass accretion rate onto the black hole (dashed curves). Circles on each curve denote the time at which the viscosity is turned on.
Middle: Total neutrino luminosity (solid curves) and total viscous heating rate (dashed curves) inside the shock wave.
Bottom: Neutrino cooling efficiency defined by the total neutrino luminosity divided by the mass accretion rate onto the black hole.
For model \texttt{BHdisk}, $t_\mathrm{pb}=0$ corresponds to the beginning of the simulation.
}
\label{fig:disk}
\end{figure}

\subsection{Tracer-particle method}
To perform the nucleosynthesis calculation, we apply our post-process tracer-particle method for the results of our simulations.
The method is the same as that in \cite{Fujibayashi2023jan}.
% \addsf{
% The unbound matter is detected at the last snapshot for each simulation with the criterion $e_\mathrm{bind}>0$.
% Several thousand tracer particles are distributed in the region that satisfies the criterion.
% The particles are then traced back in time to the initial snapshot to obtain thermodynamical evolution along each trajectory.
% }
The tracer particles are distributed for 128 polar angles in the range of $\theta=[0:\pi/2]$ on the arc with the radius of $r_\mathrm{ext}=\SI{2e4}{km}$.
The particles are continuously set with the time interval of $\Delta t_\mathrm{set} :=  r_\mathrm{ext} \Delta \theta / \langle v^r \rangle $, where $\Delta \theta = (\pi/2)/128$ and $\langle v^r \rangle$ is the average radial velocity of the ejecta at the extraction radius.
The mass of each particle is determined based on the mass flux at the extraction radius as $\Delta m={r_\mathrm{ext}}^2\Delta \Omega \rho u^r\sqrt{-g}\Delta t_\mathrm{set}$, where $\Delta \Omega$ is the solid angle element.

\section{Result} \label{sec:result}

\begin{figure*}
\epsscale{1.17}
\begin{center}
\includegraphics[width=0.40\textwidth]{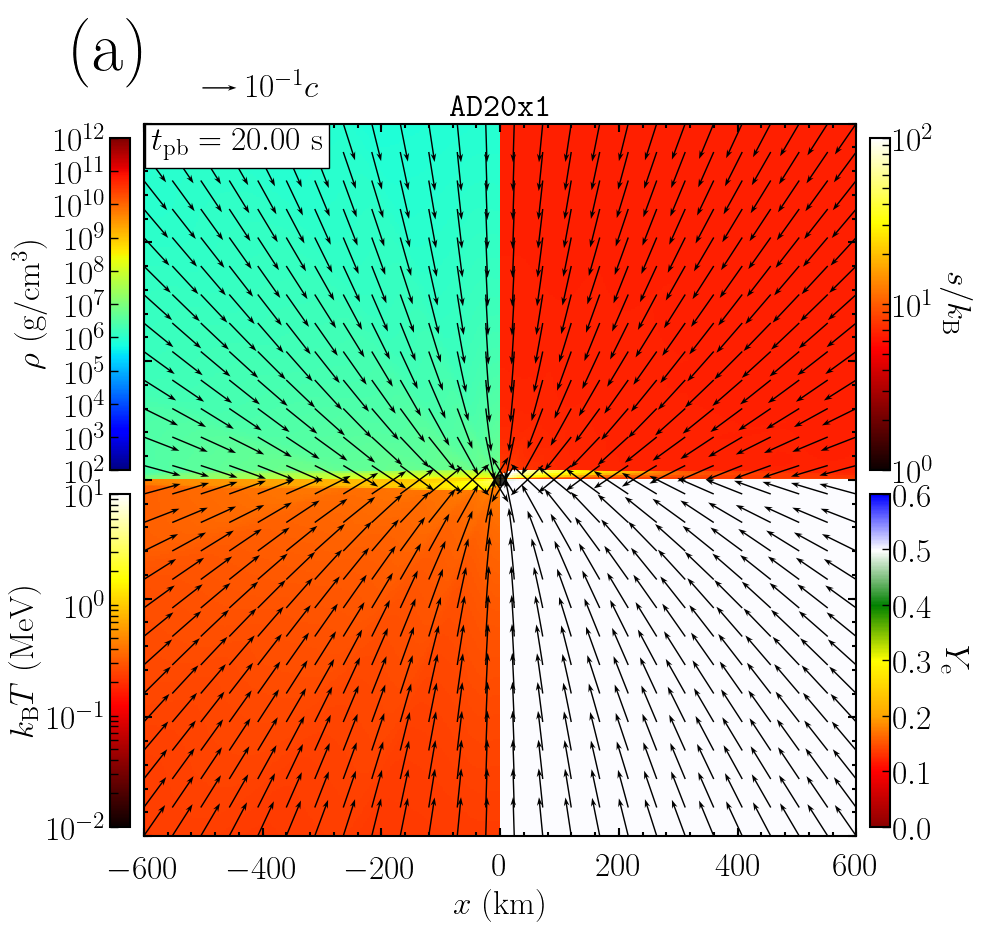}~
\includegraphics[width=0.40\textwidth]{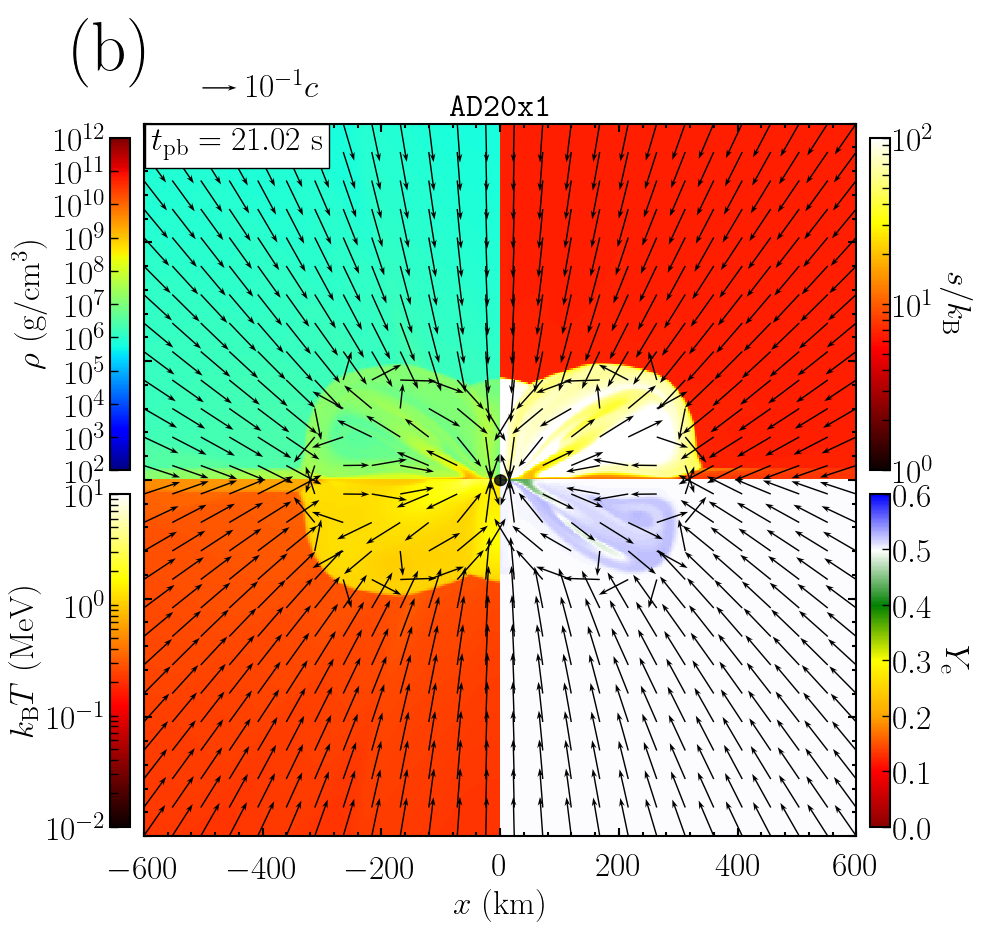}\\
\includegraphics[width=0.40\textwidth]{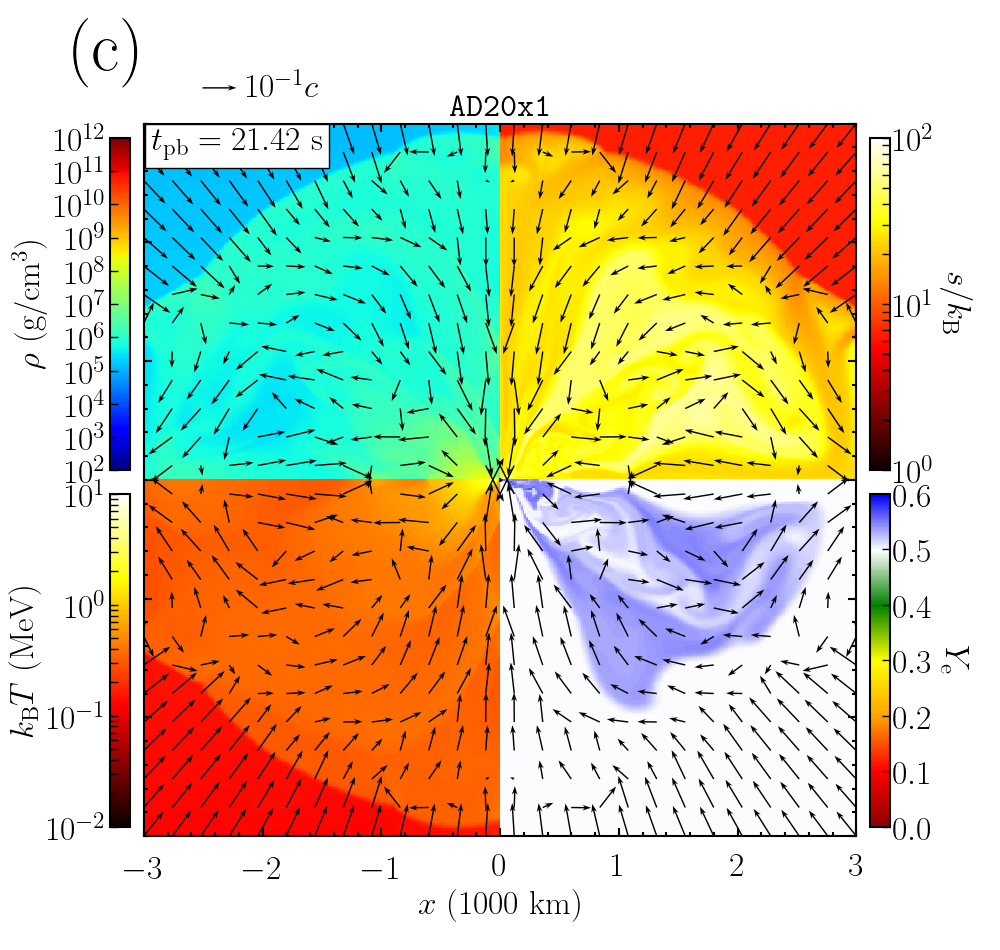}~
\includegraphics[width=0.40\textwidth]{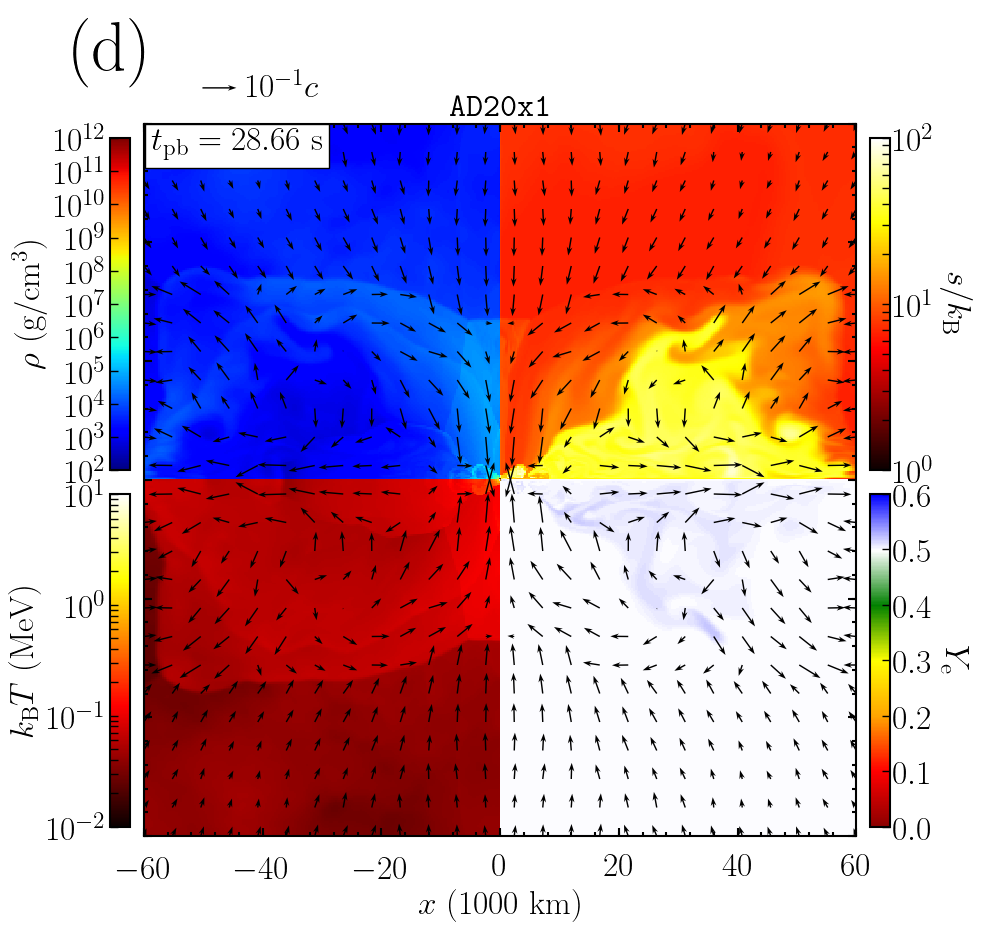}

\end{center}
\caption{
Snapshots at $t_\mathrm{pb}=20.0$ (a), 21.0 (b), 21.4 (c), and 28.7\,s (d) for model \texttt{AD20x1}.
The panels (a)--(d) show the snapshots just prior to the formation of a geometrically thick disk, at the onset of the outflow, at the expanding phase, and the final snapshot, respectively.
Each panel has four sub-panels of rest-mass density (top-left), entropy per baryon (top-right), temperature (bottom-left), and electron fraction (bottom-right). The black solid circle for the first panel shows the region inside the apparent horizon (for other panels, it is too small to be seen because the plotted region is much wider than $GM_\mathrm{BH}/c^2$). Note that the regions of the plots are different for each snapshot. 
An animation for this model is available at \url{https://www2.yukawa.kyoto-u.ac.jp/~sho.fujibayashi/share/AD20x1_multiscale.mp4}.
% \url{https://www2.yukawa.kyoto-u.ac.jp/~sho.fujibayashi/share/anim_20_hres_visc_multiscale.mp4}. 
}
\label{fig:snapshotsAD20x1}
\end{figure*}

% Animation for model \texttt{T20}:
% \url{https://www2.yukawa.kyoto-u.ac.jp/~sho.fujibayashi/share/anim_20_hres_visc_multiscale.mp4}.

\begin{figure*}
\epsscale{1.17}
\includegraphics[width=0.33\textwidth]{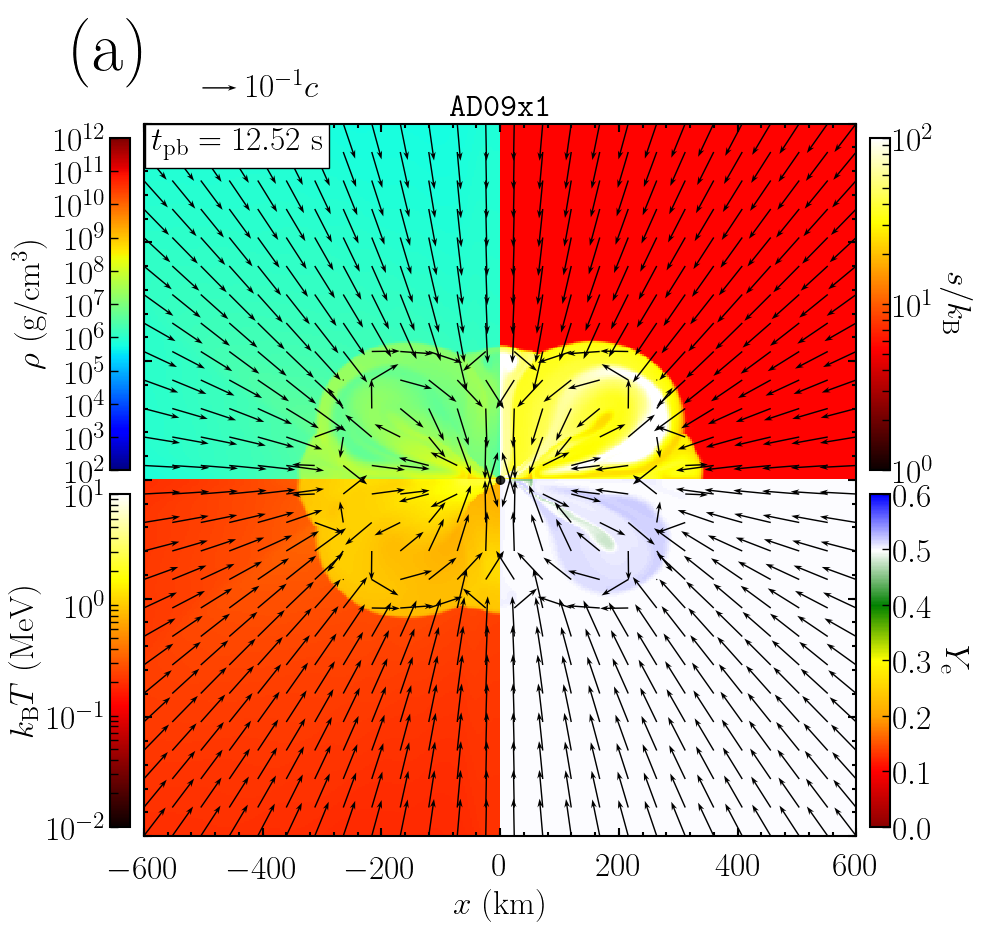}~
\includegraphics[width=0.33\textwidth]{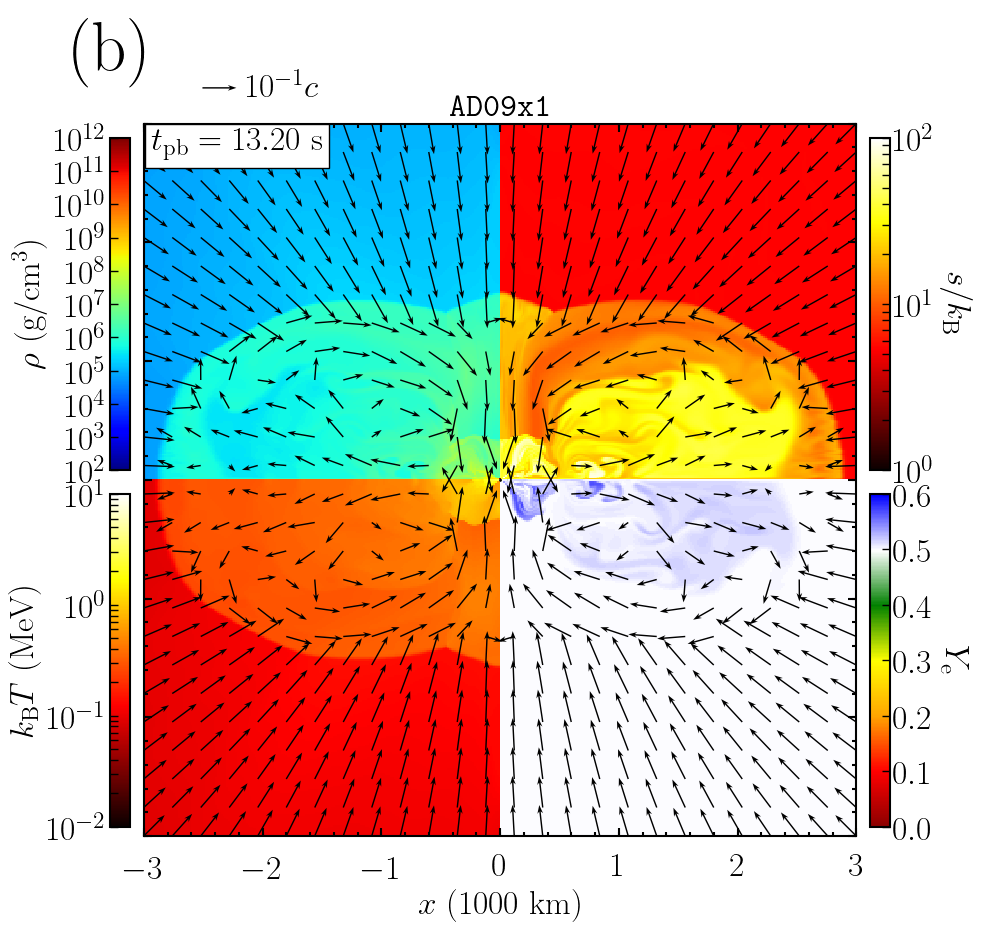}~
\includegraphics[width=0.33\textwidth]{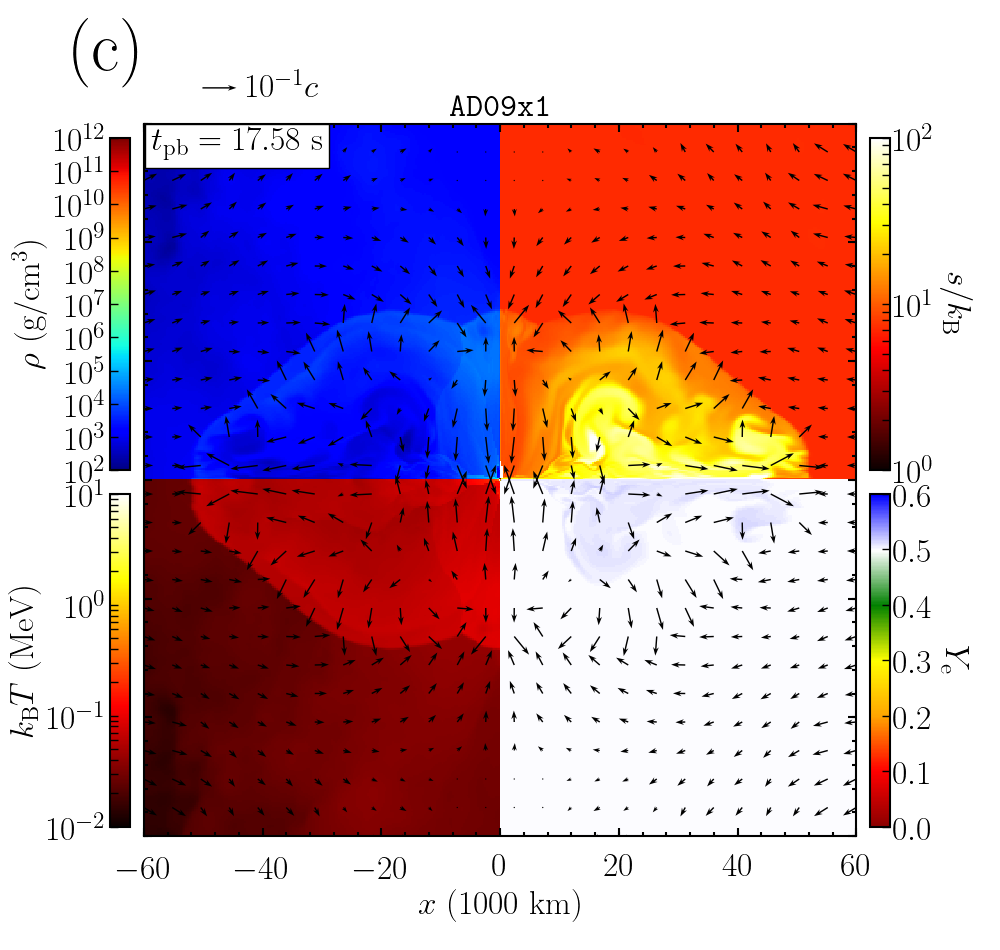}\\
\includegraphics[width=0.33\textwidth]{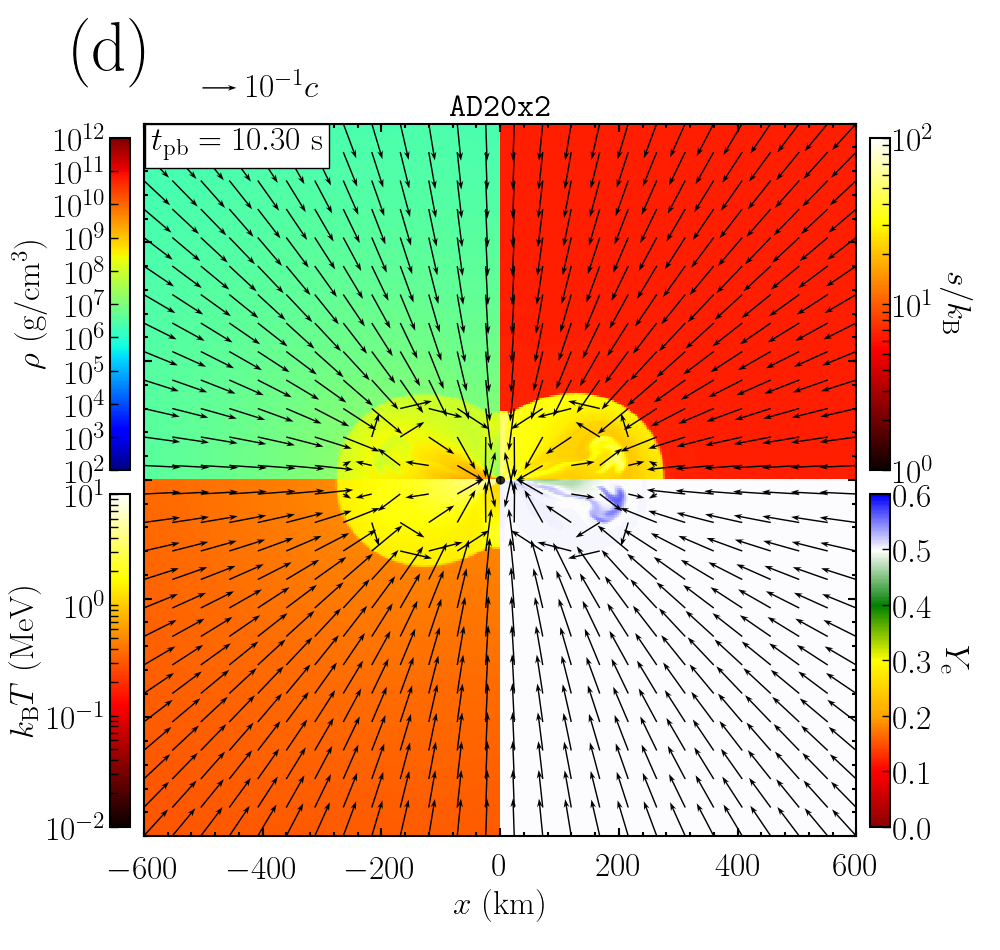}~
\includegraphics[width=0.33\textwidth]{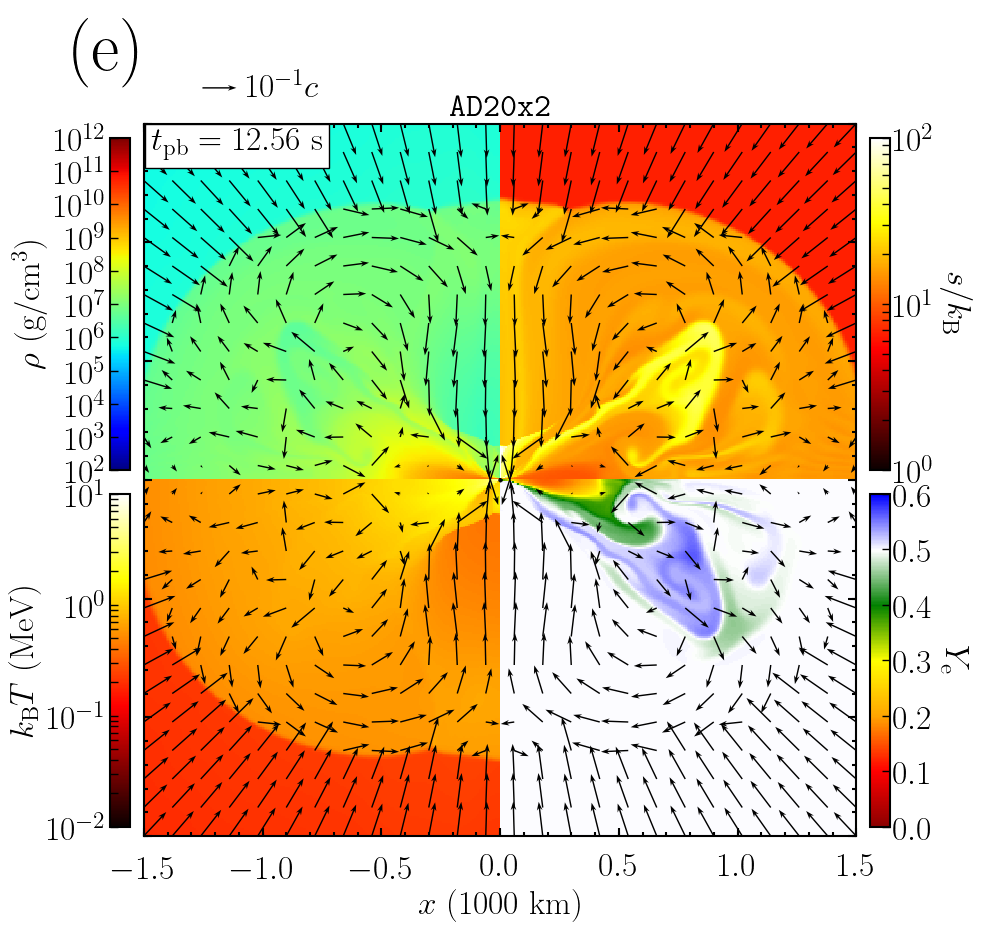}~
\includegraphics[width=0.33\textwidth]{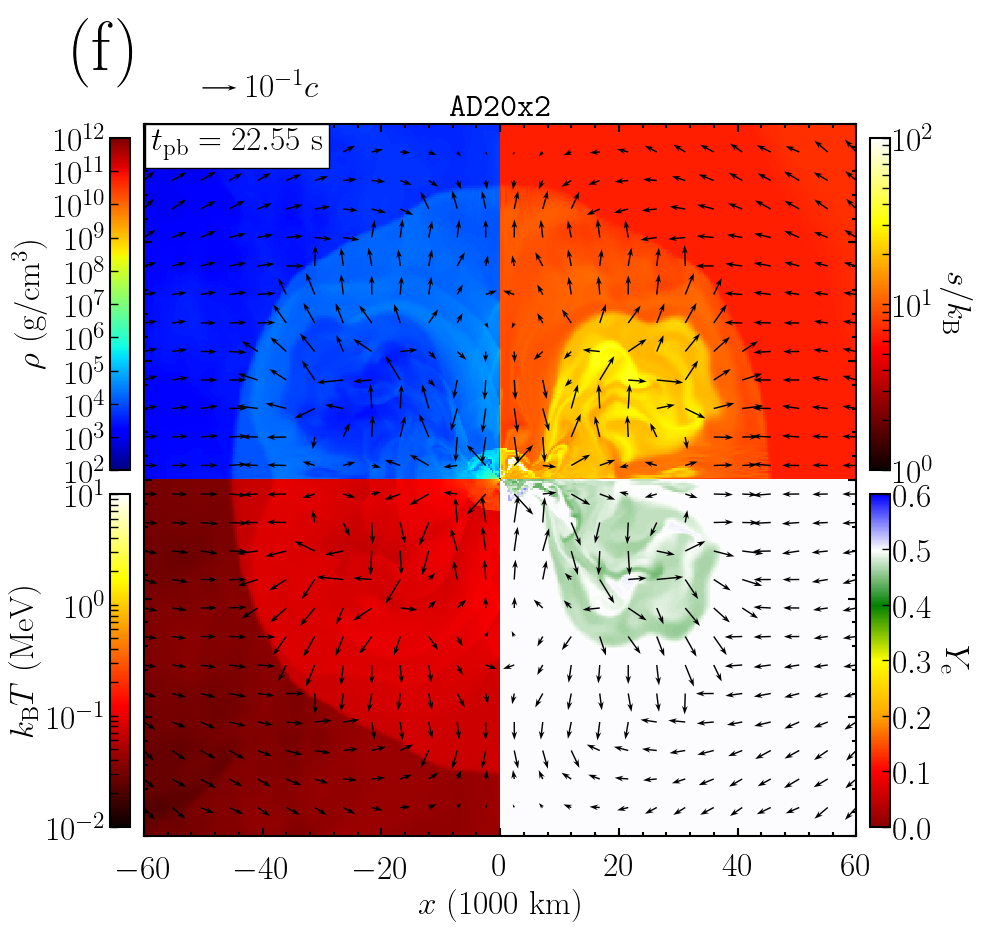}\\
\caption{The same as Fig.~\ref{fig:snapshotsAD20x1} but for models \texttt{AD09x1} (top panels) and \texttt{AD20x2} (bottom panels).
For each model, the panels from left to right show the snapshots at the formation of a geometrically thick disk, after the onset of the outflow, and the final snapshots, respectively.
}
\label{fig:snapshots-other}
\end{figure*}

\begin{figure}
\epsscale{1.17}
\includegraphics[width=0.48\textwidth]{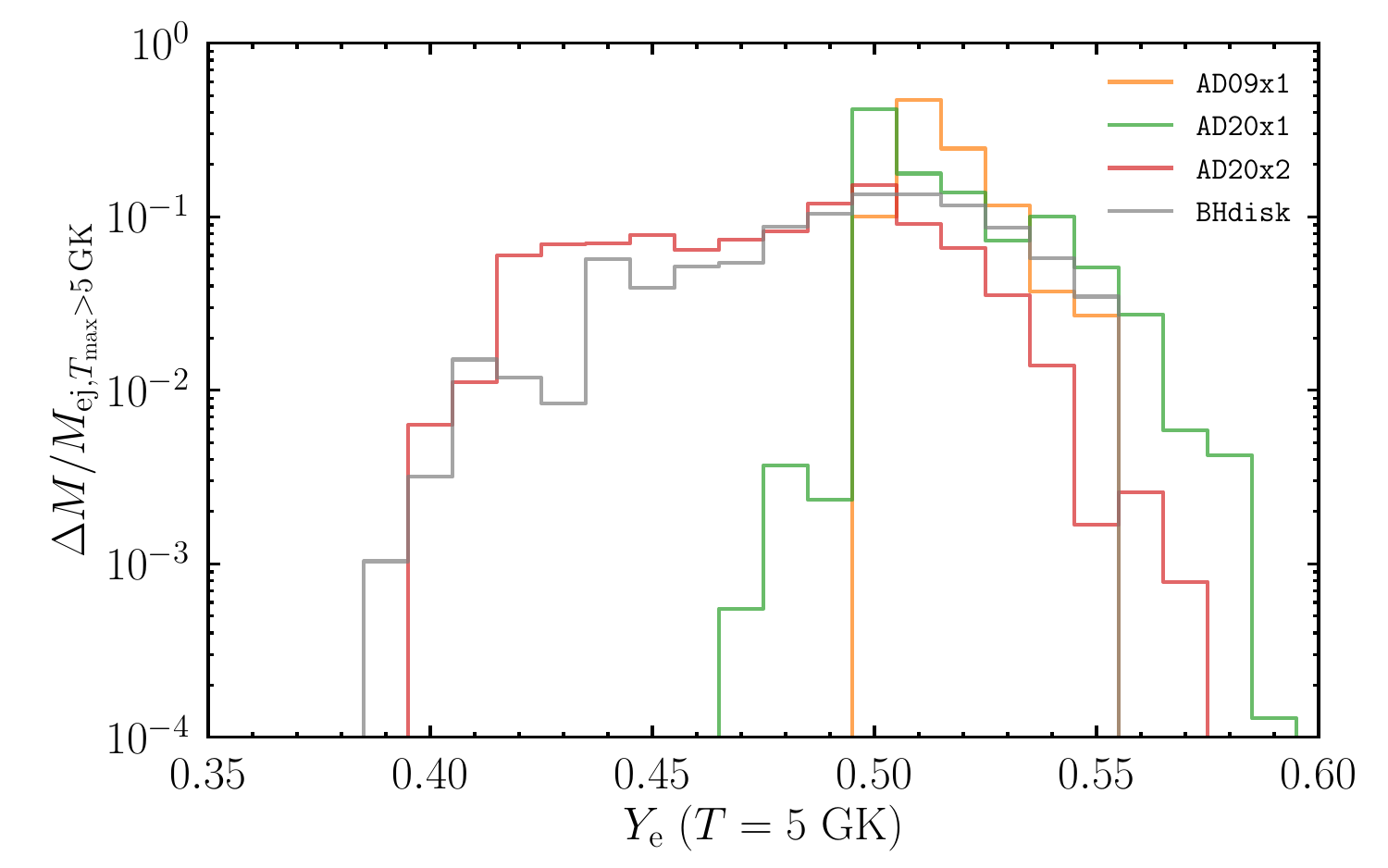}\\
\includegraphics[width=0.48\textwidth]{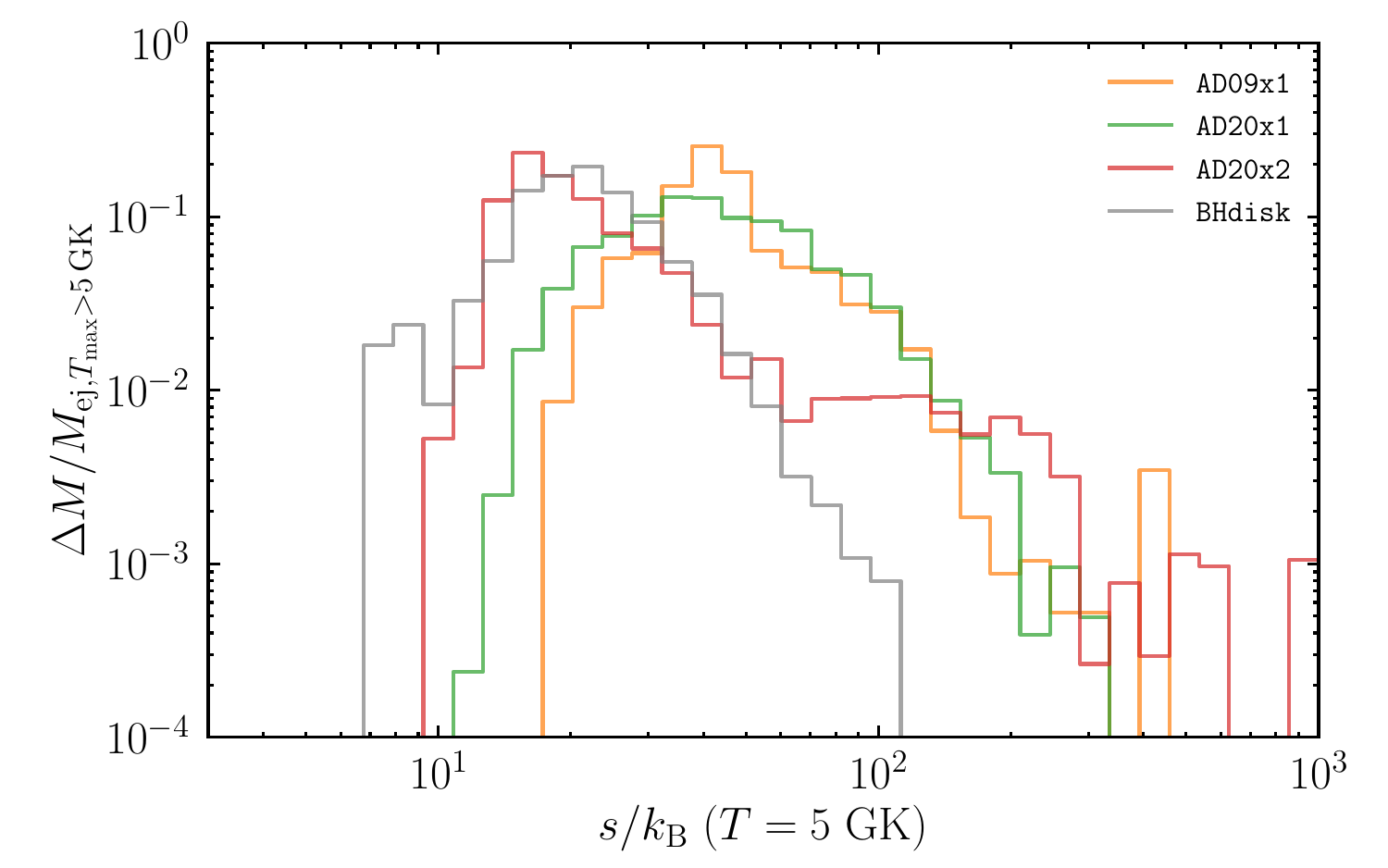}\\
\includegraphics[width=0.48\textwidth]{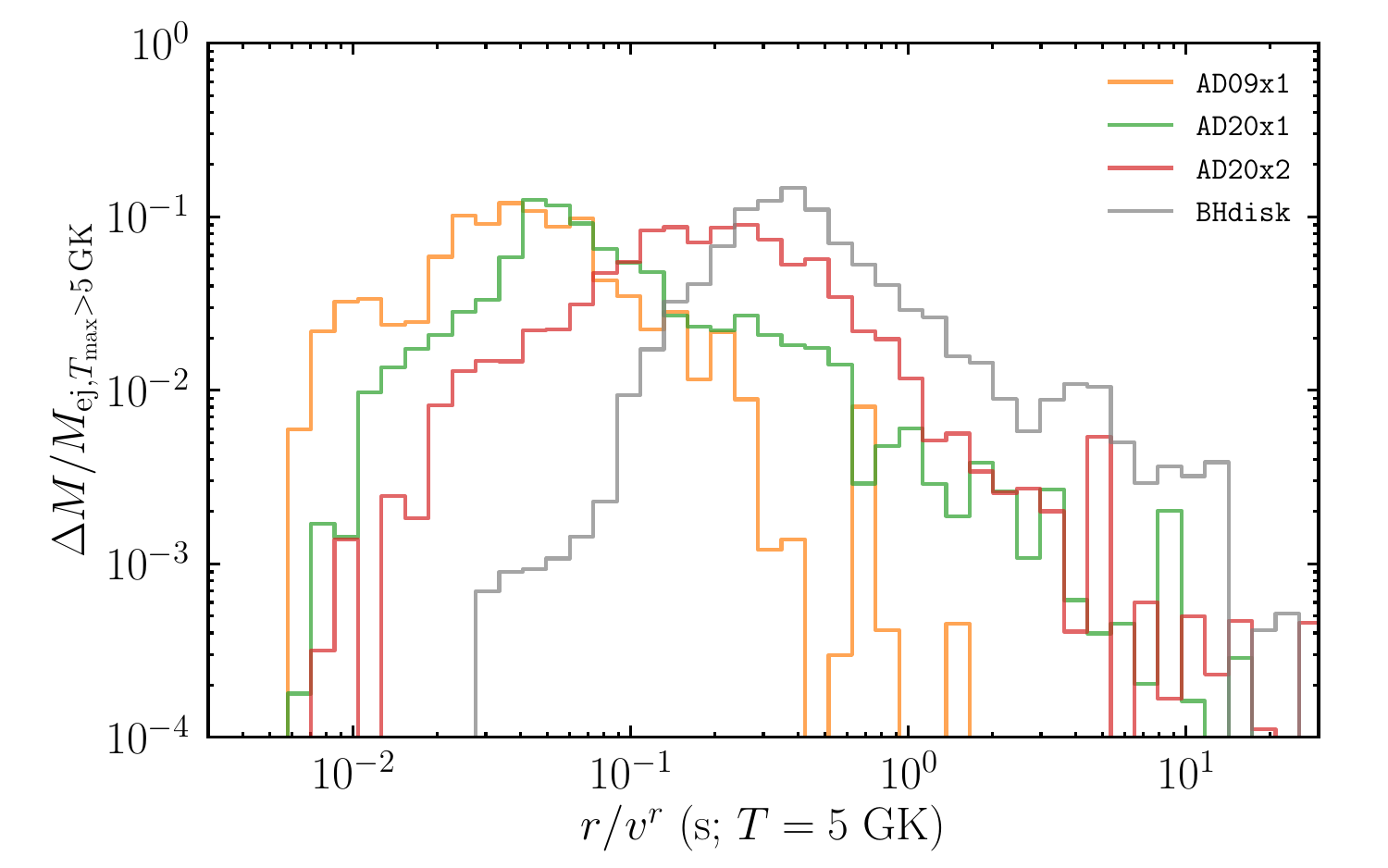}
\caption{
Mass histograms as a function of the electron fraction (top), entropy per baryon (middle), and expansion timescale ($r/v^r$; bottom) at $T=\SI{5}{GK}$ of the tracer particles that experience temperature higher than \SI{5}{GK}.
}
\label{fig:hists}
\end{figure}

\subsection{Evolution after bounce to disk formation}

For all the collapse models, a proto-neutron star is initially formed after the core bounce together with the standing accretion shock formation. Subsequently, due to the matter accretion from the outer region, the proto-neutron star collapses to a black hole. The black-hole formation time is $t_\mathrm{pb}=0.9$, 0.9, and 2.3\,s for models \texttt{AD09x1}, \texttt{AD20x1}, and \texttt{AD20x2}, respectively (denoted by $t_\mathrm{BH}$ in Table~\ref{tab:results}).
Here $t_\mathrm{pb}$ denotes the time after the core bounce. For model \texttt{AD20x2}, the formation of the black hole is delayed due to the significant centrifugal-effect associated with the rapid rotation.
These results illustrate that the formation process of the black hole (and subsequent evolution process of the black hole and disk) depends on the profiles of the density and specific angular momentum of the progenitor stars.

%\texttt{T20}, \texttt{AD20x2}, \texttt{AD09x1}, and \texttt{AD20x1}
%1.5, 5.0, 7.0, and 10.5
%3.5, 4.5, 5.0, and 7.0
%3.0, 10, 13, and 21

For seconds after the black-hole formation, the infalling matter does not have enough angular momentum to form disks, and thus, it is swallowed simply by the black hole.
A geometrically thin disk starts forming at $t_\mathrm{pb}\approx 7.0$, 10.5, and 5.0\,s at which the black-hole mass is $M_\mathrm{BH}/M_\odot \approx 5.0$, 7.0, and 4.5 for models \texttt{AD09x1}, \texttt{AD20x1}, and \texttt{AD20x2}, respectively (see, e.g., panel (a) of Fig.~\ref{fig:snapshotsAD20x1} for model \texttt{AD20x1}).
The above black-hole mass is consistent with the ones inferred from the distributions of density and specific angular momentum of the progenitor stars (see \S~\ref{subsec:model}).

The disks at their formation are geometrically thin because of the lower pressure inside the disk than the ram pressure of the infalling matter~\citep{Sekiguchi2011}.
The disks then become geometrically thick when the pressure in the outer part of the disk and ram pressure of the infalling matter become comparable. 
The vertical expansion of the disk occurs at $t_\mathrm{pb}\approx 13$, 21, and 10\,s for models \texttt{AD09x1}, \texttt{AD20x1}, and \texttt{AD20x2}, respectively (see panel (b) of Fig.~\ref{fig:snapshotsAD20x1} and panels (a) and (d) of Fig.~\ref{fig:snapshots-other}).
In the presence of the viscosity, the disk expansion occurs at time slightly later than that in the corresponding non-viscous model.
The reason for this is that the viscous angular momentum transport accelerates the accretion of the disk matter onto the black hole and the increase of the pressure in the disk is delayed.

After the vertical expansion of the disk, a shock surface between the disk and infalling matter expands (see, e.g., panel (b) of Fig.~\ref{fig:snapshotsAD20x1}). 
The geometrical cross section of the shock surface becomes large, which enhances the dissipation of the kinetic energy of the infalling matter at the shock more efficiently \citep{Sekiguchi2011}.
As a consequence, the neutrino luminosity increases during this phase.
The neutrino luminosity depends on the rate of mass supply to the disk (see Fig.~\ref{fig:disk}).
For a rapidly rotating model \texttt{AD20x2}, the neutrino luminosity is higher because the mass-infall rate is higher.

After the vertical disk expansion, the models with the original rotational profiles \texttt{AD09x1} and \texttt{AD20x1}, and a rapidly rotating model \texttt{AD20x2} show different evolution process in terms of mass infall rate and neutrino cooling efficiency.
We describe the evolution processes in the following subsections separately.

\subsection{Models \texttt{AD09x1} and \texttt{AD20x1}: Models of lower infalling rate at disk formation} \label{subsec:low-mdot}

For models \texttt{AD09x1} and \texttt{AD20x1}, the rate of mass supply to the disk after the vertical expansion of the disk, $\dot{M}_\mathrm{fall}-\dot{M}_\mathrm{BH}$, is small ($\lesssim 0.1M_\odot$/s; see the top panel of Fig.~\ref{fig:disk}).
As a result, the disk temperature cannot be high enough for the efficient cooling by neutrino emission.
Thus, the viscous heating dominates over the neutrino cooling in the entire phases after the disk expansion (the middle panel of Fig.~\ref{fig:disk}); a neutrino-dominated accretion disk is not formed in these models.
The dominance of the viscous heating leads to an early launch of the outflow (several hundreds of milliseconds) after the disk expansion for these models.
The outflow is launched mainly toward the equatorial direction (see panel (c) of Fig.~\ref{fig:snapshotsAD20x1} and (b) of Fig.~\ref{fig:snapshots-other}).
The final snapshots (see panel (d) of Fig.~\ref{fig:snapshotsAD20x1} and (c) of Fig.~\ref{fig:snapshots-other}) also show the deformed profile of shock surfaces and the outflow from the disk toward the equatorial direction.

The top panel of Fig.~\ref{fig:hists} shows the mass histogram as a function of the electron fraction of the ejecta.
For these models, the value of $Y_\mathrm{e}$ is not low (at lowest 0.47; see the top panel of Fig.~\ref{fig:hists}) because of their low disk density and low neutrino cooling efficiency throughout the disk evolution.
This leads to weak electron degeneracy of the disk matter and thus keeps higher values of $Y_\mathrm{e}$.
Some components have the electron fraction even higher than 0.5.
This reflects the fact that the positron capture proceeds in a shorter timescale than the electron capture under the condition of low electron degeneracy and mildly high temperature $k_\mathrm{B}T\gtrsim \SI{1}{MeV}$.
In this condition, the positron capture $n+e^+ \rightarrow p + \nu_e$ is energetically more preferred than the electron capture because the mass difference between a free neutron plus an electron and a free proton $(m_n + m_e - m_p)c^2$ is not negligible compared with the kinetic energy of electrons $\sim k_\mathrm{B}T$ (see \citealt{Just2022jan}, \citealt{Arcones2010nov}, and \citealt{Beloborodov2003may}).
The result of these models are similar to those found in \cite{Just2022aug} \addsf{(although our prescription of the viscous hydrodynamics is different from that of their study)}.\footnote{We note that our prescription of the viscosity may lead to smaller kinetic viscous coefficient than that in \cite{Just2022aug}.}
Because of the high electron fraction of the ejecta, we do not expect an $r$-process nucleosynthesis (see \S~\ref{subsec:rprocess}).

The middle panel of Fig.~\ref{fig:hists} shows the mass histogram as a function of entropy per baryon, $s/k_\mathrm{B}$.
These models have large values of entropy per baryon $s/k_\mathrm{B}\approx30$--50 and short expansion timescales $r/v^r\sim \SI{50}{ms}$ than a rapidly rotating model \texttt{AD20x2} (see \S~\ref{subsec:high-mdot}).
This result can be explained by the following three reasons.
First, for moderately rotating models, the typical radius of the disk is smaller at the onset of the outflow because of the shorter time of the onset of the outflow after the disk formation.
This requires a larger heating efficiency for the disk matter to be unbound and for the outflow matter to have the higher velocity.
Second, since the outflow sets in before the disk settles into a quasi-steady phase, the shear of the poloidal velocity field in addition to that of Keplerian motion leads to more efficient viscous heating.
Third, the entropy generated by the viscous heating is not efficiently lost by the neutrino emission because of the lower neutrino cooling efficiency.
These properties of moderately rotating models produce the ejecta with a higher entropy and a shorter timescale.

\subsection{Model \texttt{AD20x2}: Models of higher infalling rate at disk formation} \label{subsec:high-mdot}

\addsf{This rapidly rotating model has appreciable differences from models \texttt{AD09x1} and \texttt{AD20x1} in the following two aspects: Specific angular momentum is much larger, and as a result, the mass infalling rate after the disk formation is much higher. In particular, the latter difference significantly modifies the evolution process of the disk from the model \texttt{AD20x1}.}

For this model, a high-mass infall phase continues prior to the disk outflow for a long timescale of $\sim 10$\,s.
The reason for this is that during such a phase, the ram pressure from the infalling matter is strong and the temperature in the shocked region is high enough for enhancing the neutrino cooling, and hence, the thermal pressure generated by the viscous heating cannot be high enough for launching the disk outflow.

As a consequence, model \texttt{AD20x2} exhibits a long-term quasi-steady phase of the disk in which the viscous heating and neutrino cooling rates are comparable for $t_\mathrm{pb}\approx 10$--$\SI{16}{s}$.
In this quasi-steady phase, the neutrino cooling efficiency, defined by the neutrino luminosity $L_\nu$ divided by $\dot{M}_\mathrm{BH}c^2$, is several to ten per cent (see the bottom panel of Fig.~\ref{fig:disk}), which is sufficiently high for neutrinos to carry away the energy generated by viscous heating.
That is, a neutrino-dominated accretion flow (NDAF) is established in this phase.

After the onset of the outflow from the disk, an expanding shock with a slightly oblate shape is formed (see the panel (e) of Fig.~\ref{fig:snapshots-other}).
During the shock expansion, the matter infall onto the black hole and disk still continues in the polar and equatorial directions because the outflow from the disk is launched along the surface of the geometrically thick disk ($z\approx0.5R$).
The outflow towards the polar and equatorial directions is suppressed for different reasons.
Because the outflow is launched mainly from the surface of the inner side of the disk, the outflow is prohibited in the equatorial direction due to the presence of the dense outer disk (bound matter).
Near the polar axis, the infalling matter that passed through the expanding shock surface converges toward the polar region.
As a result, the ram pressure near the polar region is enhanced and becomes larger than that in the other direction.
This prevents the outflow toward the polar direction.
The final snapshot for this model also shows that the outflow from the disk expands in a diagonal direction (see panel (f) and (i) of Fig.~\ref{fig:snapshots-other}; the matter with higher entropy, $s/k_\mathrm{B}\gtrsim 30$, is the outflow component launched from the disk).

Although a neutron-rich region with $Y_\mathrm{e}<0.2$ is present in the inner mid-plane region of the disk reflecting the high density there (see panel (e) of Fig.~\ref{fig:snapshots-other}), the lowest value of $Y_\mathrm{e}$ of the ejecta is 0.40 for this model.
The reason for this high value is that the outflow is launched from the disk surface region, in which the electron degeneracy is not as high as that in the disk mid-plane and the value of $Y_\mathrm{e}$ is close to $0.5$. 
The component with $Y_\mathrm{e}>0.5$ is present because of the same reason as for models \texttt{AD09} and \texttt{AD20x1} (see \S~\ref{subsec:low-mdot}).
From the ejecta with $Y_\mathrm{e} \agt 0.5$ and with $T\gtrsim \SI{5}{GK}$, a substantial amount of $^{56}$Ni can be synthesized. We discuss this topic in \S\,\ref{subsec:Ni}.

The typical value of $s/k_\mathrm{B}$ is found to be 10--20 for this rapidly rotating model, which is lower than those for models \texttt{AD09x1} and \texttt{AD20x1} (see the middle panel of Fig.~\ref{fig:hists}).
This is because the outflow is developed well ($\sim \SI{10}{s}$) after the formation of quasi-steady disks in a milder manner for the rapidly rotating model than for the moderately rotating models.
This trend is also illustrated in mass histogram of the expansion timescale (see the bottom panel of Fig.~\ref{fig:hists}).
The expansion timescale for the rapidly rotating model is typically $\gtrsim \SI{0.1}{s}$, which is longer than for the moderately rotating models.
The entropy and expansion-timescale distributions for the rapidly rotating model are  similar to those for model \texttt{BHdisk}, indicating that the formation of a dense disk in this model is a key to determining the properties of the ejecta.

%It should be kept in mind that the matter with low electron fraction is present in the disk. Thus, in the presence of an efficient mass-ejection mechanism (e.g., some magnetohydrodynamical effects) that expels the disk matter may be ejected from the deeper region of the disk, in which the electron degeneracy is high  \citep{Siegel2018a,Christie2019dec,Miller2019a}. 

The disk mass is $\approx 0.2M_\odot$ for model \texttt{AD20x2} and the cooling efficiency is low at the termination of the simulation.
The electron fraction for the bulk of the disk matter is frozen to be $Y_\mathrm{e}=0.4$--0.5, because of a long weak interaction (electron/positron capture) timescale.
Therefore, the electron fraction of the matter expected to be ejected in a longer timescale is likely to be 0.4--0.5 (see also \S~\ref{subsec:bhdisk}).

As these results illustrate, the evolution of the disk, the timing for the onset of the explosion, and the electron fraction of the ejecta depend strongly on the distribution of the specific angular momentum of the progenitors and the resulting mass infall rate on the disk. 
For the models with relatively low angular momentum (\texttt{AD09x1} and \texttt{AD20x1}), the explosion occurs in a short timescale after the formation of a geometrically thick disk, at which the ram pressure of the infalling matter is low enough for launching an outflow.
In this case, the electron fraction of the ejecta resulting from the explosion cannot be very low.
By contrast, for the model with relatively high angular momentum (\texttt{AD20x2}), the explosion takes place at $\sim 10$\,s after the formation of the disk, and in this case, the electron fraction of the ejecta can be low with $Y_\mathrm{e} < 0.45$, and thus, light trans-iron nuclei can be synthesized (see \S~\ref{subsec:rprocess}). 

% For rapidly rotating models \texttt{T20} and \texttt{AD20x2} \addsf{!!!not for AD20x2, weak interaction already not efficient!!!}, a disk with mass of a few $M_\odot$ still remains at the end of our simulations and the bulk of the disk is still cooled by neutrino emission efficiently.
% When the cooling efficiency drops, a stronger outflow may be launched (e.g., \citealt{Fernandez2013a,Just2015a,Fujibayashi2020a,Just2022mar}).
% Suppose that 10\% of the disk matter becomes ejecta eventually, the matter of mass $O(0.1)M_\odot$ becomes the ejecta in a later phase.
% Since the outflow matter experiences a condition with temperature higher than \SI{5}{GK}, a non-negligible amount of $^{56}$Ni may be synthesized in the later-time ejecta of this model.

\subsection{Ejecta mass and explosion energy}\label{subsec:exp}

\begin{figure}
\epsscale{1.17}
\includegraphics[width=0.48\textwidth]{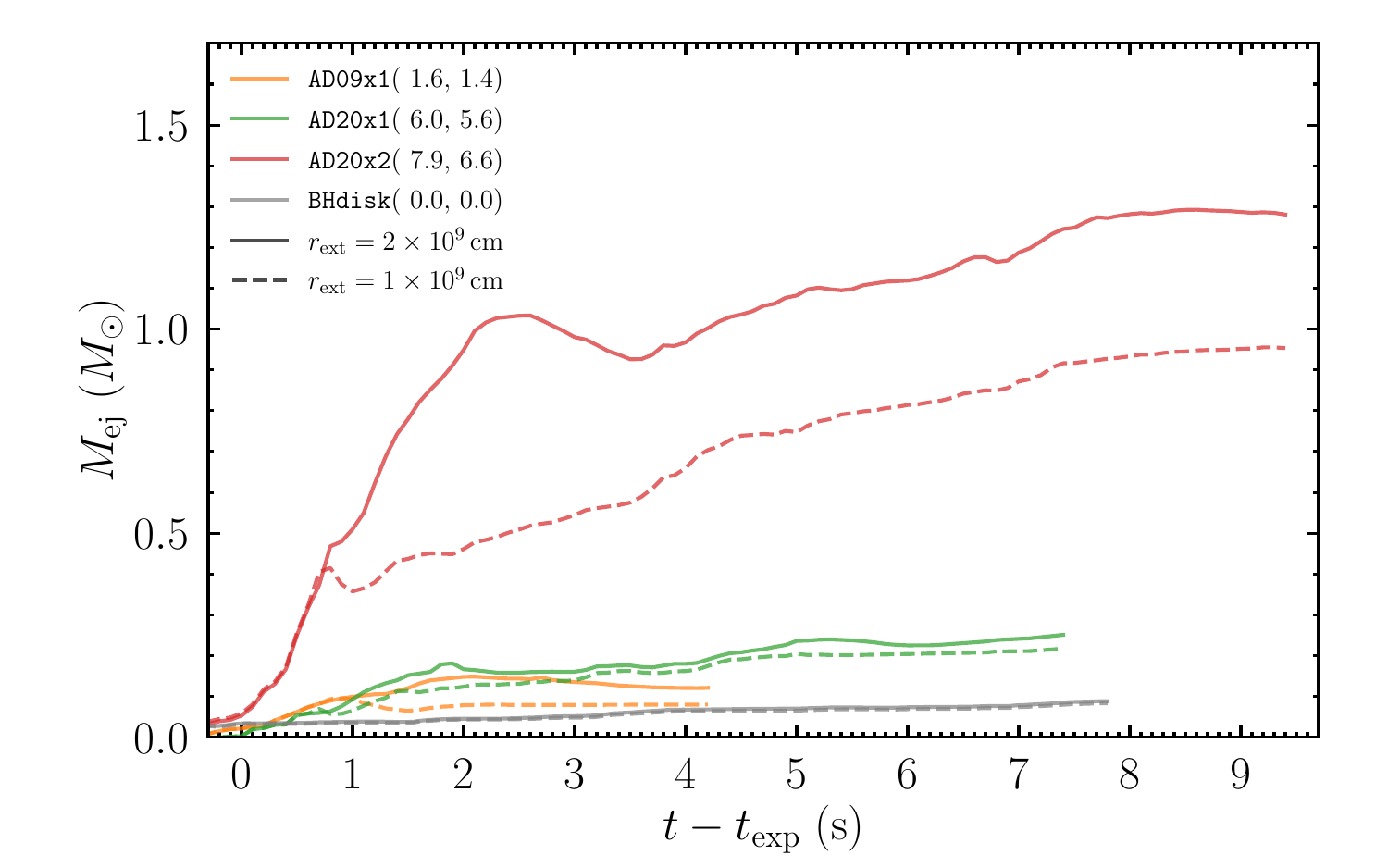}
\includegraphics[width=0.48\textwidth]{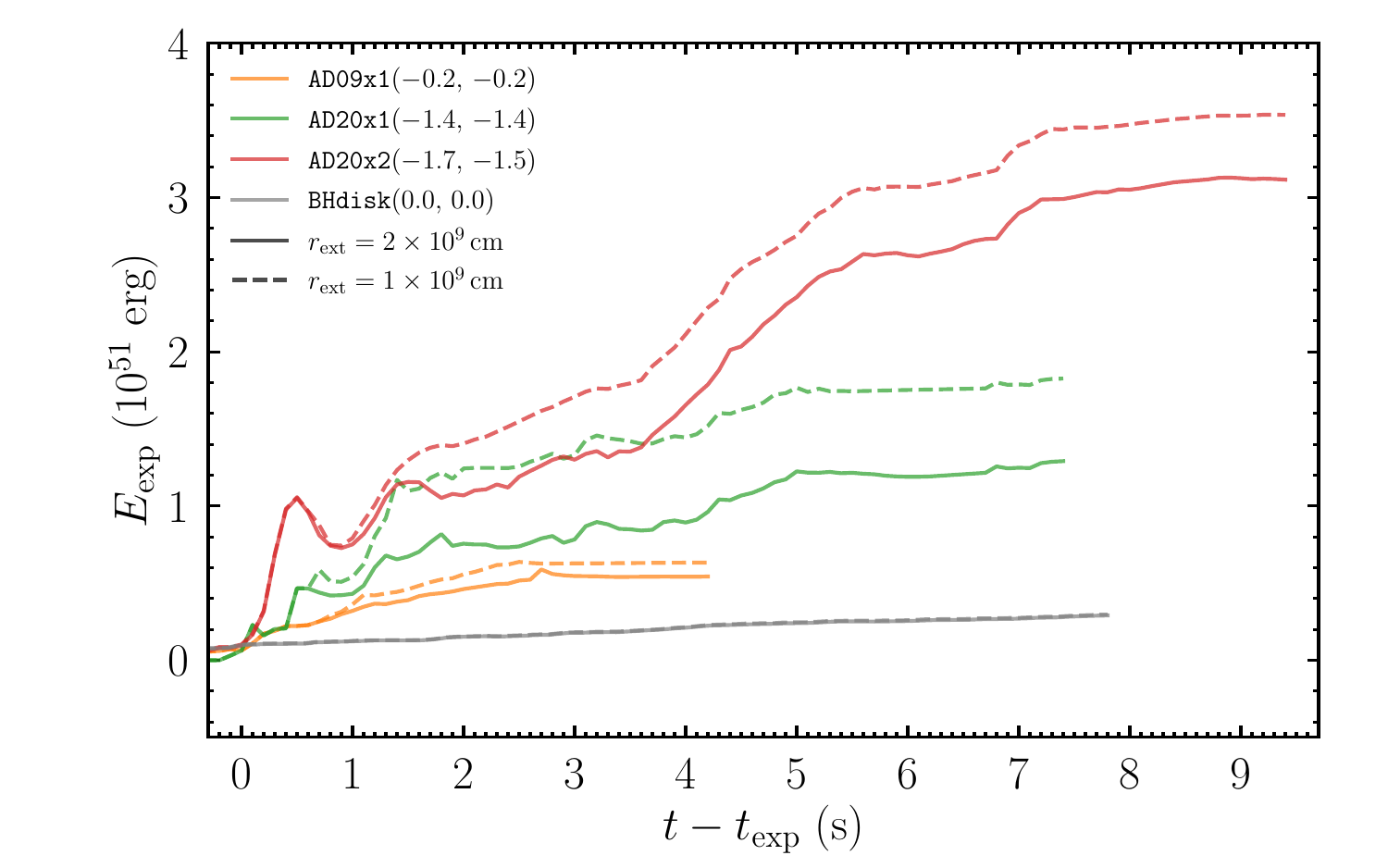}
\caption{
Time evolution of ejecta mass (top) and explosion energy (bottom) as functions of post-explosion time $t-t_\mathrm{exp}$.
The dashed and solid curves are the values calculated with $r_\mathrm{ext}=\SI{1e9}{cm}$ and $\SI{2e9}{cm}$, respectively.
The values in the legend denote the mass (in units of solar mass) and binding energy (in units of $10^{51}$\,erg) above the extraction radius when the shock wave reaches these radii (see \S~\ref{subsec:diag}).
% In the bottom panel, the final explosion energies plus binding energy outside the extraction radius are plotted with stars ($r_\mathrm{ext}=\SI{2e9}{cm}$) and circles ($\SI{1e9}{cm}$).
}
\label{fig:ejecta}
\end{figure}

Figure~\ref{fig:ejecta} shows the mass of the ejecta and the diagnostic explosion energy as functions of time after the explosion, $t-t_\mathrm{exp}$.
Here, the time of the explosion is defined as the time at which the explosion energy reaches \SI{1e50}{erg}.

For all the viscous-hydrodynamics simulations, we find an explosion by the energy injection from the disk around the black hole.
For models \texttt{AD09x1} and \texttt{AD20x1}, for which the angular momentum of the progenitor stars is relatively low, the formation of the disk is delayed, and the onset of the explosion occurs in a late phase (see Table~\ref{tab:results}).
As a consequence, only a small amount of mass remains outside the black hole at the explosion.
The small energy budget at the formation of the disk leads to a slow increase in the explosion energy.
For these models, the explosion energies amount to $\approx$ (0.5--1)$\times 10^{51}$\,erg at the termination of the simulations, which are comparable or slightly smaller than the canonical value for core-collapse supernovae.
The ejecta masses for these models are 0.1--$0.3M_\odot$ at the termination of the simulations, which are an order of magnitude smaller than that for canonical supernovae (but see below).

For the rapidly rotating model \texttt{AD20x2}, after the launch of the outflow, the explosion energy increases beyond $\SI{1e51}{erg}$ with a timescale of $\sim\SI{1}{s}$.
For this model, the explosion energy at the termination of the simulation reaches $\gtrsim \SI{3e51}{erg}$.
At the termination of the simulations, disks with $0.2M_\odot$ is present around the black holes for this model.
In addition, matter infall still continues around the central region.
Therefore, an energy budget to provide more explosion energy is still present.
The result suggests that if the massive progenitors are rapidly rotating, a high-energy supernova-like explosion could occur from the disk outflow triggered by viscous heating.
We discuss this point in \S \ref{subsec:transient}.

For all the collapse models, appreciable stellar matter is still present above the extraction radius.
This can contribute additionally to the ejecta mass and explosion energy.
%The masses outside the shock wave are 3.3, 0.8, 2.9, and $3.4M_\odot$ for models \texttt{T20}, \texttt{AD09x1}, \texttt{AD20x1}, and \texttt{AD20x2}, respectively.
The masses outside the extraction radius are larger than those of the ejecta, and thus, they may have an impact on the ejecta velocity.
The terminal average velocity of the ejecta depends on the amount of the stellar matter that becomes ejecta in the later phase.

% We estimate the binding energy above the shock wave for each model by analyzing that of the matter above the enclosed mass to which the shock wave reaches for the progenitor star. 
% The values are $\approx -\SI{7e50}{}$, $-\SI{1e50}{}$, $-\SI{7e50}{}$, and $-\SI{1.1e51}{erg}$ for models \texttt{T20}, \texttt{AD09x1}, \texttt{AD20x1}, and \texttt{AD20x2}, respectively (see the parenthesis in $E_\mathrm{exp}$ column of Table~\ref{tab:results}).
% We estimate the binding energy of the matter outside the extraction radius as in \S~\ref{subsec:diag}.
The mass and binding energy of the matter outside the extraction radii $r_\mathrm{ext}=\SI{1e9}{cm}$ and $\SI{2e9}{cm}$ are listed in Table~\ref{tab:results} (see the values in parentheses for columns of $M_\mathrm{ej}$ and $E_\mathrm{exp}$; also see Fig.~\ref{fig:ejecta}).
At the termination of the simulations, the absolute values of the binding energy are below the explosion energies for models \texttt{AD09x1}, and \texttt{AD20x2}.
Thus, the explosion is likely successful for these models.
For model \texttt{AD20x1}, on the other hand, the value is slightly above the explosion energy for $r_\mathrm{ext}=\SI{2e9}{cm}$.
Thus, it is not clear whether this model results in successful explosion.
Note however that the bound matter outside the extraction radius at the termination of the simulation has a sufficiently large angular momentum to circularize around the black hole (see Fig.~\ref{fig:ang}).
Thus, a possible increase may be expected in the explosion energy for the later phase if such a matter falls into the central region to power an additional outflow.
We plan to perform a longer-term simulation in our future work to address the possibility of the explosion for these models.

As already mentioned in \S~\ref{subsec:diag}, the composition of the computational domain is assumed to be that in NSE in our simulation.
In reality, the outer region of the star is composed mainly of $^{12}$C, $^{16}$O, and $^{20}$Ne, and hence, the nuclear burning of them in the stellar mantle swept by the shock wave can provide an additional energy.
% If an amount of matter composed either of $^{12}$C or $^{16}$O burn into $^{56}$Ni, released energy are, respectively,
% \begin{align}
% E_{{}^{12}\mathrm{C}\rightarrow {}^{56}\mathrm{Ni}} = \SI{1.8e51}{erg} \biggl(\frac{M_{{}^{12}\mathrm{C}}}{1M_\odot}\biggr),\\
% E_{{}^{16}\mathrm{O}\rightarrow {}^{56}\mathrm{Ni}} = \SI{1.3e51}{erg} \biggl(\frac{M_{{}^{16}\mathrm{O}}}{1M_\odot}\biggr),
% \end{align}
% which are proportional to the mass of the burning matter.
% If we suppose that the matter in the stellar mantle is composed purely of $^{12}$C and that the matter experienced the temperature higher than \SI{5}{GK} burns into $^{56}$Ni completely, the released energy are \SI{1.3e51}, \SI{8.2e49}, \SI{2.5e50}, and \SI{1.1e51}{erg} for models \texttt{T20}, \texttt{AD09x1}, \texttt{AD20x1}, and \texttt{AD20x2}, respectively.
% Note that these values would be an overestimated one because only a part of the matter burns into $^{56}$Ni usually and a part of the released energy can be lost by neutrino emission if the temperature becomes sufficiently large ($k_\mathrm{B}T\gtrsim\SI{1}{MeV}$).
% The above estimates indicate that the effect of nuclear burning may significantly contribute to the explosion energy.
In \S~\ref{subsec:Ni}, we will investigate this more quantitatively.

% \addsf{
% The nuclear burning of carbon and oxygen into iron group nuclei is not taken into account in our simulation and is potentially an additional energy source.
% As we will present in \S~\ref{subsec:Ni}, the contribution of the energy released by the nuclear burning is up to $\sim 10\%$, which is not negligible but does not change the explosion energy presented in this study.
% }

% For models \texttt{AD09x1} and \texttt{AD20x1}, for which the angular momentum of the progenitor stars is relatively low, the formation of the disk is delayed, and the onset of the explosion occurs in a late phase (see Table~\ref{tab:results}).
% As a consequence, only a small amount of mass remains outside the black hole at the explosion.
% The small energy budget at the formation of the disk leads to early saturation of the explosion energy at $t-t_\mathrm{exp}\approx 3$ and \SI{6}{s}, respectively.
% The saturated energy for these models are $\approx \SI{5e50}{}$, which are smaller than the canonical value for core-collapse supernovae.
% The ejecta masses for these models, on the other hand, are the order of $0.1M_\odot$, which are an order of magnitude smaller than that for canonical supernovae.
% Thus, the ejecta velocity ($\agt 2 \times 10^9$\,cm/s) is higher than that for canonical supernovae, and hence, this can lead to a rapidly evolving transient as found in \S~\ref{subsec:transient}.

\subsection{Comparison of the collapse models with model \texttt{BHdisk}} \label{subsec:bhdisk}

For model \texttt{BHdisk}, which consists of a $3M_\odot$ disk around a $10M_\odot$ black hole with the dimensionless spin of $0.6$, the disk matter accretes onto the black hole in a quasi-steady manner in the first 5 seconds.
The outflow is then launched at $t\approx\SI{5}{s}$ because the neutrino cooling rate has dropped far below the viscous heating rate (see the middle panel of Fig.~\ref{fig:disk}). 
The increase rate in the explosion energy for model \texttt{BHdisk} is much lower than for the collapsing star models (see bottom panel of Fig.~\ref{fig:ejecta}).
The primarily reason for this is that, for the collapsing stars, a velocity shear is present not only in the accretion disk with the nearly Keplerian motion but also on the surface of the disk resulting from the infalling matter onto the disk, which is absent for model \texttt{BHdisk}.
The strong shear on the disk surface significantly enhances the viscous heating rate, which results in the higher increase rate of the explosion energy.

The origin of the velocity shear associated with the infalling matter is different from the Keplerian motion of the disk.
Here, for the latter we suppose that the MRI turbulence \citep{Balbus1991a} is the origin of the effective viscosity.
For the surface region of the disk, on the other hand, we suppose that the shear region on the disk surface should induce the Kelvin-Helmholtz instability.
In such regions, magnetic fields are supposed to be enhanced significantly leading to the development of turbulence and dissipating the kinetic energy of the infalling matter (e.g., \citealt{Zhang2009feb, Obergaulinger2010jun, Rembiasz2016mar, Vigano2020jun}).
Thus, it is natural to consider that the effective viscosity on such a region is also high.\footnote{For this case, the kinetic viscosity is likely to be proportional to the infall velocity, not to the sound velocity, as $\nu=\ell_\mathrm{tur} v_\mathrm{infall}$, where $v_\mathrm{infall}$ is comparable to or larger than $c_s$, and thus, our present treatment for the viscous coefficient may be conservative if $\ell_\mathrm{tur}/H=O(10^{-2})$.}
However, to clarify this process in the first-principle way, magnetohydrodynamics simulation is necessary.
In the future work, we plan to perform this to confirm that our assumption is indeed correct.

For model \texttt{BHdisk}, we stopped the simulation at $t\approx \SI{21}{s}$.
For this model, the mass ejection still continues with the ejection rate slightly higher than the mass accretion rate onto the black hole.
This suggests that a large fraction of the disk matter will be eventually ejected from the system.
At the termination of the simulation, the total ejecta mass is only $\sim 0.1M_\odot$.
However, the disk mass is still $\approx 2M_\odot$.
Extrapolating the mass ejection rate at the final time $\sim 10^{-2}M_\odot$/s, we infer that the mass ejection continues for more than $100$\,s for this model.

The mass ejection for model \texttt{BHdisk} sets in after the neutrino cooling efficiency of the bulk of the disk drops (cf. the middle panel of Fig.~\ref{fig:disk}).
The ejecta for model \texttt{BHdisk} has a low-$Y_\mathrm{e}$ component down to $Y_\mathrm{e}\approx 0.4$.
The value is determined by the electron fraction in electron/positron capture equilibrium when the timescale of these reactions becomes comparable to that of the viscous expansion \citep[see][]{Fujibayashi2020c,Just2022jan}.
Thus, the mass ejection mechanism for rapidly rotating model \texttt{AD20x2} is qualitatively similar to that for model \texttt{BHdisk}.
% This is the likely reason for the result that the electron fraction of the matter after the cooling efficiency drops in the bulk of the disk for model \texttt{T20} is  similar to that for model \texttt{BHdisk}.
%, which resembles that for model \texttt{AD20x2}.
%The similarity for these two models is that a part of the ejecta experiences a condition of mild electron degeneracy during its expansion.
%Although the typical entropy and expansion timescale for model \texttt{BHdisk} are similar to those for rapidly rotating models \texttt{T20} and \texttt{AD20x2}, .

\subsection{Production of \texorpdfstring{$^{56}$}{56}Ni}\label{subsec:Ni}

Using the time evolution of the temperature and density along tracer particles, the post-process nucleosynthesis calculations are performed with the nuclear reaction network code \texttt{rNET} (\citealt{Wanajo2018a}).
The initial composition of the nucleosynthesis calculation depends on the thermal history of the tracer particles.
If the maximum temperature along a particle is higher than \SI{10}{GK}, we start the nucleosynthesis calculation at the time that the temperature decreases to $T=\SI{10}{GK}$ with the mass fraction of free protons and nucleons, $Y_\mathrm{e}$ and $1-Y_\mathrm{e}$, respectively.
For the tracer particles with the maximum temperature lower than \SI{10}{GK}, we start the nucleosynthesis calculation at $t=0$ of the simulation with the composition depending on the position of the particle in the progenitor star (mostly consisting of $^{16}$O and $^{20}$Ne).

The resulting mass of $^{56}$Ni is listed in Table~\ref{tab:results}.
For models \texttt{AD09x1} and \texttt{AD20x1}, the $^{56}$Ni masses are smaller than $0.1M_\odot$ (0.01 and $0.06M_\odot$, respectively), because of their smaller ejecta masses.
Thus for moderately rotating progenitor models, the $^{56}$Ni mass is likely to be comparable to or smaller than that for an ordinary supernova. 
These models predict the presence of moderately bright, but rapidly varying  optical transients as found in \S~\ref{subsec:transient}.

It is found that for model \texttt{AD20x2} the mass of $^{56}$Ni amounts to $0.15M_\odot$, reflecting the large mass of the ejecta that experiences high temperature $\gtrsim \SI{5}{GK}$.
Thus, the mass of $^{56}$Ni found for this model could be high enough for explaining high-energy supernovae such as broad-lined type Ic (type Ic-BL) supernovae, considering that the $^{56}$Ni mass inferred with the so-called ``Arnett's rule" (\citealt{Arnett1982}) is possibly overestimated (e.g., \citealt{Meza2020sep}, suggesting that the $^{56}$Ni masses inferred from the radioactive tail luminosity for nearby two type Ic-BL supernovae, SN2009bb and SN2016coi, are 0.08 and 0.10$M_\odot$, respectively).
These results suggest that massive and rapidly rotating stars leading to a black hole and massive disk are candidates for the progenitors of type Ic-BL supernovae. 

We also note that the numerical simulation for this model underestimates the total ejecta mass because we stopped the simulations at a time when the ejecta mass and explosion energy are still increasing.
Our results here indicate the lower bound for the $^{56}$Ni mass.

It is important to note that, as can be found in Table~\ref{tab:results}, the produced $^{56}$Ni mass fraction relative to the mass of ejecta exceeding 5~GK varies from 24\% (\texttt{AD20x2}) to 43\% (\texttt{AD20x1}) depending on the electron fraction, entropy, and expansion timescale of the outflowing matter for each model.
The conditions of $Y_\mathrm{e} \gtrsim 0.5$, low entropy, and slow expansion are favored for the efficient production of $^{56}$Ni.
Therefore, the mass of ejecta with  $\geq 5$~GK (as frequently used in the literature) only serves as a loose upper limit for the produced amount of $^{56}$Ni.

Suppose that $^{56}$Ni is synthesized from $^{16}$O, the rest-mass energy released into the internal energy due to the nuclear burning is $\SI{1.6e49}{}$, $\SI{7.5e49}{}$, and $\SI{1.9e50}{erg}$ for models \texttt{AD09x1}, \texttt{AD20x1}, and \texttt{AD20x2}, respectively.
These contributions are several to tens percents of the explosion energy estimated in \S~\ref{subsec:diag} and can have notable effects.
Especially for model \texttt{AD20x1}, the explosion energy plus binding energy above the extraction radius $r_\mathrm{ext}=\SI{2e9}{cm}$ becomes negative but the value is comparable to the energy generated by nuclear burning.
Thus, to clarify whether such a marginal model explodes successfully, feedback of the nuclear reaction has to be taken into account in hydrodynamics simulations (see \citealt{Bollig2020} and \citealt{Navo2022arxiv} for a recent attempt).

\section{Discussion} \label{sec:discussion}

\subsection{Optical transients} \label{subsec:transient}

Using the ejecta properties obtained in the present study, we analytically calculate bolometric luminosity models for photons following \cite{Arnett1982}.
The thermalization efficiency for gamma-rays is estimated following \cite{Colgate1997} with the optical depth for non-thermal gamma-rays $\kappa_\gamma = \SI{0.03}{cm^2/g}$.
For the optical depth of thermal photons, we simply set $\kappa=\SI{0.1}{cm^2/g}$.
We note that the ``Arnett" model tends to infer a larger $^{56}$Ni mass by a factor of a few than that inferred by the ``radioactive tail" luminosity of a supernova, the latter being less ambiguous (e.g., \citealt{Meza2020sep}, \citealt{Afsariardchi2021sep}, and \citealt{Rodriguez2022}).
This indicates that the luminosity predicted by the Arnett model for a given $^{56}$Ni mass may be underestimated by a factor of a few (see \citealt{Dessart2015oct,Dessart2016may} and \citealt{Khatami2019jun}).
We also note that for the rapidly rotating model \texttt{AD20x2}, light trans-iron elements could be synthesized in the ejecta (see \S~\ref{subsec:rprocess}), and hence, the opacity for optical wavelengths may be higher than $\SI{0.1}{cm^2/g}$.
For more quantitative study, we obviously need a radiation transfer simulation for photons taking into account a realistic opacity table.

\begin{figure}
\epsscale{1.17}
\includegraphics[width=0.48\textwidth]{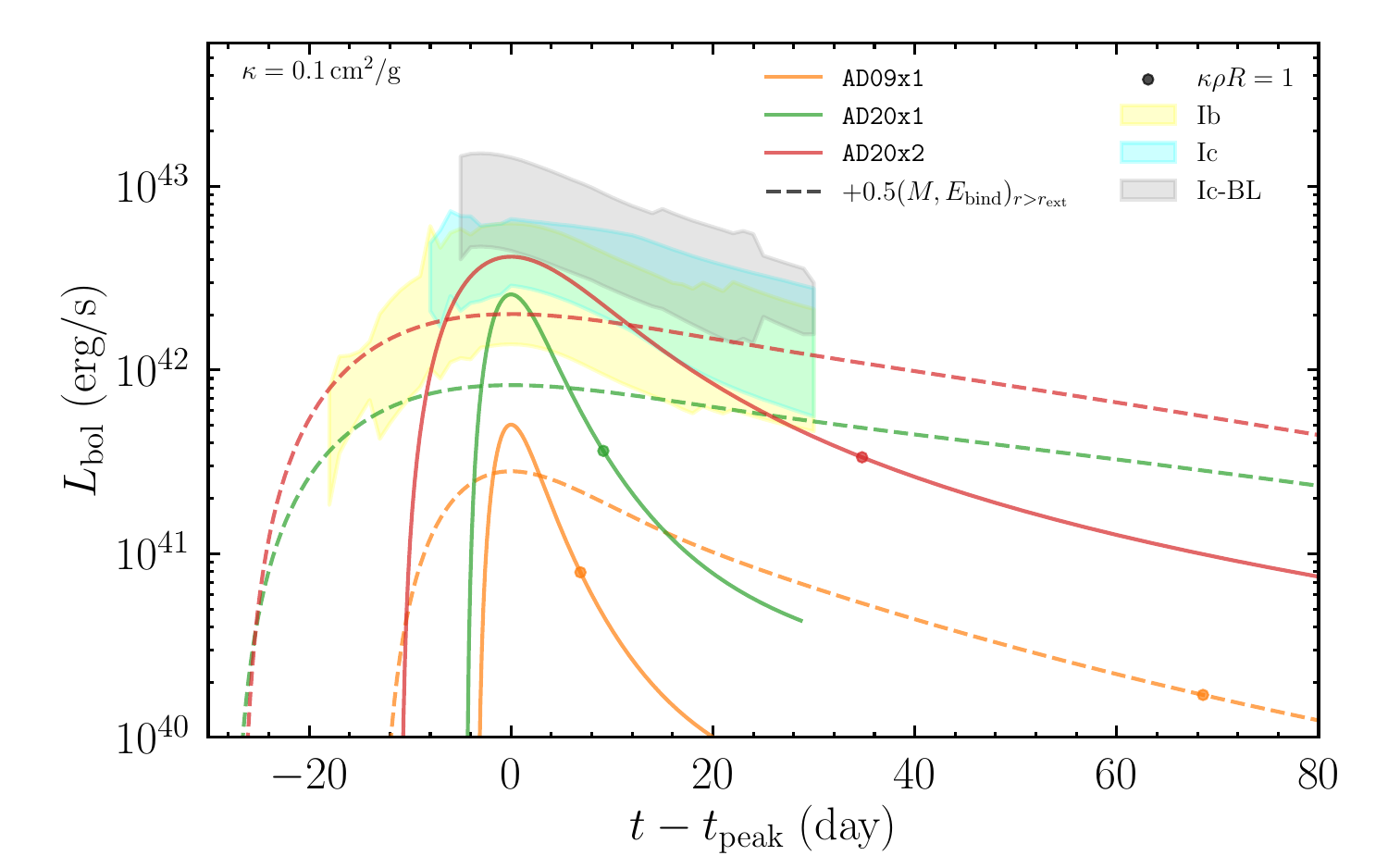}
\caption{
Bolometric light curve models.
The time origin is chosen to be the peak time for each curve.
The solid curves denote the light curves with ejecta mass and explosion energy extracted for $r_\mathrm{ext}=\SI{2e9}{cm}$ (see Table~\ref{tab:results} for the values).
The dashed curves denote the ones with the assumption that a half of the mass and binding energy outside the extraction radius contribute to the ejecta properties.
The point shows the time at which the ejecta becomes optically thin, $\tau=\kappa \rho R = 1$, to thermal photons for each model.
The shaded regions denote the templates of the bolometric light curves with standard deviations for type Ib, Ic, and Ic-BL supernovae taken from \cite{Lyman2016mar}.}

\label{fig:lbol}
\end{figure}

Figure~\ref{fig:lbol} shows the bolometric light curves for all the models investigated in this paper.
For deriving the solid curves of Fig.~\ref{fig:lbol}, we take into account only the ejecta mass and explosion energy extracted at $r_\mathrm{ext}=\SI{2e9}{cm}$.
For models \texttt{AD09x1} and \texttt{AD20x1}, the luminosity evolves rapidly.
The rise times, defined by the time until the maximum luminosity is reached, are $t_\mathrm{rise} \approx 3.3$ and 4.4\,d, respectively, for these models (see Table~\ref{tab:results}).
Such fast transients may be discovered in the future high-cadence transient surveys.
On the other hand, for model \texttt{AD20x2}, $t_\mathrm{rise}\approx\SI{10}{d}$, which is consistent with that of type Ib/c supernovae (see, e.g., \citealt{Taddia2018feb}).
This indicates that there could be a possible subclass of type Ib/c supernovae driven by the disk outflow in the black hole-forming core collapses of rotating massive stars.

To investigate the possible effects of the stellar matter outside the extraction radius, we also calculate the light curve for each model assuming that a half of the mass and binding energy of the matter for $r>r_\mathrm{ext}=\SI{2e9}{cm}$ contribute to the ejecta mass and explosion energy (see the dashed curves in Fig.~\ref{fig:lbol}).
Because of the increased ejecta mass and decreased explosion energy, peak luminosity and timescale for all the light curves become smaller and longer, respectively.

For rapidly rotating model \texttt{AD20x2}, the timescale of the light curve is longer than those typically found for type Ib/c supernovae.
\cite{Karamehmetoglu2022} reported such long-timescale type Ib/c supernovae recently.
The long-timescale supernovae are reported as the explosions of massive ($M_\mathrm{ZAMS}\gtrsim 25M_\odot$) stars with the explosion energies comparable to typical type Ib/c supernovae, but they are infered to have a larger amount ($\gtrsim 0.1M_\odot$) of $^{56}$Ni mass.
These facts indicate that our rapidly rotating model can explain such a subclass of type Ib/c supernovae.

For the moderately rotating models \texttt{AD09x1} and \texttt{AD20x1}, on the other hand, the timescales become $\approx 13$ and 28\,d, respectively, comparable to or longer than that of typical type Ib/c supernovae.
The peak luminosity is, however, about 10 times dimmer than that of typical type Ib/c supernovae.
This indicates that rotating massive stars exploded by outflows from black hole-disk systems may produce a variety of transients depending on the rotation profile of the progenitors and the presence of the stellar envelope, although the explosion mechanism is qualitatively universal.

As found in the comparison of the solid and dashed curves in Fig.~\ref{fig:lbol}, the features of the bolometric light curves depend on the possible contribution of the matter outside the extraction radius.
The quantitative prediction of the optical transients requires us to perform simulations for entire stars until the outer layer of the star is swept by the shock wave.

Another possible astrophysical transients can be powered by the interaction of the ejecta with a circum-stellar medium that can result from the strong mass loss of their progenitor prior to the stellar core collapse.
The progenitor models provided by \cite{Aguilera-Dena2018} are likely to be surrounded by a dense, massive ($\sim0.1$--$1M_\odot$ within $\sim10^{15}$\,cm) circum-stellar medium at the core collapse.
%%%%%%%% Simple estimation
Since the ejecta mass is an order of $0.1$--$1M_\odot$ for models \texttt{AD09x1} and \texttt{AD20x1}, the ejecta will be significantly decelerated in the circum-stellar medium, releasing a substantial fraction of its kinetic energy $\alt 10^{51}$\,erg.
The optical depth of the circum-stellar medium is estimated as
\begin{align}
\tau_\mathrm{CSM} \sim \frac{3\kappa M_\mathrm{CSM}}{4\pi {R_\mathrm{CSM}}^2} \approx &\, 170 \biggl(\frac{\kappa}{\SI{0.35}{cm^2/g}}\biggr)\notag\\
&\times \biggl(\frac{M_\mathrm{CSM}}{1M_\odot}\biggr)\biggl(\frac{R_\mathrm{CSM}}{10^{15}\,\mathrm{cm}}\biggr)^{-2}.
\end{align}
Here, the opacity for photons is assumed to be dominated by the Thomson scattering of fully ionized medium with $Y_\mathrm{e}=0.875$ (i.e., hydrogen and helium with mass fractions 0.75 and 0.25, respectively).
Since the circum-stellar medium is optically thick, the released energy diffuses out from the circum-stellar medium with the diffusion time.
The luminosity is then estimated as
\begin{align}
L & \sim \frac{\epsilon E_\mathrm{exp}}{t_\mathrm{diff}}\cdot \frac{R_\mathrm{CSM}}{v_\infty t_\mathrm{diff}} \notag \\
& \approx \SI{5.5e43}{erg/s}\biggl(\frac{\kappa}{\SI{0.35}{cm^2/g}}\biggr)^{-1}\notag\\
& ~~~\times \biggl(\frac{\epsilon}{0.1}\biggr)\biggl(\frac{v_\infty}{\SI{1e9}{cm/s}}\biggr)^2\biggl(\frac{R_\mathrm{CSM}}{10^{15}\,\mathrm{cm}}\biggr), \label{eq:lum_shock}
\end{align}
where $v_\infty = \sqrt{ 2 E_\mathrm{exp}/(M_\mathrm{ej}+M_\mathrm{CSM})}$ is the terminal velocity of the ejecta plus circum-stellar medium, $\epsilon$ is the radiation efficiency, and
\begin{align}
t_\mathrm{diff} &\sim \sqrt{\frac{\kappa(M_\mathrm{ej}+M_\mathrm{CSM})}{4\pi v_\infty c}}\notag\\
&\approx \SI{26}{d}\biggl(\frac{\kappa}{\SI{0.35}{cm^2/g}}\biggr)^{1/2}\notag\\
&~~~~~\times \biggl(\frac{M_\mathrm{ej}+M_\mathrm{CSM}}{2M_\odot}\biggr)^{3/4}\biggl(\frac{E_\mathrm{exp}}{10^{51}\,\mathrm{erg}}\biggr)^{-1/4}, \label{eq:difftime}
\end{align}
is the diffusion time of the expanding ejecta plus circum-stellar medium (see, e.g., \citealt{Matsumoto2022sep} for similar expression).
The second factor of the first line in Eq.~\eqref{eq:lum_shock} is the contribution of the adiabatic cooling.
%This can lead to an optical transient as bright as superluminous supernovae (see, e.g., \citealt{Moriya2018mar} for a review) with a timescale of $\lesssim$\,a month.
This can naturally lead to an optical transient like superluminous supernovae (see, e.g., \citealt{Moriya2018mar} for a review).
We note, however, that the properties of the transient depend not only on the mass and radius but also on the density profile of the circum-stellar medium (see, e.g., \citealt{Chevalier2011mar} and \citealt{Suzuki2020aug} for the curcum-stellar medium like a stationary wind).

\subsection{Possible synthesis of light trans-iron nuclei}\label{subsec:rprocess}

\begin{figure}[t]
\epsscale{1.17}
\includegraphics[width=0.48\textwidth]{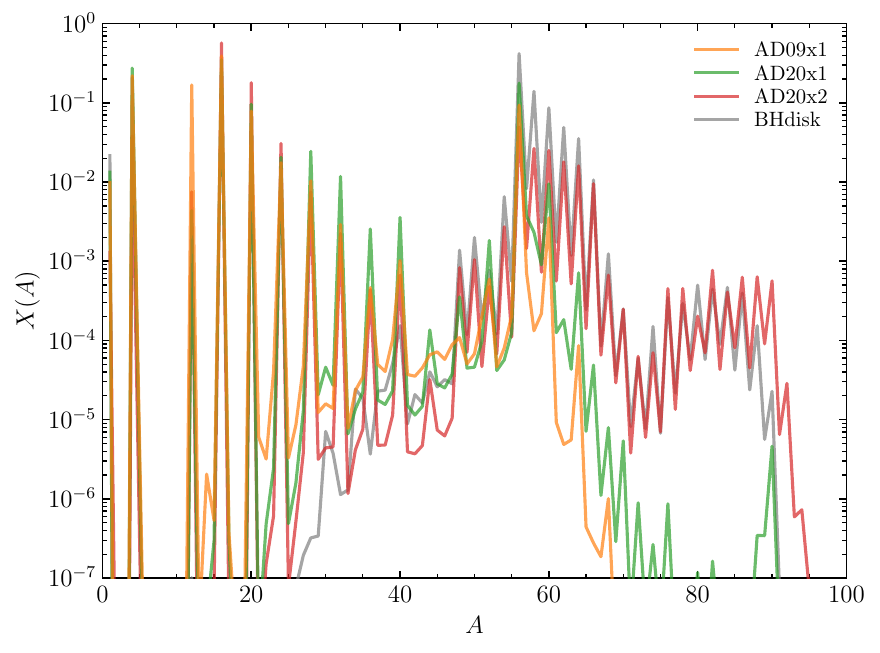}
\caption{
Isobaric mass fraction for our models.
Nuclei at $A = 12$ and 16 are, respectively, predominantly unprocessed $^{12}$C and $^{16}$O in the progenitor stars.
}
\label{fig:aabunx}
\end{figure}

Figure~\ref{fig:aabunx} shows isobaric mass fractions obtained by the nucleosynthesis calculations for our models.
%Since most of the ejecta has the electron fraction close to 0.5, there are peaks at alpha-nuclei and $A=56$.
We find prominent peaks at alpha-nuclei (with $A$ multiple of 4) and $A=56$.
The peaks at $A=12$ and 16 reflect the initial composition of $^{12}$C and $^{16}$O, respectively, in progenitor stars.
The bulk of these nuclei remains unprocessed owing to relatively low temperature achieved.
On the other hand, the nuclei at peaks of $A=20$--40 (corresponding to elements Ne, Mg, Si, S, Ar, and Ca) are synthesized from $^{12}$C and $^{16}$O in the ejecta that experience higher temperature but lower than that required for achieving NSE.
The peak at $A=56$ corresponds to $^{56}$Ni, which is synthesized predominantly in NSE with $Y_\mathrm{e}\gtrsim0.5$.
Interestingly, a certain amount of nuclei heavier than the iron group ($A\gtrsim 60$), up to $A\approx 90$, is found to be synthesized for a rapidly rotating model \texttt{AD20x2} (as well as \texttt{BHdisk}).
It is known that in slightly neutron rich ($Y_\mathrm{e}\gtrsim 0.4$) ejecta, such trans-iron nuclei are synthesized predominantly in quasi-nuclear statistical equilibrium (QSE, \citealt{Meyer1998may,Wanajo2018a}) under an alpha-rich condition.
Since the neutron-richness is not very high, heavy $r$-process elements with $A > 100$ are not synthesized for any of the present models.

The first peak nuclei of $r$-process, especially Zr and Y (synthesized in QSE here), are known to have opacities higher than those of iron-group elements~\citep{Kawaguchi2021jun}.
Therefore, if such elements are appreciably synthesized, the resulting optical transients may have longer timescales than those without such elements.
In addition, the peak luminosity will be lower and the spectrum could be redder.
To quantify the light curve and spectrum, a radiation transfer simulation is needed in future work.

% for AD20x2, X(Zr+Y) ~ 0.0030.

\subsection{Implications for gamma-ray bursts}
\begin{figure}[t]
\epsscale{1.17}
\includegraphics[width=0.48\textwidth]{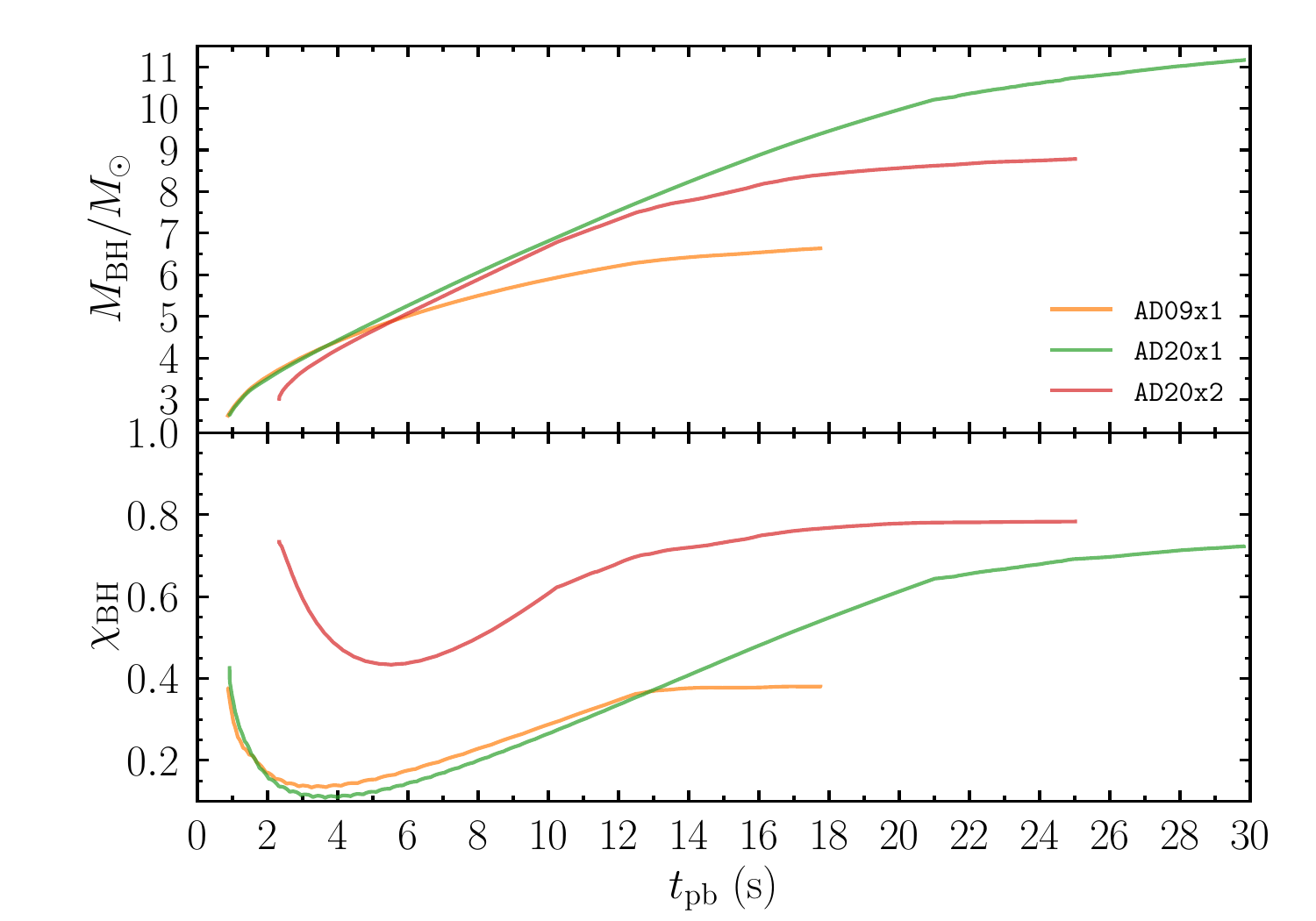}
\caption{
Masses (top) and dimensionless spins (bottom) of black holes for our models as functions of post-bounce time.
}
\label{fig:bh}
\end{figure}

Figure~\ref{fig:bh} shows the masses (top) and dimensionless spins (bottom) of the black holes for our models as functions of post-bounce time.
Here, the mass and dimensionless spin of the black holes are estimated from the equatorial and polar circumference radii of the apparent horizon (e.g., \citealt{Shibata2016a}).

For models \texttt{AD09x1} and \texttt{AD20x1}, in which we use the original angular momentum distribution of the stellar evolution simulations, the mass accretion and thus the spin up of the black hole are suppressed because of the late-time disk formation and the quicker launch of the outflow from the disk.
The dimensionless spins for these models are $\sim 0.4$ and 0.7 respectively.
For these models, the Blandford-Znajek mechanism \citep{Blandford1977}, which is one of the most promising mechanisms to power gamma-ray bursts (e.g., \citealt{Gottlieb2022mar}), may provide only moderately large Poynting luminosity for launching intense electromagnetic waves because of its strong dependence on the black-hole spin.
Moreover, because of the low density of the disk for these models, strong magnetic fields may not be sustained in the vicinity of the black hole.
This indicates that more rapidly rotating progenitors than those of \texttt{AD09} and \texttt{AD20} would be preferred for generating powerful relativistic jets.

For the rapidly rotating model \texttt{AD20x2}, the mass accretion from the disk onto the black hole continues for a long timescale ($\agt 20$\,s) because of the presence of the long-term high-mass-accretion rate phase.
This leads to a rapidly spinning black hole, the dimensionless spin of which is $\sim 0.8$ at the termination of the simulations.
During the long-term evolution of the black hole, the accretion disk is likely to be in a turbulent state associated with magnetohydrodynamical instabitlities such as the MRI \citep{Balbus1991a} in a realistic situation.
By this, the magnetic-field strength should be enhanced, and the magnetic flux penetrating the black hole is increased as a result of the mass accretion (and magnetic flux accretion) onto the black hole.
Such a highly spinning black hole penetrated by a strong magnetic field could be a promising central engine for relativistic jet via the Blandford-Znajek process (e.g., \cite{Christie2019dec,Hayashi2022} for a related topic).

It is known that at least a fraction of long-duration gamma-ray bursts are accompanied by type Ic-BL supernovae (e.g., \citealt{Cano2017a}). 
As found in \S~\ref{subsec:Ni}, rapidly rotating models synthesize a large amount of $^{56}$Ni ($>0.1M_\odot$), which is consistent with required amount to explain the light curves of type Ic-BL supernovae.
In \S~\ref{subsec:exp}, it is also found that the rapidly rotating models show a large average velocity of the ejecta $\gtrsim \SI{2e9}{cm/s}$, which is necessary for the broad-line features for these supernovae.
In addition, the large explosion energy $\gtrsim 10^{52}$\,erg observationally inferred from type Ic-BL supernovae is likely achieved if we consider longer evolution of the system than simulated in this study.
Therefore, the rapidly rotating models in this study may reasonably represent the supernovae accompanying long-duration gamma-ray bursts.

\subsection{Possible effects of relativistic jet}
We briefly discuss possible effects of a relativistic jet that may be launched in the polar direction by some mechanisms, accompanied with the formation of a rapidly rotating black hole. 
Here, we suppose that the jet is launched during the phase of a high mass accretion rate onto the black hole.

If a jet is powerful enough, it may partly prevent the infall of the stellar matter~\citep{Tominaga2009jan}.
If so, the ram pressure of the infalling matter decreases, and as a result, the outflow from the disk surface is launched earlier.
The decrease of the matter infall also prevents the matter supply to the disk, and thus, the total energy for the outflow may be reduced.
%Therefore, the outflow energy may also be reduced.

Another possible effect of the relativistic jet is that the energy injection by the jet can be a source of $^{56}$Ni production~(see, e.g., \citealt{Tominaga2007}, \citealt{Barnes2018jun}, and \citealt{Leung2023arxiv}).
If the jet luminosity is high enough ($\gtrsim 10^{53}$\,erg/s; \citealt{Tominaga2007}), a significant amount of $^{56}$Ni ($\gtrsim 0.1M_\odot$) may be synthesized and the optical transient may become more luminous.
The energy injection by the jet may also modify the morphology of the ejecta, which can affect the features of the optical transient.

\section{Summary} \label{sec:summary}

In this paper, we studied the explosion in the rotating massive-star collapse leading to a black hole and a massive disk in fully general relativistic radiation-viscous-hydrodynamics simulations with an approximate neutrino radiation transfer, employing evolved stars with a compact core (\citealt{Aguilera-Dena2020oct}) as the initial conditions.
We adopt the original or doubled angular velocity for models \texttt{AD09} and \texttt{AD20}.
For all the models investigated in this paper, we found the formation of an accretion disk after a proto-neutron star collapsed into a black hole although the time at the onset of the disk formation depends strongly on the rotational profile of the progenitor stars.
The evolution after the disk formation was also found to depend strongly on the degree of rotation of the progenitor stars.

For moderately rotating models, \texttt{AD09x1} and \texttt{AD20x1}, for which the original angular momentum profiles obtained in stellar evolution calculations are employed, the rest-mass density and the neutrino cooling efficiency were already low at the disk formation.
For these models, the viscous heating efficiency is always higher than the neutrino cooling one after the disk formation, and thus, no NDAF phase is established.
As a result, the outflow is launched before a massive disk is formed, and the explosion occurs at several hundreds of millisecond after the disk formation.
Because the disk starts forming at a late stage of the stellar collapse (at $t_\mathrm{pb} > 10$\,s), the mass of the envelop is relatively small, and hence, the energy budget is small.
As a consequence, the ejecta mass and explosion energy are relatively small as $\sim 0.1M_\odot$ and $\approx$ (0.5--1) $\times 10^{51}$\,erg, respectively. 
The electron fraction of the ejecta is always higher than 0.47, and thus, an $r$-process nucleosynthesis cannot proceed in the ejecta for these models.
%\delsf{Their small ejecta mass, as well as large ejecta velocity ($\agt 2\times 10^9$\,cm/s), can lead to a rapid transient with timescales of several days.}

For the model with a rapidly rotating progenitor \texttt{AD20x2}, the mass infall rate to the disk and the black hole are high ($\dot M_\mathrm{fall} \agt 0.3M_\odot/{\rm s}$).
As a result, the disk settles to an NDAF phase and evolves quasi-steadily prior to the onset of the outflow, which starts $\sim \SI{10}{s}$ after the disk formation.
The outflow is launched from the surface of the disk after the mass infall rate decreases (i.e., the ram pressure of the infalling matter drops).
At the launch of the outflow, the neutrino cooling efficiency around the disk surface is lower than that deeper in the disk.
The ejecta mass and explosion energy amount to $\agt 1M_\odot$ and $\agt 3\times10^{51}$\,erg, respectively.
The electron fraction of the ejecta is at lowest $\approx 0.4$, which is still not sufficient for an $r$-process nucleosynthesis.
Indeed, the nucleosynthesis calculation shows that heavy nuclei are synthesized at most up to $A\approx 100$.
However, there is low-electron fraction ($Y_\mathrm{e}<0.2$) matter deep inside the disk in which the rest-mass density is high enough to enhance the electron degeneracy.
If such components were ejected by a very efficient mass ejection process with a shorter timescale, e.g., by magnetohydrodynamics processes, $r$-process elements might be synthesized. Magnetohydrodynamics simulation for this problem is one of our future issues.

For moderately rotating models, the synthesized $^{56}$Ni mass is less than $0.1M_\odot$, and the luminosity of the supernova-like explosion is inferred to be comparable to those of the ordinary supernovae.
By contrast, for the rapidly rotating model \texttt{AD20x2}, the synthesized $^{56}$Ni mass is larger than $0.1M_\odot$, and hence, a luminous supernova-like explosion may be expected.
The bolometric light curve for this model is suitable for a light-curve model of type Ib/c supernovae.
This suggests that there might be a possible subclass of bright stripped-envelope supernovae driven by the outflow from a massive disk around a rapidly spinning black hole formed from the collapse of a massive rotating star.
The possible existence of high-opacity trans-iron elements (such as Y and Zr) may lead to a longer timescale and redder transient.

Depending on the effect of the mass in the outer layer of the star, the resulting optical transient can have a very short timescale (a few days) or that comparable to normal supernovae.
If dense and massive circum-stellar media are present as predicted in \cite{Aguilera-Dena2018}, a very bright ($\sim 10^{44}$\,erg/s) transient with a timescale of months is expected due to the interaction of ejecta with the circum-stellar medium.

To more rigorously predict observational features (photometric luminosity and spectra) of optical transients  based on our present results, we need to perform a photon-radiation transfer simulation. The inclusion of high-opacity trans-iron elements, which could exist in the ejecta, may drastically change the observational feature, which will be investigated in our future work.

We employed viscous hydrodynamics to incorporate angular momentum transport and dissipation of kinetic energy to internal energy in the region in which a velocity shear or differential rotation is present. This enables us to approximately capture the effective viscosity induced by the magnetohydrodynamical turbulence.
However, it is obviously necessary to perform first-principle magnetohydrodynamics simulations in order to strictly explore the effects of the angular momentum transport and (effectively) viscous dissipation. Thus, three-dimensional radiation-magnetohydrodynamics simulation is necessary in future work. 

A missing but potentially important ingredient of the scenario presented in this work is the possible existence of a relativistic jet.
The disk evolution may be affected by this because the history of mass supply is modified by the feedback of the jet.
The relativistic jet, if powerful enough, can also synthesize a significant amount of $^{56}$Ni, which makes the optical transient more luminous.
It can also modify the ejecta morphology, and as a result, may affect the features of the optical transient. These possible effects are also the issues to be investigated in our future work.

\acknowledgements

We thank Kunihito Ioka, Keiichi Maeda (at Kyoto University), and Takashi Moriya for helpful discussions and Koh Takahashi and David Aguilera-Dena for providing their stellar evolution models.
Numerical computation was performed on Sakura, Cobra, and Raven clusters at Max Planck Computing and Data Facility. 
This work was in part supported by Grant-in-Aid for Scientific Research (grant Nos.~20H00158 and 23H04900) of Japanese MEXT/JSPS.

\appendix

\section{A high mass supply rate case}\label{appA}

\begin{figure*}
\epsscale{1.17}
(a)\includegraphics[width=0.33\textwidth]{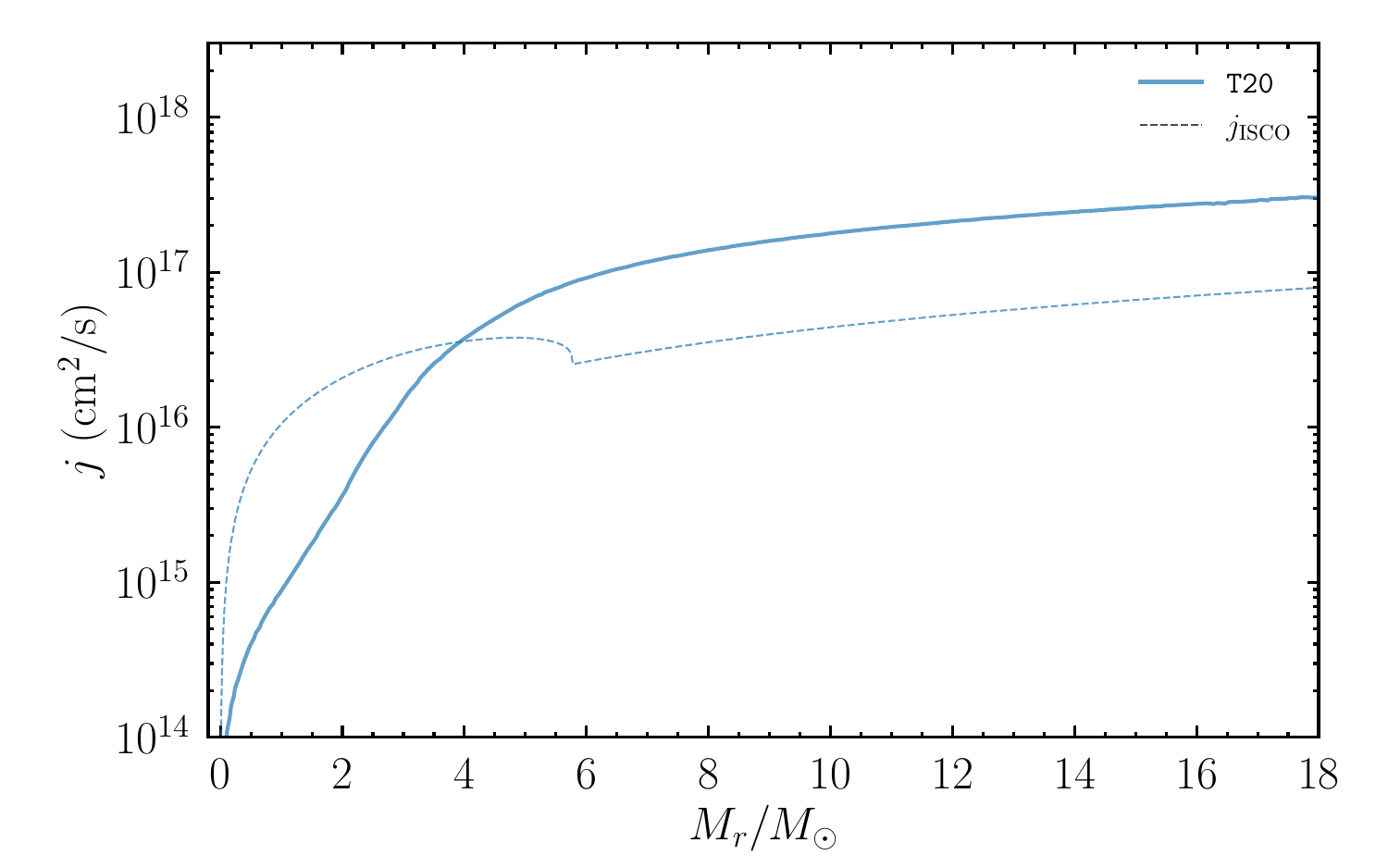}
(b)\includegraphics[width=0.33\textwidth]{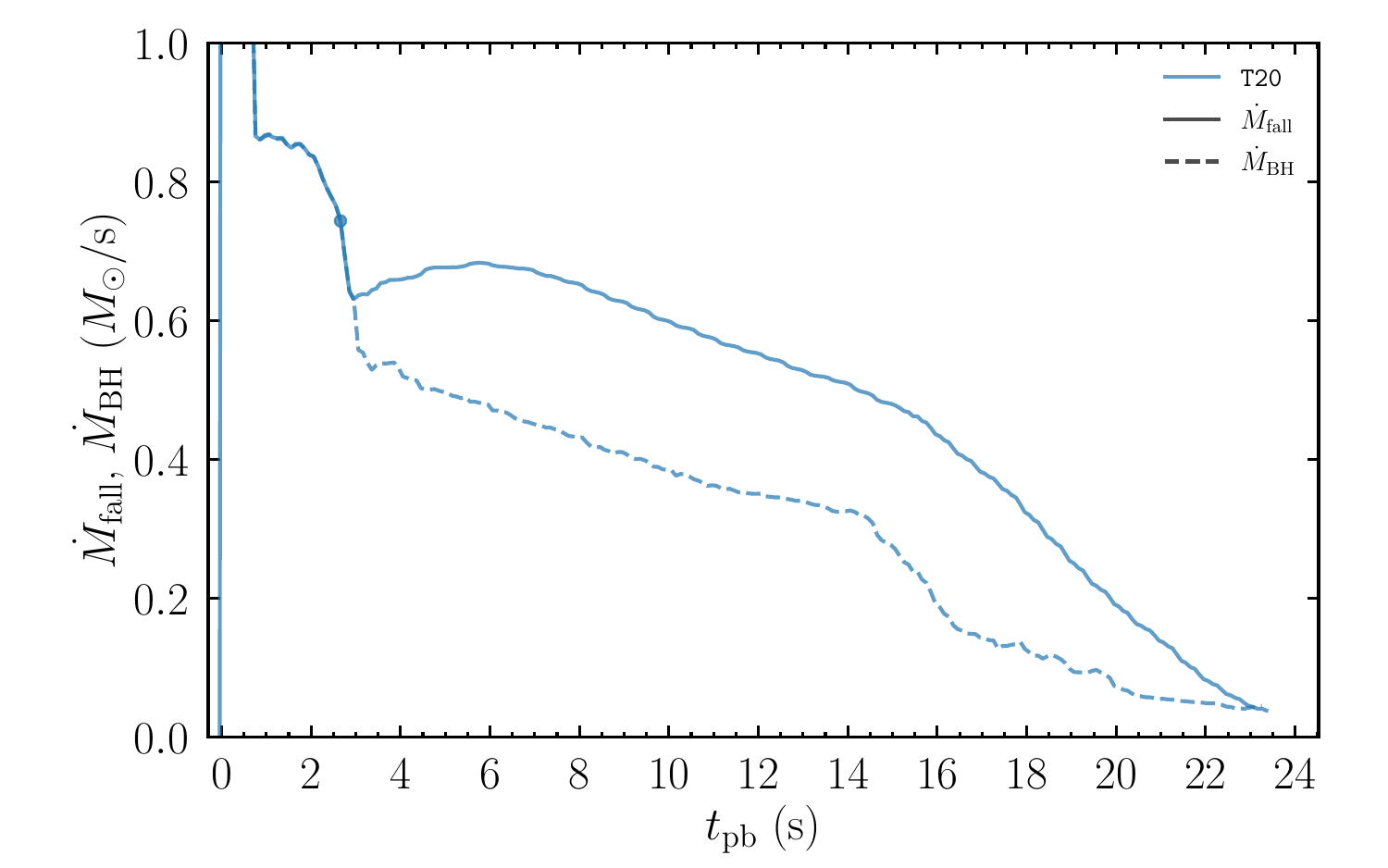}
(c)\includegraphics[width=0.33\textwidth]{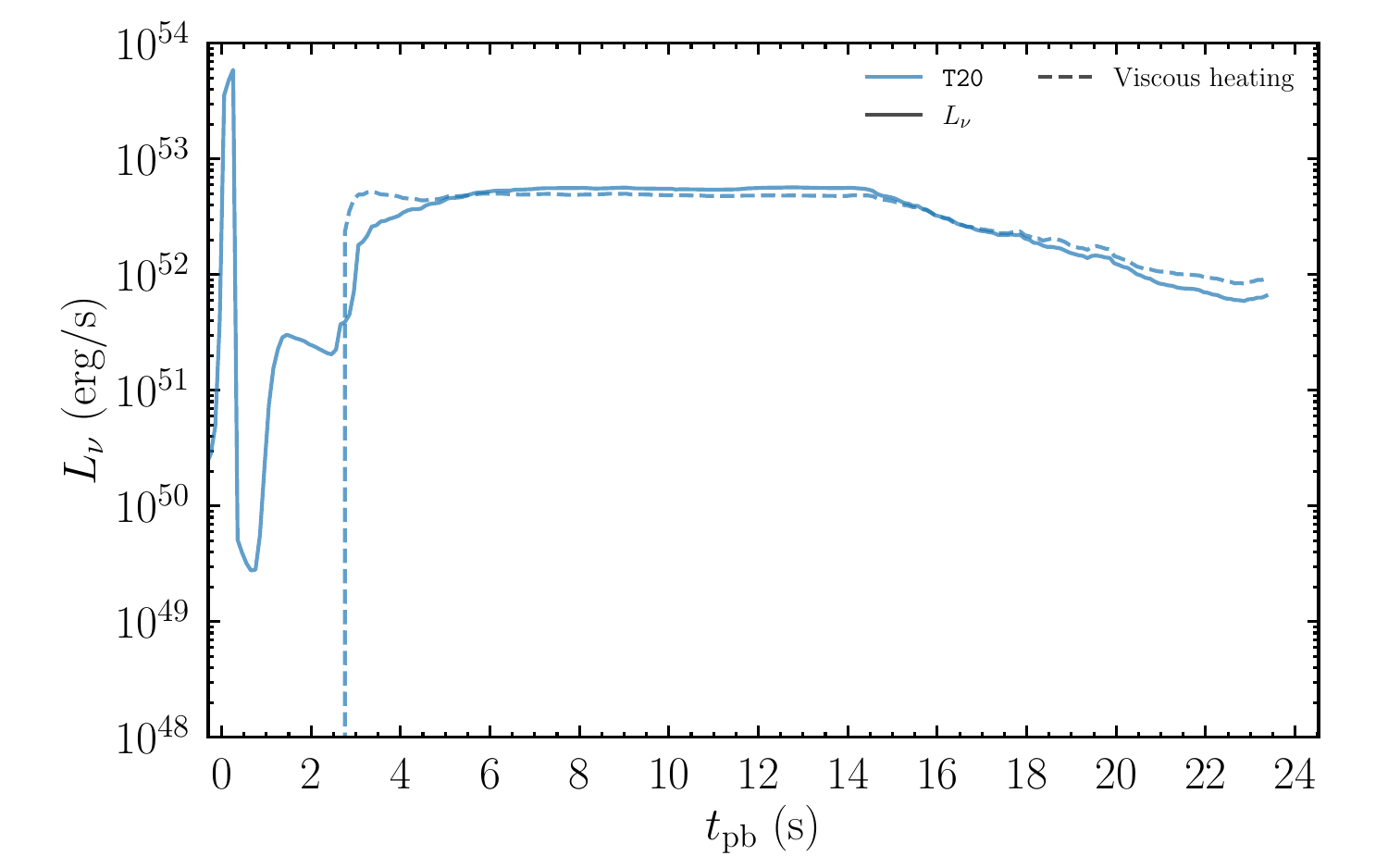}\\
(d)\includegraphics[width=0.33\textwidth]{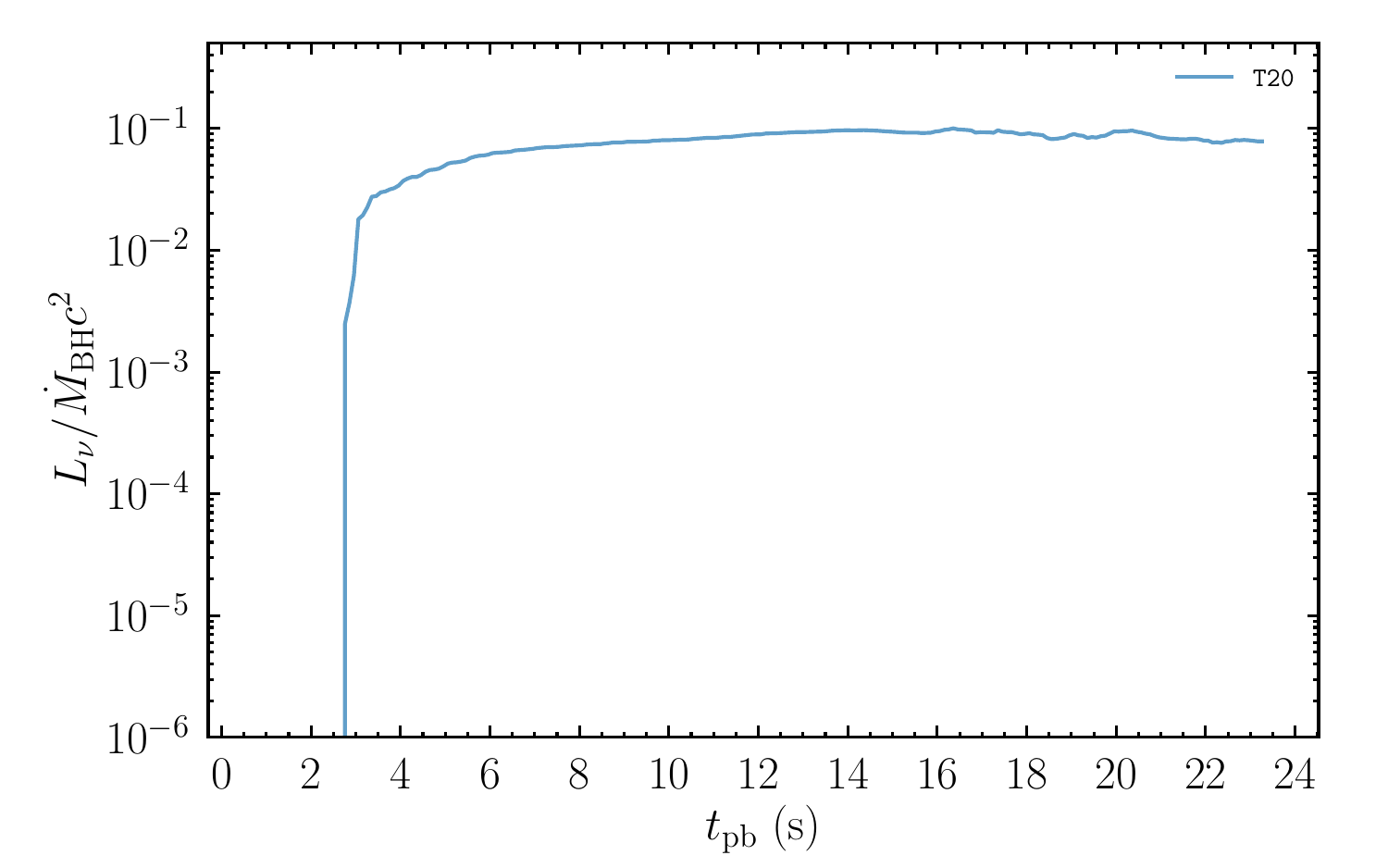}
(e)\includegraphics[width=0.33\textwidth]{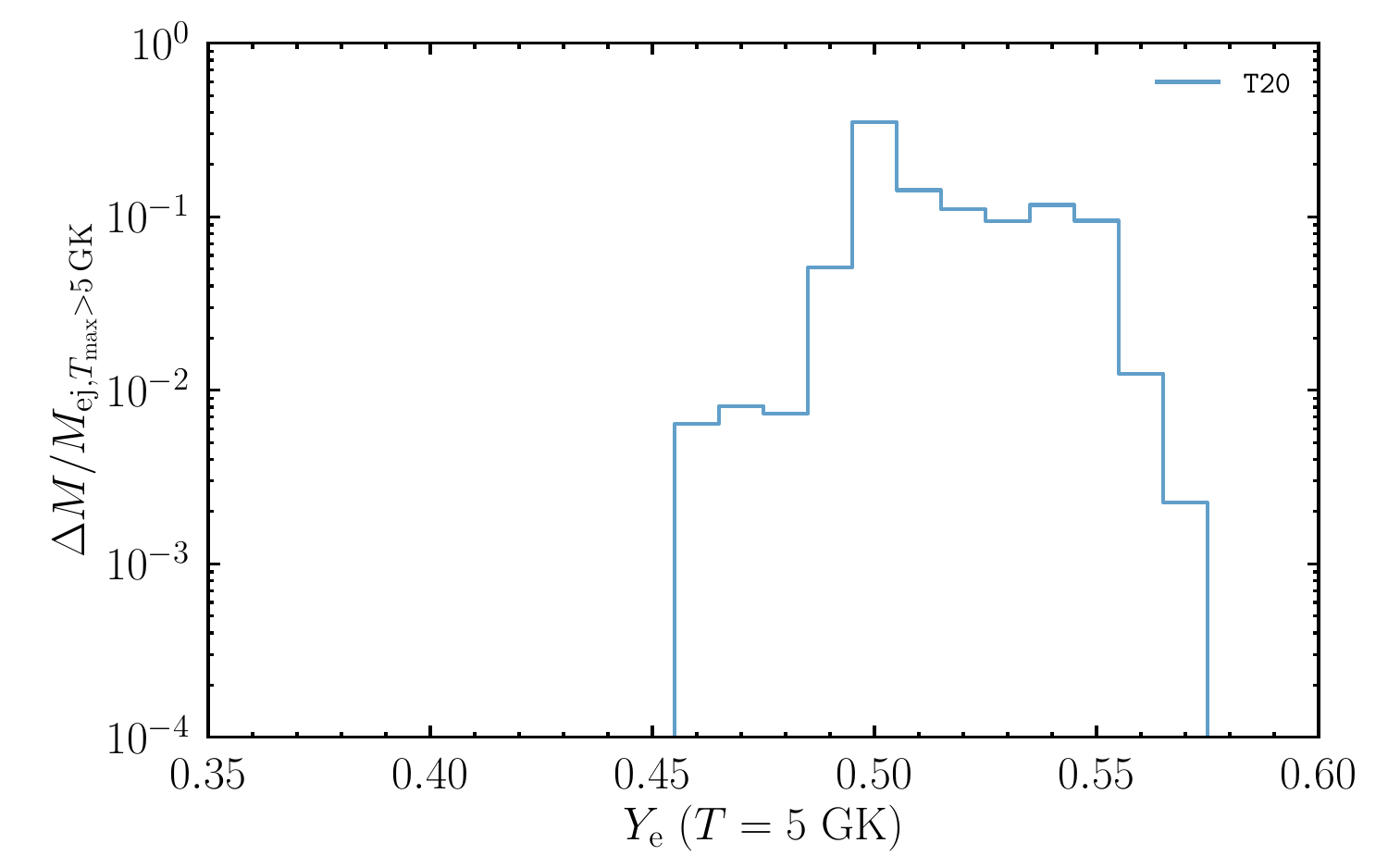}
(f)\includegraphics[width=0.33\textwidth]{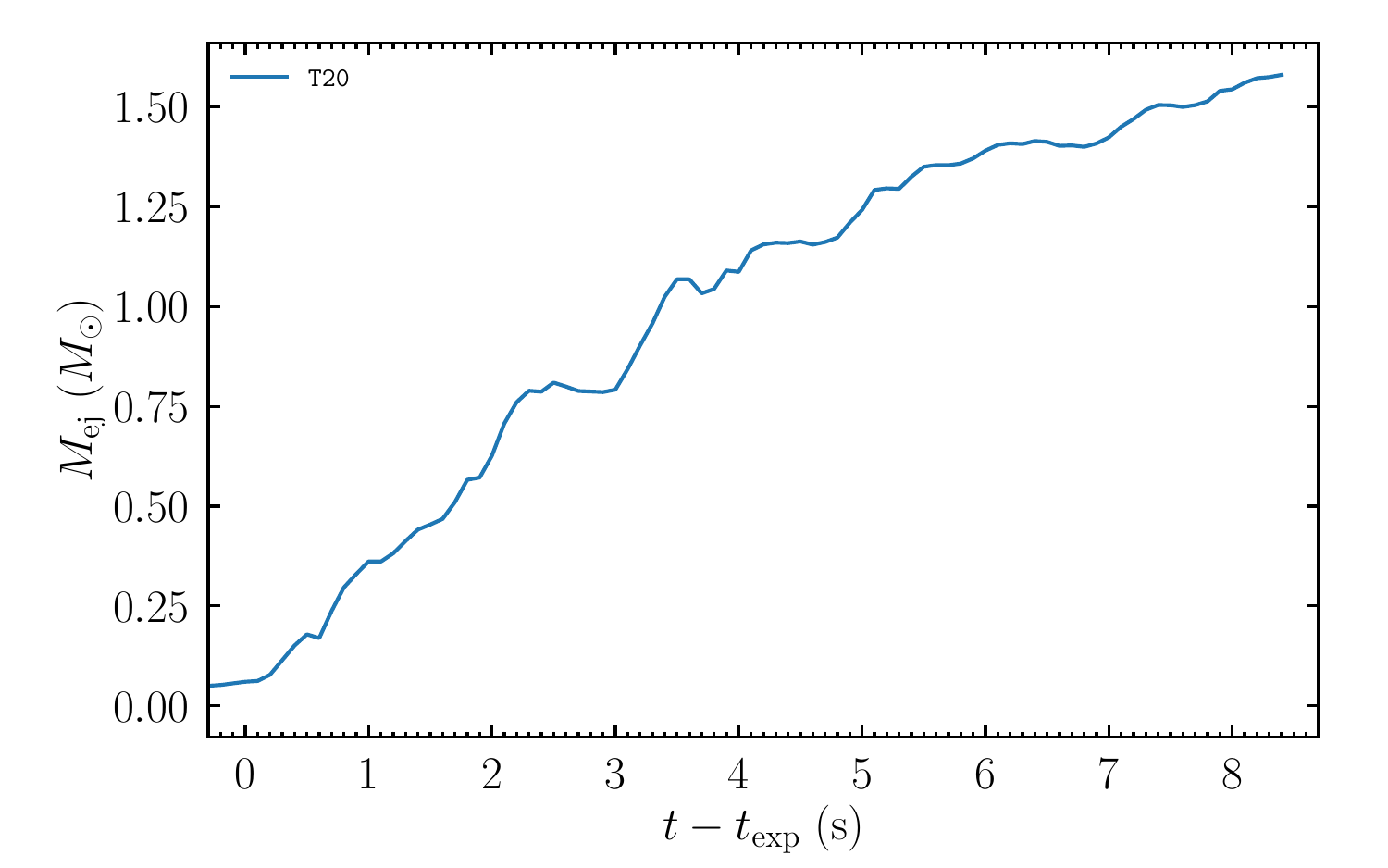}\\
(g)\includegraphics[width=0.33\textwidth]{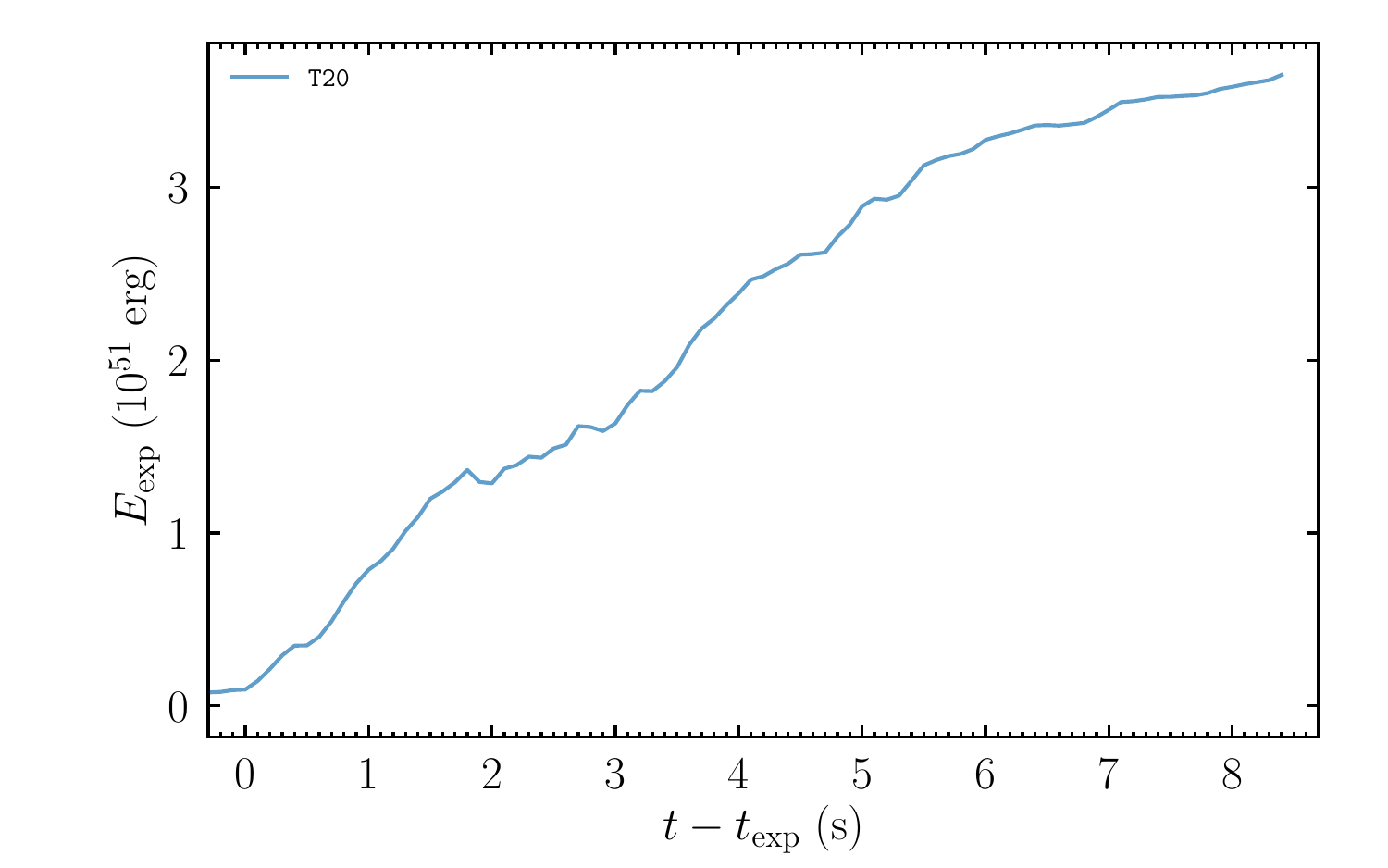}
(i)\includegraphics[width=0.33\textwidth]{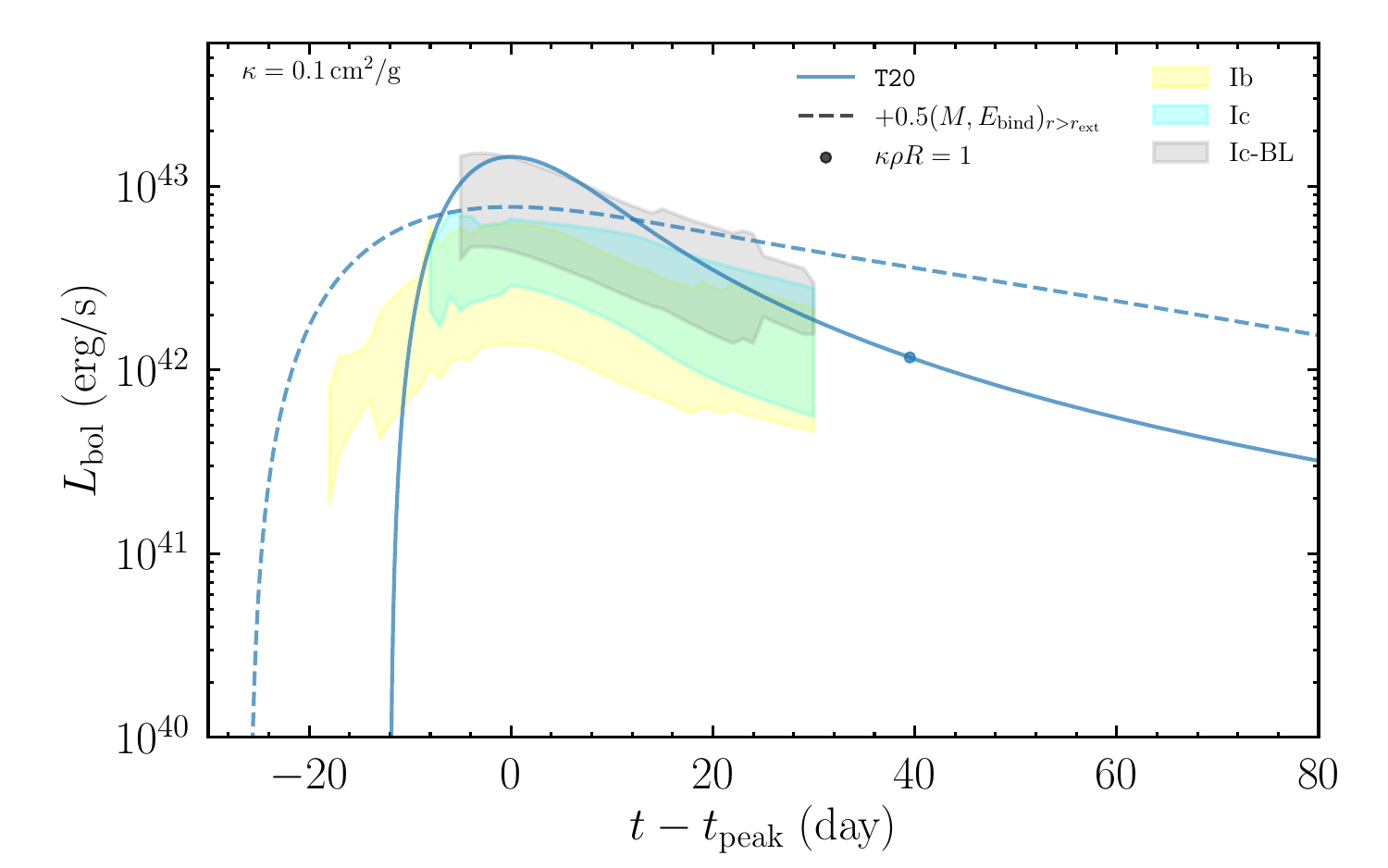}
(h)\includegraphics[width=0.33\textwidth]{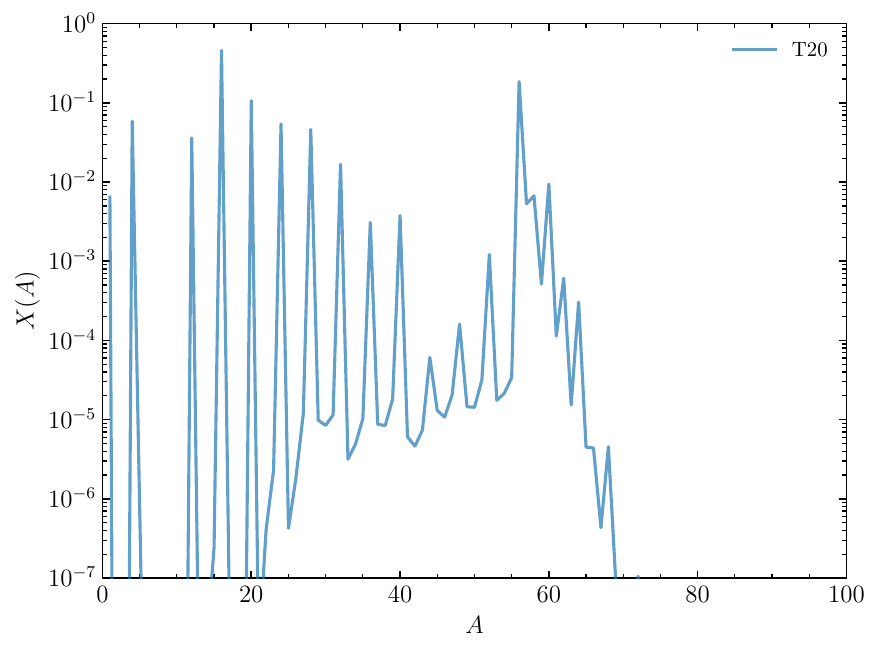}
\caption{
The same figures as Figs.~\ref{fig:ang}, \ref{fig:disk}, \ref{fig:hists}--\ref{fig:aabunx} but for model \texttt{T20}.
}
\label{fig:T20}
\end{figure*}

\begin{figure*}
\epsscale{1.17}
\includegraphics[width=0.5\textwidth]{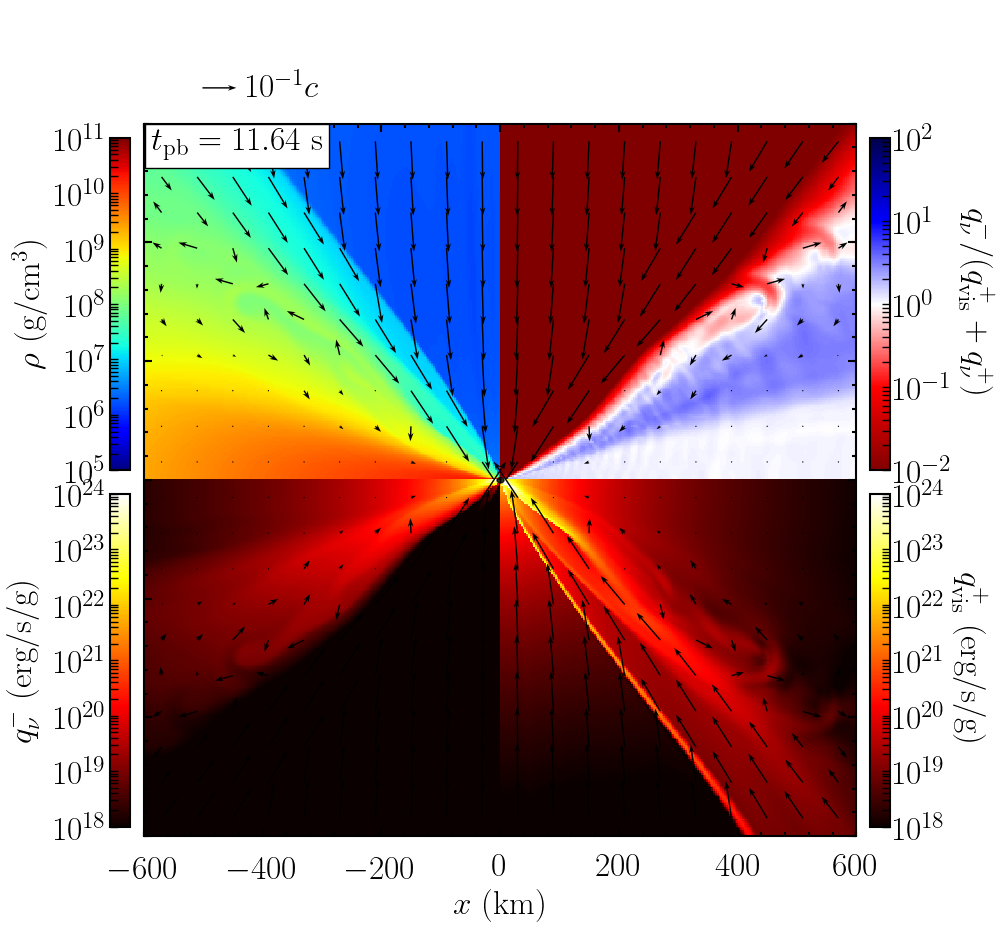}
\includegraphics[width=0.5\textwidth]{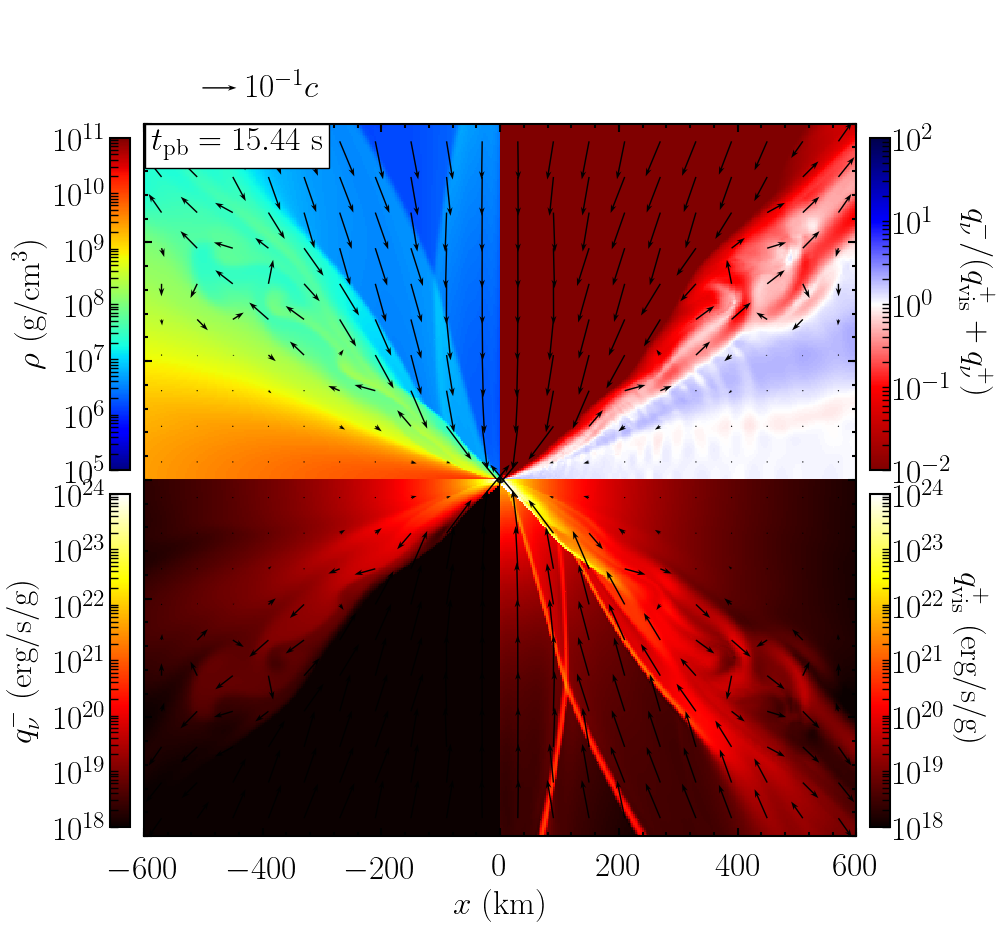}
\caption{
Density (top-left), neutrino cooling rate density (bottom-left), cooling efficiency (top-right), and viscous heating rate density (bottom-right) for model \texttt{T20} at $t_\mathrm{bp}\approx 11.6$ (left panel) and \SI{15.4}{s} (right panel), respectively. 
}
\label{fig:efflocal}
\end{figure*}

We here present an additional result for a model in which we assume a hypothetically rapid rotation. The motivation to consider this additional model is (i) to explore the case for which a disk is formed in a short timescale after the formation of a black hole and (ii) to understand the evolution process of the system in which the matter infalling rate onto the disk is very high.
%We employ the final states of massive stars evolved in stellar evolution calculations as the initial conditions of our numerical-relativity simulations. 

For these purposes, we employ the star evolved from a helium star with the initial mass of $M_\mathrm{He}=20M_\odot$, which is taken from \cite{Takahashi2018} (hereafter referred to as model \texttt{T20}).
This progenitor star also has a very compact core with $\xi_{2.5}=0.74$.
Because the progenitor is non-rotating, we add an artificial rotation profile (although this is not self-consistent to the stellar model but we attempt to mimic the profile of more realistic models of \citealt{Aguilera-Dena2020oct}) as
\begin{align}
\Omega(r) = \frac{1}{1/\Omega_0 + 1/\Omega_{r_\mathrm{K}=r_0}(r)}, \label{eq:omega_ana}
\end{align}
where $\Omega_0 = \SI{0.1}{rad/s}$ and 
\begin{align}
\Omega_{r_\mathrm{K}=r_0}(r) = \frac{\sqrt{Gm(r)\cdot r_0}}{r^2}
\end{align}
is the angular velocity for which the fluid element settles into a Keplerian motion with the cylindrical radius of $r_0$.
Note that the shellular rotation is assumed and the initial angular velocity is given as a function only of the spherical radius $r$.
In this study, we set $r_0=\SI{1000}{km}$.
This implies that the maximum radius at which the infalling matter is circularized is \SI{1000}{km}.
The panel (a) of Fig.~\ref{fig:T20} shows the specific angular momentum profile of this model.
From the comparison between the profile and $j_\mathrm{ISCO}$, the disk is expected to be formed when the black hole mass becomes $\approx 4M_\odot$. This mass is smaller than those for the other models shown in this paper.
In this model, we intentionally prepare a hypothetically rapidly rotating progenitor star to investigate the effect of earlier disk formation around a formed black hole on the subsequent evolution of the system.
A soft EOS, SFHo~\citep{steiner2013a}, is employed to accelerate the black hole formation.
The grid structure is similar to those of the other models, but with $\Delta x_0=\SI{150}{m}$, $\delta=0.01$, and $N=991$, which results in $L=\SI{1.0e10}{cm}$.

As in model \texttt{AD20x2}, for this model, the disk evolves in a quasi-steady manner for $\approx \SI{10}{s}$ after the vertical expansion of the disk.
Thus, the evolution process of the disk is similar to that of model \texttt{AD20x2}; the explosion also sets in at $\sim 10$\,s after the vertical expansion of the disk.
In the quasi-steady phase, however, the disk mass increases with a rate $\dot{M}_\mathrm{disk} = \dot{M}_\mathrm{fall}-\dot{M}_\mathrm{BH}\approx 0.2M_\odot$/s by the matter infall (see panel (b) of Fig.~\ref{fig:T20}), which is much higher than other models. This high-mass infall rate stems from the earlier formation of the disk and characterizes this model.

The panel (c) of Fig.~\ref{fig:T20} shows that the neutrino luminosity is slightly higher than the viscous heating rate.
This situation is possible because of the presence of an additional heating by the infalling matter, which releases its gravitational binding energy to settle into a nearly Keplerian rotation, with the rate of
\begin{align}
L_\mathrm{infall} &\sim \frac{GM_\mathrm{BH}\dot{M}_\mathrm{disk}}{2r_\mathrm{disk}} \notag\\
&\approx \SI{3e51}{erg/s}
\biggl(\frac{M_\mathrm{BH}}{6M_\odot}\biggr)\biggl(\frac{\dot{M}_\mathrm{disk}}{0.2M_\odot}\biggr)\biggl(\frac{r_\mathrm{disk}}{\SI{500}{km}}\biggr)^{-1},
\end{align}
where the infalling matter is assumed to settle into a Keplerian motion at a radius $r_\mathrm{disk}$.
Interestingly, the neutrino luminosity is approximately constant in time for the quasi-steady NDAF phase, $\SI{4}{s} \lesssim t_\mathrm{pb} \lesssim \SI{14}{s}$.

After the infalling rate $\dot{M}_\mathrm{fall}$ starts decreasing at $t_\mathrm{pb}\agt \SI{14}{s}$, the viscosity-driven outflow is launched because the viscous heating plus shock heating dominate over the neutrino cooling on the surface region of the disk.
Figure~\ref{fig:efflocal} shows the rest-mass density, local neutrino cooling efficiency, neutrino cooling rate, and viscous heating rate  before ($t_\mathrm{pb}=\SI{11.6}{s}$: left) and after ($t_\mathrm{pb}=\SI{15.4}{s}$: right) the onset of the outflow.
Here, the local neutrino cooling efficiency is defined by $q^{-}_\nu/(q^{+}_\mathrm{vis}+q^{+}_\nu)$, where $q^-_\nu$ and $q^+_\nu$ are neutrino cooling and heating rate densities and $q^{+}_\mathrm{vis}$ is the viscous heating rate density (see \citealt{fujibayashi2018a} for its definition).
The balance between the neutrino cooling and the viscous heating is established in the mid-plane region (white regions) at both phases.
For a given cylindrical radius, this region extends over $z\lesssim H$, where $H:= c_s/\Omega$ is the disk scale height evaluated at $z=0$.
The scale height is found to be $0.2$--$0.3R$, where $R$ is the cylindrical radius.
In this region, the timescale of weak interaction, $\approx \SI{0.03}{s} (k_\mathrm{B}T/\SI{2}{MeV})^{-5}$, is much shorter than the timescale of radial motion, $r/v^r\approx 1$--10\,s.
This indicates that the region may be well modeled by the NDAF solution.
On the surface region of the disk, which extends over $z\gtrsim H$, a velocity shear between the disk matter and the infalling matter is clearly present and produces a high viscous heating rate (see the bottom-right plot in each panel).
In addition, it is found that the neutrino cooling is not efficient on the disk surface region, $z\sim2H\approx0.5R$, because of the low temperature there.
This results in a launch of the outflow from the surface region of the disk once the ram pressure of the infalling matter becomes weak.

For model \texttt{T20}, the lowest value of $Y_\mathrm{e}$ for the ejecta is $\approx0.46$ (see panel (e) of Fig.~\ref{fig:T20}), which is higher than that for model \texttt{AD20x2}, although the disk is more massive and $Y_\mathrm{e}$ at the mid-plane is lower for model \texttt{T20}.
This is because, as mentioned above, the outflow is launched mainly from the surface region, which is the only place the neutrino cooling is not efficient (see Fig.~\ref{fig:efflocal}).
This results in the ejection of the matter with a weak electron degeneracy, and thus, with a high-$Y_\mathrm{e}$.

For model \texttt{T20}, a disk with mass of $\approx 2M_\odot$ still remains at the termination of our simulations and the bulk of the disk is still cooled by neutrino emission efficiently.
When the cooling efficiency drops, a stronger outflow may be launched (e.g., \citealt{Fernandez2013a,Just2015a,Fujibayashi2020a,Just2022jan}).
Suppose that 10\% of the disk matter becomes ejecta eventually, the matter with mass of $O(0.1)M_\odot$ becomes the ejecta in a later phase.
Since the outflow matter experiences a condition with temperature higher than \SI{5}{GK}, a non-negligible amount of $^{56}$Ni may be synthesized in the later-time ejecta of this model.
Also, the explosion energy at the termination of the simulation reaches $\gtrsim \SI{3e51}{erg}$, and is still increasing (see panel (g) of Fig.~\ref{fig:T20}).
Since the timescale of the energy increase appears long, we stopped the simulation by $t=30$\,s.
Thus, the ejecta mass and explosion energy which we describe below give the lower limit.
Therefore, this model would represent more energetic transients than typical supernovae such as type Ic-BL (see also explosion energy for the similar scenario found in \citealt{Just2022aug})

%% Ni
For this model, the mass of $^{56}$Ni is found to be $\approx 0.56M_\odot$, reflecting the large mass of the ejecta that experiences high temperature $\gtrsim \SI{5}{GK}$.
This amount of $^{56}$Ni is high enough for explaining a typical $^{56}$Ni mass inferred from the observations of type Ic-BL supernovae $\approx 0.4M_\odot$ (see, e.g., \citealt{Cano2017a} and \citealt{Anderson2019aug}).
These results suggest that very massive and rapidly rotating stars leading to black hole plus massive disk formation are candidates for the progenitors of type Ic-BL supernovae.
We also note that, since we stopped the simulation at a time when the ejecta mass and explosion energy are still increasing, it is the lower bound for the $^{56}$Ni mass.

\bibliography{reference}

\begin{thebibliography}{}
\expandafter\ifx\csname natexlab\endcsname\relax\def\natexlab#1{#1}\fi
\providecommand{\url}[1]{\href{#1}{#1}}
\providecommand{\dodoi}[1]{doi:~\href{http://doi.org/#1}{\nolinkurl{#1}}}
\providecommand{\doeprint}[1]{\href{http://ascl.net/#1}{\nolinkurl{http://ascl.net/#1}}}
\providecommand{\doarXiv}[1]{\href{https://arxiv.org/abs/#1}{\nolinkurl{https://arxiv.org/abs/#1}}}

\bibitem[{{Afsariardchi} {et~al.}(2021){Afsariardchi}, {Drout}, {Khatami},
  {Matzner}, {Moon}, \& {Ni}}]{Afsariardchi2021sep}
{Afsariardchi}, N., {Drout}, M.~R., {Khatami}, D.~K., {et~al.} 2021, \apj, 918,
  89, \dodoi{10.3847/1538-4357/ac0aeb}

\bibitem[{{Aguilera-Dena} {et~al.}(2020){Aguilera-Dena}, {Langer},
  {Antoniadis}, \& {M{\"u}ller}}]{Aguilera-Dena2020oct}
{Aguilera-Dena}, D.~R., {Langer}, N., {Antoniadis}, J., \& {M{\"u}ller}, B.
  2020, \apj, 901, 114, \dodoi{10.3847/1538-4357/abb138}

\bibitem[{{Aguilera-Dena} {et~al.}(2018){Aguilera-Dena}, {Langer}, {Moriya}, \&
  {Schootemeijer}}]{Aguilera-Dena2018}
{Aguilera-Dena}, D.~R., {Langer}, N., {Moriya}, T.~J., \& {Schootemeijer}, A.
  2018, \apj, 858, 115, \dodoi{10.3847/1538-4357/aabfc1}

\bibitem[{{Alcubierre} {et~al.}(2001){Alcubierre}, {Br{\"u}gmann}, {Holz},
  {Takahashi}, {Brandt}, {Seidel}, {Thornburg}, \&
  {Ashtekar}}]{Alcubierre2001a}
{Alcubierre}, M., {Br{\"u}gmann}, B., {Holz}, D., {et~al.} 2001, International
  Journal of Modern Physics D, 10, 273, \dodoi{10.1142/S0218271801000834}

\bibitem[{{Aloy} \& {Obergaulinger}(2021)}]{Aloy2021jan}
{Aloy}, M.~{\'A}., \& {Obergaulinger}, M. 2021, \mnras, 500, 4365,
  \dodoi{10.1093/mnras/staa3273}

\bibitem[{{Anderson}(2019)}]{Anderson2019aug}
{Anderson}, J.~P. 2019, \aap, 628, A7, \dodoi{10.1051/0004-6361/201935027}

\bibitem[{{Arcones} {et~al.}(2010){Arcones}, {Mart{\'\i}nez-Pinedo}, {Roberts},
  \& {Woosley}}]{Arcones2010nov}
{Arcones}, A., {Mart{\'\i}nez-Pinedo}, G., {Roberts}, L.~F., \& {Woosley},
  S.~E. 2010, \aap, 522, A25, \dodoi{10.1051/0004-6361/201014276}

\bibitem[{{Arnett}(1982)}]{Arnett1982}
{Arnett}, W.~D. 1982, \apj, 253, 785, \dodoi{10.1086/159681}

\bibitem[{{Balbus} \& {Hawley}(1991)}]{Balbus1991a}
{Balbus}, S.~A., \& {Hawley}, J.~F. 1991, \apj, 376, 214,
  \dodoi{10.1086/170270}

\bibitem[{Balbus \& Hawley(1998)}]{Balbus:1998ja}
Balbus, S.~A., \& Hawley, J.~F. 1998, Rev. Mod. Phys., 70, 1,
  \dodoi{10.1103/RevModPhys.70.1}

\bibitem[{{Banik} {et~al.}(2014){Banik}, {Hempel}, \&
  {Bandyopadhyay}}]{banik2014a}
{Banik}, S., {Hempel}, M., \& {Bandyopadhyay}, D. 2014, \apjs, 214, 22,
  \dodoi{10.1088/0067-0049/214/2/22}

\bibitem[{{Barnes} {et~al.}(2018){Barnes}, {Duffell}, {Liu}, {Modjaz},
  {Bianco}, {Kasen}, \& {MacFadyen}}]{Barnes2018jun}
{Barnes}, J., {Duffell}, P.~C., {Liu}, Y., {et~al.} 2018, \apj, 860, 38,
  \dodoi{10.3847/1538-4357/aabf84}

\bibitem[{{Baumgarte} \& {Shapiro}(1998)}]{baumgarte1998a}
{Baumgarte}, T.~W., \& {Shapiro}, S.~L. 1998, \apj, 504, 431,
  \dodoi{10.1086/306067}

\bibitem[{{Beloborodov}(2003)}]{Beloborodov2003may}
{Beloborodov}, A.~M. 2003, \apj, 588, 931, \dodoi{10.1086/374217}

\bibitem[{{Blandford} \& {Znajek}(1977)}]{Blandford1977}
{Blandford}, R.~D., \& {Znajek}, R.~L. 1977, \mnras, 179, 433,
  \dodoi{10.1093/mnras/179.3.433}

\bibitem[{{Bollig} {et~al.}(2020){Bollig}, {Yadav}, {Kresse}, {Janka},
  {Mueller}, \& {Heger}}]{Bollig2020}
{Bollig}, R., {Yadav}, N., {Kresse}, D., {et~al.} 2020, arXiv e-prints,
  arXiv:2010.10506.
\newblock \doarXiv{2010.10506}

\bibitem[{{Burrows} {et~al.}(2019){Burrows}, {Radice}, \&
  {Vartanyan}}]{Burrows2019}
{Burrows}, A., {Radice}, D., \& {Vartanyan}, D. 2019, \mnras, 485, 3153,
  \dodoi{10.1093/mnras/stz543}

\bibitem[{{Cano} {et~al.}(2017){Cano}, {Wang}, {Dai}, \& {Wu}}]{Cano2017a}
{Cano}, Z., {Wang}, S.-Q., {Dai}, Z.-G., \& {Wu}, X.-F. 2017, Advances in
  Astronomy, 2017, 8929054, \dodoi{10.1155/2017/8929054}

\bibitem[{{Chevalier} \& {Irwin}(2011)}]{Chevalier2011mar}
{Chevalier}, R.~A., \& {Irwin}, C.~M. 2011, \apjl, 729, L6,
  \dodoi{10.1088/2041-8205/729/1/L6}

\bibitem[{{Christie} {et~al.}(2019){Christie}, {Lalakos}, {Tchekhovskoy},
  {Fern{\'a}ndez}, {Foucart}, {Quataert}, \& {Kasen}}]{Christie2019dec}
{Christie}, I.~M., {Lalakos}, A., {Tchekhovskoy}, A., {et~al.} 2019, \mnras,
  490, 4811, \dodoi{10.1093/mnras/stz2552}

\bibitem[{{Colgate} {et~al.}(1997){Colgate}, {Fryer}, \& {Hand}}]{Colgate1997}
{Colgate}, S.~A., {Fryer}, C.~L., \& {Hand}, K.~P. 1997, in NATO Advanced Study
  Institute (ASI) Series C, Vol. 486, Thermonuclear Supernovae, ed.
  P.~{Ruiz-Lapuente}, R.~{Canal}, \& J.~{Isern}, 273,
  \dodoi{10.1007/978-94-011-5710-0_19}

\bibitem[{{Dessart} {et~al.}(2015){Dessart}, {Hillier}, {Woosley}, {Livne},
  {Waldman}, {Yoon}, \& {Langer}}]{Dessart2015oct}
{Dessart}, L., {Hillier}, D.~J., {Woosley}, S., {et~al.} 2015, \mnras, 453,
  2189, \dodoi{10.1093/mnras/stv1747}

\bibitem[{{Dessart} {et~al.}(2016){Dessart}, {Hillier}, {Woosley}, {Livne},
  {Waldman}, {Yoon}, \& {Langer}}]{Dessart2016may}
---. 2016, \mnras, 458, 1618, \dodoi{10.1093/mnras/stw418}

\bibitem[{Di~Matteo {et~al.}(2002)Di~Matteo, Perna, \&
  Narayan}]{DiMatteo:2002iex}
Di~Matteo, T., Perna, R., \& Narayan, R. 2002, Astrophys. J., 579, 706,
  \dodoi{10.1086/342832}

\bibitem[{{Drout} {et~al.}(2014){Drout}, {Chornock}, {Soderberg}, {Sanders},
  {McKinnon}, {Rest}, {Foley}, {Milisavljevic}, {Margutti}, {Berger},
  {Calkins}, {Fong}, {Gezari}, {Huber}, {Kankare}, {Kirshner}, {Leibler},
  {Lunnan}, {Mattila}, {Marion}, {Narayan}, {Riess}, {Roth}, {Scolnic},
  {Smartt}, {Tonry}, {Burgett}, {Chambers}, {Hodapp}, {Jedicke}, {Kaiser},
  {Magnier}, {Metcalfe}, {Morgan}, {Price}, \& {Waters}}]{Drout2014oct}
{Drout}, M.~R., {Chornock}, R., {Soderberg}, A.~M., {et~al.} 2014, \apj, 794,
  23, \dodoi{10.1088/0004-637X/794/1/23}

\bibitem[{{Eisenberg} {et~al.}(2022){Eisenberg}, {Gottlieb}, \&
  {Nakar}}]{Eisenberg2022nov}
{Eisenberg}, M., {Gottlieb}, O., \& {Nakar}, E. 2022, \mnras, 517, 582,
  \dodoi{10.1093/mnras/stac2184}

\bibitem[{{Fern{\'a}ndez} \& {Metzger}(2013)}]{Fernandez2013a}
{Fern{\'a}ndez}, R., \& {Metzger}, B.~D. 2013, \mnras, 435, 502,
  \dodoi{10.1093/mnras/stt1312}

\bibitem[{{Frank} {et~al.}(2002){Frank}, {King}, \& {Raine}}]{Frank2002}
{Frank}, J., {King}, A., \& {Raine}, D.~J. 2002, {Accretion Power in
  Astrophysics: Third Edition}

\bibitem[{{Fujibayashi} {et~al.}(2018){Fujibayashi}, {Kiuchi}, {Nishimura},
  {Sekiguchi}, \& {Shibata}}]{fujibayashi2018a}
{Fujibayashi}, S., {Kiuchi}, K., {Nishimura}, N., {Sekiguchi}, Y., \&
  {Shibata}, M. 2018, \apj, 860, 64, \dodoi{10.3847/1538-4357/aabafd}

\bibitem[{{Fujibayashi} {et~al.}(2023){Fujibayashi}, {Kiuchi}, {Wanajo},
  {Kyutoku}, {Sekiguchi}, \& {Shibata}}]{Fujibayashi2023jan}
{Fujibayashi}, S., {Kiuchi}, K., {Wanajo}, S., {et~al.} 2023, \apj, 942, 39,
  \dodoi{10.3847/1538-4357/ac9ce0}

\bibitem[{{Fujibayashi} {et~al.}(2017){Fujibayashi}, {Sekiguchi}, {Kiuchi}, \&
  {Shibata}}]{fujibayashi2017a}
{Fujibayashi}, S., {Sekiguchi}, Y., {Kiuchi}, K., \& {Shibata}, M. 2017, \apj,
  846, 114, \dodoi{10.3847/1538-4357/aa8039}

\bibitem[{{Fujibayashi} {et~al.}(2020{\natexlab{a}}){Fujibayashi}, {Shibata},
  {Wanajo}, {Kiuchi}, {Kyutoku}, \& {Sekiguchi}}]{Fujibayashi2020a}
{Fujibayashi}, S., {Shibata}, M., {Wanajo}, S., {et~al.} 2020{\natexlab{a}},
  \prd, 101, 083029, \dodoi{10.1103/PhysRevD.101.083029}

\bibitem[{{Fujibayashi} {et~al.}(2020{\natexlab{b}}){Fujibayashi}, {Shibata},
  {Wanajo}, {Kiuchi}, {Kyutoku}, \& {Sekiguchi}}]{Fujibayashi2020b}
---. 2020{\natexlab{b}}, \prd, 102, 123014, \dodoi{10.1103/PhysRevD.102.123014}

\bibitem[{{Fujibayashi} {et~al.}(2021){Fujibayashi}, {Takahashi}, {Sekiguchi},
  \& {Shibata}}]{fujibayashi2021oct}
{Fujibayashi}, S., {Takahashi}, K., {Sekiguchi}, Y., \& {Shibata}, M. 2021,
  \apj, 919, 80, \dodoi{10.3847/1538-4357/ac10cb}

\bibitem[{{Fujibayashi} {et~al.}(2020{\natexlab{c}}){Fujibayashi}, {Wanajo},
  {Kiuchi}, {Kyutoku}, {Sekiguchi}, \& {Shibata}}]{Fujibayashi2020c}
{Fujibayashi}, S., {Wanajo}, S., {Kiuchi}, K., {et~al.} 2020{\natexlab{c}},
  \apj, 901, 122, \dodoi{10.3847/1538-4357/abafc2}

\bibitem[{{Gottlieb} {et~al.}(2022{\natexlab{a}}){Gottlieb}, {Lalakos},
  {Bromberg}, {Liska}, \& {Tchekhovskoy}}]{Gottlieb2022mar}
{Gottlieb}, O., {Lalakos}, A., {Bromberg}, O., {Liska}, M., \& {Tchekhovskoy},
  A. 2022{\natexlab{a}}, \mnras, 510, 4962, \dodoi{10.1093/mnras/stab3784}

\bibitem[{{Gottlieb} {et~al.}(2022{\natexlab{b}}){Gottlieb}, {Tchekhovskoy}, \&
  {Margutti}}]{Gottlieb2022jul}
{Gottlieb}, O., {Tchekhovskoy}, A., \& {Margutti}, R. 2022{\natexlab{b}},
  \mnras, 513, 3810, \dodoi{10.1093/mnras/stac910}

\bibitem[{Hawley {et~al.}(2013)Hawley, Richers, Guan, \&
  Krolik}]{Hawley:2013lga}
Hawley, J.~F., Richers, S.~A., Guan, X., \& Krolik, J.~H. 2013, Astrophys. J.,
  772, 102, \dodoi{10.1088/0004-637X/772/2/102}

\bibitem[{{Hayakawa} \& {Maeda}(2018)}]{Hayakawa2018feb}
{Hayakawa}, T., \& {Maeda}, K. 2018, \apj, 854, 43,
  \dodoi{10.3847/1538-4357/aaa76c}

\bibitem[{{Hayashi} {et~al.}(2022){Hayashi}, {Fujibayashi}, {Kiuchi},
  {Kyutoku}, {Sekiguchi}, \& {Shibata}}]{Hayashi2022}
{Hayashi}, K., {Fujibayashi}, S., {Kiuchi}, K., {et~al.} 2022, \prd, 106,
  023008, \dodoi{10.1103/PhysRevD.106.023008}

\bibitem[{Held \& Mamatsashvili(2022)}]{Held:2022gds}
Held, L.~E., \& Mamatsashvili, G. 2022, Mon. Not. Roy. Astron. Soc., 517, 2309,
  \dodoi{10.1093/mnras/stac2656}

\bibitem[{{Hilditch} {et~al.}(2013){Hilditch}, {Bernuzzi}, {Thierfelder},
  {Cao}, {Tichy}, \& {Br{\"u}gmann}}]{Hilditch2013a}
{Hilditch}, D., {Bernuzzi}, S., {Thierfelder}, M., {et~al.} 2013, \prd, 88,
  084057, \dodoi{10.1103/PhysRevD.88.084057}

\bibitem[{{Israel} \& {Stewart}(1979)}]{Israel1979a}
{Israel}, W., \& {Stewart}, J.~M. 1979, Annals of Physics, 118, 341,
  \dodoi{10.1016/0003-4916(79)90130-1}

\bibitem[{{Janka} {et~al.}(2012){Janka}, {Hanke}, {H{\"u}depohl}, {Marek},
  {M{\"u}ller}, \& {Obergaulinger}}]{Janka2012b}
{Janka}, H.-T., {Hanke}, F., {H{\"u}depohl}, L., {et~al.} 2012, Progress of
  Theoretical and Experimental Physics, 2012, 01A309,
  \dodoi{10.1093/ptep/pts067}

\bibitem[{{Just} {et~al.}(2022{\natexlab{a}}){Just}, {Aloy}, {Obergaulinger},
  \& {Nagataki}}]{Just2022aug}
{Just}, O., {Aloy}, M.~A., {Obergaulinger}, M., \& {Nagataki}, S.
  2022{\natexlab{a}}, \apjl, 934, L30, \dodoi{10.3847/2041-8213/ac83a1}

\bibitem[{{Just} {et~al.}(2015){Just}, {Bauswein}, {Ardevol Pulpillo},
  {Goriely}, \& {Janka}}]{Just2015a}
{Just}, O., {Bauswein}, A., {Ardevol Pulpillo}, R., {Goriely}, S., \& {Janka},
  H.~T. 2015, \mnras, 448, 541, \dodoi{10.1093/mnras/stv009}

\bibitem[{{Just} {et~al.}(2022{\natexlab{b}}){Just}, {Goriely}, {Janka},
  {Nagataki}, \& {Bauswein}}]{Just2022jan}
{Just}, O., {Goriely}, S., {Janka}, H.~T., {Nagataki}, S., \& {Bauswein}, A.
  2022{\natexlab{b}}, \mnras, 509, 1377, \dodoi{10.1093/mnras/stab2861}

\bibitem[{{Karamehmetoglu} {et~al.}(2022){Karamehmetoglu}, {Sollerman},
  {Taddia}, {Barbarino}, {Feindt}, {Fremling}, {Gal-Yam}, {Kasliwal},
  {Petrushevska}, {Schulze}, {Stritzinger}, \& {Zapartas}}]{Karamehmetoglu2022}
{Karamehmetoglu}, E., {Sollerman}, J., {Taddia}, F., {et~al.} 2022, arXiv
  e-prints, arXiv:2210.09402.
\newblock \doarXiv{2210.09402}

\bibitem[{{Kawaguchi} {et~al.}(2021){Kawaguchi}, {Fujibayashi}, {Shibata},
  {Tanaka}, \& {Wanajo}}]{Kawaguchi2021jun}
{Kawaguchi}, K., {Fujibayashi}, S., {Shibata}, M., {Tanaka}, M., \& {Wanajo},
  S. 2021, \apj, 913, 100, \dodoi{10.3847/1538-4357/abf3bc}

\bibitem[{{Khatami} \& {Kasen}(2019)}]{Khatami2019jun}
{Khatami}, D.~K., \& {Kasen}, D.~N. 2019, \apj, 878, 56,
  \dodoi{10.3847/1538-4357/ab1f09}

\bibitem[{{Kiuchi} {et~al.}(2018){Kiuchi}, {Kyutoku}, {Sekiguchi}, \&
  {Shibata}}]{kiuchi2018a}
{Kiuchi}, K., {Kyutoku}, K., {Sekiguchi}, Y., \& {Shibata}, M. 2018, \prd, 97,
  124039, \dodoi{10.1103/PhysRevD.97.124039}

\bibitem[{Kohri \& Mineshige(2002)}]{Kohri:2002kz}
Kohri, K., \& Mineshige, S. 2002, Astrophys. J., 577, 311,
  \dodoi{10.1086/342166}

\bibitem[{{Kohri} {et~al.}(2005){Kohri}, {Narayan}, \& {Piran}}]{Kohri2005aug}
{Kohri}, K., {Narayan}, R., \& {Piran}, T. 2005, \apj, 629, 341,
  \dodoi{10.1086/431354}

\bibitem[{{Leung} {et~al.}(2023){Leung}, {Nomoto}, \&
  {Suzuki}}]{Leung2023arxiv}
{Leung}, S.-C., {Nomoto}, K., \& {Suzuki}, T. 2023, arXiv e-prints,
  arXiv:2304.14935, \dodoi{10.48550/arXiv.2304.14935}

\bibitem[{{Lyman} {et~al.}(2016){Lyman}, {Bersier}, {James}, {Mazzali},
  {Eldridge}, {Fraser}, \& {Pian}}]{Lyman2016mar}
{Lyman}, J.~D., {Bersier}, D., {James}, P.~A., {et~al.} 2016, \mnras, 457, 328,
  \dodoi{10.1093/mnras/stv2983}

\bibitem[{{MacFadyen} \& {Woosley}(1999)}]{Macfadyen1999}
{MacFadyen}, A.~I., \& {Woosley}, S.~E. 1999, \apj, 524, 262,
  \dodoi{10.1086/307790}

\bibitem[{{Margutti} {et~al.}(2019){Margutti}, {Metzger}, {Chornock}, {Vurm},
  {Roth}, {Grefenstette}, {Savchenko}, {Cartier}, {Steiner}, {Terreran},
  {Margalit}, {Migliori}, {Milisavljevic}, {Alexander}, {Bietenholz},
  {Blanchard}, {Bozzo}, {Brethauer}, {Chilingarian}, {Coppejans}, {Ducci},
  {Ferrigno}, {Fong}, {G{\"o}tz}, {Guidorzi}, {Hajela}, {Hurley}, {Kuulkers},
  {Laurent}, {Mereghetti}, {Nicholl}, {Patnaude}, {Ubertini}, {Banovetz},
  {Bartel}, {Berger}, {Coughlin}, {Eftekhari}, {Frederiks}, {Kozlova},
  {Laskar}, {Svinkin}, {Drout}, {MacFadyen}, \& {Paterson}}]{Margutti2019feb}
{Margutti}, R., {Metzger}, B.~D., {Chornock}, R., {et~al.} 2019, \apj, 872, 18,
  \dodoi{10.3847/1538-4357/aafa01}

\bibitem[{{Matsumoto} \& {Metzger}(2022)}]{Matsumoto2022sep}
{Matsumoto}, T., \& {Metzger}, B.~D. 2022, \apj, 936, 114,
  \dodoi{10.3847/1538-4357/ac892c}

\bibitem[{{Meyer} {et~al.}(1998){Meyer}, {Krishnan}, \&
  {Clayton}}]{Meyer1998may}
{Meyer}, B.~S., {Krishnan}, T.~D., \& {Clayton}, D.~D. 1998, \apj, 498, 808,
  \dodoi{10.1086/305562}

\bibitem[{{Meza} \& {Anderson}(2020)}]{Meza2020sep}
{Meza}, N., \& {Anderson}, J.~P. 2020, \aap, 641, A177,
  \dodoi{10.1051/0004-6361/201937113}

\bibitem[{{Miller} {et~al.}(2020){Miller}, {Sprouse}, {Fryer}, {Ryan},
  {Dolence}, {Mumpower}, \& {Surman}}]{Miller2020oct}
{Miller}, J.~M., {Sprouse}, T.~M., {Fryer}, C.~L., {et~al.} 2020, \apj, 902,
  66, \dodoi{10.3847/1538-4357/abb4e3}

\bibitem[{{Moriya} {et~al.}(2018){Moriya}, {Sorokina}, \&
  {Chevalier}}]{Moriya2018mar}
{Moriya}, T.~J., {Sorokina}, E.~I., \& {Chevalier}, R.~A. 2018, \ssr, 214, 59,
  \dodoi{10.1007/s11214-018-0493-6}

\bibitem[{{Nagataki} {et~al.}(2007){Nagataki}, {Takahashi}, {Mizuta}, \&
  {Takiwaki}}]{Nagataki2007apr}
{Nagataki}, S., {Takahashi}, R., {Mizuta}, A., \& {Takiwaki}, T. 2007, \apj,
  659, 512, \dodoi{10.1086/512057}

\bibitem[{{Nav{\'o}} {et~al.}(2022){Nav{\'o}}, {Reichert}, {Obergaulinger}, \&
  {Arcones}}]{Navo2022arxiv}
{Nav{\'o}}, G., {Reichert}, M., {Obergaulinger}, M., \& {Arcones}, A. 2022,
  arXiv e-prints, arXiv:2210.11848, \dodoi{10.48550/arXiv.2210.11848}

\bibitem[{{Obergaulinger} \& {Aloy}(2020)}]{Obergaulinger2020}
{Obergaulinger}, M., \& {Aloy}, M.~{\'A}. 2020, \mnras, 492, 4613,
  \dodoi{10.1093/mnras/staa096}

\bibitem[{{Obergaulinger} \& {Aloy}(2021)}]{Obergaulinger2021}
---. 2021, \mnras, 503, 4942, \dodoi{10.1093/mnras/stab295}

\bibitem[{{Obergaulinger} \& {Aloy}(2022)}]{Obergaulinger2022may}
---. 2022, \mnras, 512, 2489, \dodoi{10.1093/mnras/stac613}

\bibitem[{{Obergaulinger} {et~al.}(2010){Obergaulinger}, {Aloy}, \&
  {M{\"u}ller}}]{Obergaulinger2010jun}
{Obergaulinger}, M., {Aloy}, M.~A., \& {M{\"u}ller}, E. 2010, \aap, 515, A30,
  \dodoi{10.1051/0004-6361/200913386}

\bibitem[{{O'Connor} \& {Ott}(2011)}]{Oconnor2011apr}
{O'Connor}, E., \& {Ott}, C.~D. 2011, \apj, 730, 70,
  \dodoi{10.1088/0004-637X/730/2/70}

\bibitem[{{Perley} {et~al.}(2019){Perley}, {Mazzali}, {Yan}, {Cenko}, {Gezari},
  {Taggart}, {Blagorodnova}, {Fremling}, {Mockler}, {Singh}, {Tominaga},
  {Tanaka}, {Watson}, {Ahumada}, {Anupama}, {Ashall}, {Becerra}, {Bersier},
  {Bhalerao}, {Bloom}, {Butler}, {Copperwheat}, {Coughlin}, {De}, {Drake},
  {Duev}, {Frederick}, {Gonz{\'a}lez}, {Goobar}, {Heida}, {Ho}, {Horst},
  {Hung}, {Itoh}, {Jencson}, {Kasliwal}, {Kawai}, {Khanam}, {Kulkarni},
  {Kumar}, {Kumar}, {Kutyrev}, {Lee}, {Maeda}, {Mahabal}, {Murata}, {Neill},
  {Ngeow}, {Penprase}, {Pian}, {Quimby}, {Ramirez-Ruiz}, {Richer},
  {Rom{\'a}n-Z{\'u}{\~n}iga}, {Sahu}, {Srivastav}, {Socia}, {Sollerman},
  {Tachibana}, {Taddia}, {Tinyanont}, {Troja}, {Ward}, {Wee}, \&
  {Yu}}]{Perley2019mar}
{Perley}, D.~A., {Mazzali}, P.~A., {Yan}, L., {et~al.} 2019, \mnras, 484, 1031,
  \dodoi{10.1093/mnras/sty3420}

\bibitem[{{Piran} {et~al.}(2019){Piran}, {Nakar}, {Mazzali}, \&
  {Pian}}]{Piran2019feb}
{Piran}, T., {Nakar}, E., {Mazzali}, P., \& {Pian}, E. 2019, \apjl, 871, L25,
  \dodoi{10.3847/2041-8213/aaffce}

\bibitem[{{Prentice} {et~al.}(2018){Prentice}, {Maguire}, {Smartt}, {Magee},
  {Schady}, {Sim}, {Chen}, {Clark}, {Colin}, {Fulton}, {McBrien}, {O'Neill},
  {Smith}, {Ashall}, {Chambers}, {Denneau}, {Flewelling}, {Heinze}, {Holoien},
  {Huber}, {Kochanek}, {Mazzali}, {Prieto}, {Rest}, {Shappee}, {Stalder},
  {Stanek}, {Stritzinger}, {Thompson}, \& {Tonry}}]{Prentice2018sep}
{Prentice}, S.~J., {Maguire}, K., {Smartt}, S.~J., {et~al.} 2018, \apjl, 865,
  L3, \dodoi{10.3847/2041-8213/aadd90}

\bibitem[{{Pruet} {et~al.}(2003){Pruet}, {Woosley}, \&
  {Hoffman}}]{Pruet2003apr}
{Pruet}, J., {Woosley}, S.~E., \& {Hoffman}, R.~D. 2003, \apj, 586, 1254,
  \dodoi{10.1086/367957}

\bibitem[{{Rembiasz} {et~al.}(2016){Rembiasz}, {Obergaulinger},
  {Cerd{\'a}-Dur{\'a}n}, {M{\"u}ller}, \& {Aloy}}]{Rembiasz2016mar}
{Rembiasz}, T., {Obergaulinger}, M., {Cerd{\'a}-Dur{\'a}n}, P., {M{\"u}ller},
  E., \& {Aloy}, M.~A. 2016, \mnras, 456, 3782, \dodoi{10.1093/mnras/stv2917}

\bibitem[{{Rodr{\'\i}guez} {et~al.}(2022){Rodr{\'\i}guez}, {Maoz}, \&
  {Nakar}}]{Rodriguez2022}
{Rodr{\'\i}guez}, {\'O}., {Maoz}, D., \& {Nakar}, E. 2022, arXiv e-prints,
  arXiv:2209.05552.
\newblock \doarXiv{2209.05552}

\bibitem[{{Sekiguchi}(2010)}]{Sekiguchi2010a}
{Sekiguchi}, Y. 2010, Progress of Theoretical Physics, 124, 331,
  \dodoi{10.1143/PTP.124.331}

\bibitem[{{Sekiguchi} \& {Shibata}(2011)}]{Sekiguchi2011}
{Sekiguchi}, Y., \& {Shibata}, M. 2011, \apj, 737, 6,
  \dodoi{10.1088/0004-637X/737/1/6}

\bibitem[{{Shakura} \& {Sunyaev}(1973)}]{Shakura1973a}
{Shakura}, N.~I., \& {Sunyaev}, R.~A. 1973, \aap, 500, 33

\bibitem[{Shi {et~al.}(2016)Shi, Stone, \& Huang}]{Shi:2015mvh}
Shi, J.-M., Stone, J.~M., \& Huang, C.~X. 2016, Mon. Not. Roy. Astron. Soc.,
  456, 2273, \dodoi{10.1093/mnras/stv2815}

\bibitem[{{Shibata}(2000)}]{Shibata2000a}
{Shibata}, M. 2000, Progress of Theoretical Physics, 104, 325,
  \dodoi{10.1143/PTP.104.325}

\bibitem[{Shibata(2007)}]{Shibata2007}
Shibata, M. 2007, Phys. Rev. D, 76, 064035, \dodoi{10.1103/PhysRevD.76.064035}

\bibitem[{{Shibata}(2016)}]{Shibata2016a}
{Shibata}, M. 2016, {Numerical Relativity}, \dodoi{10.1142/9692}

\bibitem[{{Shibata} {et~al.}(2011){Shibata}, {Kiuchi}, {Sekiguchi}, \&
  {Suwa}}]{shibata2011a}
{Shibata}, M., {Kiuchi}, K., {Sekiguchi}, Y., \& {Suwa}, Y. 2011, Progress of
  Theoretical Physics, 125, 1255, \dodoi{10.1143/PTP.125.1255}

\bibitem[{{Shibata} {et~al.}(2017){Shibata}, {Kiuchi}, \&
  {Sekiguchi}}]{shibata2017b}
{Shibata}, M., {Kiuchi}, K., \& {Sekiguchi}, Y.-i. 2017, \prd, 95, 083005,
  \dodoi{10.1103/PhysRevD.95.083005}

\bibitem[{{Shibata} \& {Nakamura}(1995)}]{shibata1995a}
{Shibata}, M., \& {Nakamura}, T. 1995, \prd, 52, 5428,
  \dodoi{10.1103/PhysRevD.52.5428}

\bibitem[{{Siegel} {et~al.}(2019){Siegel}, {Barnes}, \&
  {Metzger}}]{Siegel2019may}
{Siegel}, D.~M., {Barnes}, J., \& {Metzger}, B.~D. 2019, \nat, 569, 241,
  \dodoi{10.1038/s41586-019-1136-0}

\bibitem[{{Steiner} {et~al.}(2013){Steiner}, {Hempel}, \&
  {Fischer}}]{steiner2013a}
{Steiner}, A.~W., {Hempel}, M., \& {Fischer}, T. 2013, \apj, 774, 17,
  \dodoi{10.1088/0004-637X/774/1/17}

\bibitem[{{Surman} {et~al.}(2006){Surman}, {McLaughlin}, \&
  {Hix}}]{Surman2006jun}
{Surman}, R., {McLaughlin}, G.~C., \& {Hix}, W.~R. 2006, \apj, 643, 1057,
  \dodoi{10.1086/501116}

\bibitem[{{Suzuki} {et~al.}(2020){Suzuki}, {Moriya}, \&
  {Takiwaki}}]{Suzuki2020aug}
{Suzuki}, A., {Moriya}, T.~J., \& {Takiwaki}, T. 2020, \apj, 899, 56,
  \dodoi{10.3847/1538-4357/aba0ba}

\bibitem[{Suzuki \& Inutsuka(2014)}]{Suzuki:2013rka}
Suzuki, T.~K., \& Inutsuka, S.-i. 2014, Astrophys. J., 784, 121,
  \dodoi{10.1088/0004-637X/784/2/121}

\bibitem[{{Taddia} {et~al.}(2018){Taddia}, {Stritzinger}, {Bersten}, {Baron},
  {Burns}, {Contreras}, {Holmbo}, {Hsiao}, {Morrell}, {Phillips}, {Sollerman},
  \& {Suntzeff}}]{Taddia2018feb}
{Taddia}, F., {Stritzinger}, M.~D., {Bersten}, M., {et~al.} 2018, \aap, 609,
  A136, \dodoi{10.1051/0004-6361/201730844}

\bibitem[{Takahashi {et~al.}(2018)Takahashi, Yoshida, \& Umeda}]{Takahashi2018}
Takahashi, K., Yoshida, T., \& Umeda, H. 2018, The Astrophysical Journal, 857,
  111, \dodoi{10.3847/1538-4357/aab95f}

\bibitem[{{Tampo} {et~al.}(2020){Tampo}, {Tanaka}, {Maeda}, {Yasuda},
  {Tominaga}, {Jiang}, {Moriya}, {Morokuma}, {Suzuki}, {Takahashi}, {Kokubo},
  \& {Kawana}}]{Tampo2020may}
{Tampo}, Y., {Tanaka}, M., {Maeda}, K., {et~al.} 2020, \apj, 894, 27,
  \dodoi{10.3847/1538-4357/ab7ccc}

\bibitem[{{Thorne}(1981)}]{Thorne1981a}
{Thorne}, K.~S. 1981, \mnras, 194, 439, \dodoi{10.1093/mnras/194.2.439}

\bibitem[{{Tominaga}(2009)}]{Tominaga2009jan}
{Tominaga}, N. 2009, \apj, 690, 526, \dodoi{10.1088/0004-637X/690/1/526}

\bibitem[{{Tominaga} {et~al.}(2007){Tominaga}, {Maeda}, {Umeda}, {Nomoto},
  {Tanaka}, {Iwamoto}, {Suzuki}, \& {Mazzali}}]{Tominaga2007}
{Tominaga}, N., {Maeda}, K., {Umeda}, H., {et~al.} 2007, \apjl, 657, L77,
  \dodoi{10.1086/513193}

\bibitem[{{Vigan{\`o}} {et~al.}(2020){Vigan{\`o}}, {Aguilera-Miret},
  {Carrasco}, {Mi{\~n}ano}, \& {Palenzuela}}]{Vigano2020jun}
{Vigan{\`o}}, D., {Aguilera-Miret}, R., {Carrasco}, F., {Mi{\~n}ano}, B., \&
  {Palenzuela}, C. 2020, \prd, 101, 123019, \dodoi{10.1103/PhysRevD.101.123019}

\bibitem[{{Wanajo} {et~al.}(2018){Wanajo}, {M{\"u}ller}, {Janka}, \&
  {Heger}}]{Wanajo2018a}
{Wanajo}, S., {M{\"u}ller}, B., {Janka}, H.-T., \& {Heger}, A. 2018, \apj, 852,
  40, \dodoi{10.3847/1538-4357/aa9d97}

\bibitem[{{Woosley}(1993)}]{Woosley1993}
{Woosley}, S.~E. 1993, \apj, 405, 273, \dodoi{10.1086/172359}

\bibitem[{{Zhang} {et~al.}(2009){Zhang}, {MacFadyen}, \& {Wang}}]{Zhang2009feb}
{Zhang}, W., {MacFadyen}, A., \& {Wang}, P. 2009, \apjl, 692, L40,
  \dodoi{10.1088/0004-637X/692/1/L40}

\end{thebibliography}
\end{document}